\documentclass[12pt,english,floatfix,nofootinbib,superscriptaddress,aps,prd,preprint]{revtex4}
\usepackage[utf8]{inputenc}       
\usepackage[english]{babel}       
\usepackage[utf8]{inputenc}
\usepackage{textcomp}
\usepackage{amsmath,amsthm,amssymb,amsfonts,mathrsfs,amsbsy} 
\usepackage{tensor}               
\usepackage{slashed}  
\usepackage{bbm}
\usepackage{esint}                
\usepackage{cancel}               
\usepackage{graphicx}             
\usepackage{natbib}
\usepackage{float}                
\usepackage{subfig}               
\usepackage[font=small,labelfont=bf]{caption} 
\usepackage{multirow}             
\usepackage{array}                
\usepackage{tikz}                 
\usetikzlibrary{quotes,angles,arrows,decorations.markings} 
\usepackage{makecell} 

\usepackage{lipsum}

\usepackage{xcolor}               
\usepackage{color}                
\usepackage{textcomp}             
\usepackage{units}                

\newcommand{\be}{\begin{equation}}
\newcommand{\ee}{\end{equation}}
\newcommand{\Be}{\begin{eqnarray}}
\newcommand{\Ee}{\end{eqnarray}}

\newcommand{\mincir}{\raise
-3.truept\hbox{\rlap{\hbox{$\sim$}}\raise4.truept\hbox{$<$}\ }}
\newcommand{\magcir}{\raise
-3.truept\hbox{\rlap{\hbox{$\sim$}}\raise4.truept\hbox{$>$}\ }}

\newcolumntype{Y}{>{\centering\arraybackslash}X}
\providecommand{\U}[1]

\usepackage[dvips]{epsfig}        
\usepackage[dvips]{graphicx}      
\usepackage{hyperref}             
\hypersetup{
    colorlinks=true,              
    breaklinks=true,              
    citecolor=blue,               
    linkcolor=[rgb]{0,0.5,0.9},   
    urlcolor=red,                 
    filecolor=green               
}

\newcommand{\ie}{\begin{equation}}
\newcommand{\fe}{\end{equation}}
\newcommand{\se}{\begin{eqnarray}}
\newcommand{\ff}{\end{eqnarray}}

\newcommand{\nl}{\text{NLED}}
\newcommand{\eff}{\text{eff}}

\begin{document}

\title{Spin effects on particle creation and evaporation in $f(R,T)$ gravity}


\author{A. A. Ara\'{u}jo Filho}
\email{dilto@fisica.ufc.br}
\affiliation{Departamento de Física, Universidade Federal da Paraíba, Caixa Postal 5008, 58051--970, João Pessoa, Paraíba,  Brazil.}
\affiliation{Departamento de Física, Universidade Federal de Campina Grande Caixa Postal 10071, 58429-900 Campina Grande, Paraíba, Brazil.}
\affiliation{Center for Theoretical Physics, Khazar University, 41 Mehseti Street, Baku, AZ-1096, Azerbaijan.}


\author{N. Heidari}
\email{heidari.n@gmail.com}

\affiliation{Departamento de Física, Universidade Federal de Campina Grande Caixa Postal 10071, 58429-900 Campina Grande, Paraíba, Brazil.}
\affiliation{Center for Theoretical Physics, Khazar University, 41 Mehseti Street, Baku, AZ-1096, Azerbaijan.}
\affiliation{School of Physics, Damghan University, Damghan, 3671641167, Iran.}


\author{Francisco S. N. Lobo} \email{fslobo@ciencias.ulisboa.pt}
\affiliation{Instituto de Astrof\'{i}sica e Ci\^{e}ncias do Espa\c{c}o, Faculdade de Ci\^{e}ncias da Universidade de Lisboa, Edifício C8, Campo Grande, P-1749-016 Lisbon, Portugal}
\affiliation{Departamento de F\'{i}sica, Faculdade de Ci\^{e}ncias da Universidade de Lisboa, Edif\'{i}cio C8, Campo Grande, P-1749-016 Lisbon, Portugal}


\date{\today}

\begin{abstract}

In this work, we study how the spin of particle modes influences particle creation, greybody factors, absorption, and evaporation of a black hole within the framework of modified electrodynamics in $f(R,T)$ gravity, recently proposed in Ref. \cite{Rois:2024iiu}. All spin sectors—scalar, vector, tensor, and spinorial—are analyzed to obtain the corresponding features. For particle creation, we consider massless bosonic and fermionic perturbations to determine the respective particle densities. Analytical expressions for the greybody factors are derived, with suitable approximations for the tensor and spinorial cases. The absorption cross section is computed numerically, and using the \textit{Stefan–Boltzmann} law, we estimate the black hole evaporation lifetime. The associated energy and particle emission rates are also discussed, along with the correspondence between quasinormal modes and greybody factors. 

\end{abstract}


\maketitle

\tableofcontents


\section{Introduction}

In modern cosmology, increasing attention has been directed toward gravitational models that deviate from standard General Relativity (GR). This interest is driven by mounting observational tensions that the $\Lambda$CDM framework struggles to accommodate, coupled with persistent conceptual challenges within GR itself. A prominent case is the discrepancy in measurements of the Hubble expansion rate: early--universe probes such as the cosmic microwave background yield values of $H_{0}$ that are inconsistent with those inferred from late--time observables like Type~Ia supernovae \cite{Hu:2023jqc,CosmoVerseNetwork:2025alb}. Similarly, the amplitude of matter clustering inferred from weak--lensing surveys suggests an $S_8$ value lower than that predicted by the standard cosmological model \cite{Joseph:2022jsf}. Recent DESI results further indicate that the dark energy component may evolve with cosmic time, challenging the assumption of a constant $\Lambda$ \cite{Cortes:2024lgw,DESI:2025zgx}.

Beyond these observational discrepancies, several theoretical issues further justify reconsidering the gravitational framework underlying cosmology. Classical GR predicts the formation of spacetime singularities—such as those associated with black holes and the Big Bang—where curvature quantities diverge and the theory loses predictive power, indicating that GR cannot be complete at all energy scales \cite{Senovilla:1998oua}. 
Additionally, while dark energy effectively accounts for the observed late--time accelerated expansion, its physical origin remains unclear. Interpreting dark energy as a cosmological constant leads to the vacuum energy problem, in which theoretical estimates of vacuum fluctuations exceed the observed value by many orders of magnitude, requiring extreme fine--tuning \cite{Frusciante:2019xia}. These issues motivate the exploration of alternative gravitational models that may naturally explain cosmic acceleration without invoking such finely tuned parameters.

A further point of theoretical concern relates to the mechanism underlying the inflationary phase of the early Universe. Although the inflationary paradigm has proven remarkably successful in accounting for the observed large--scale uniformity of the cosmos and in providing a natural origin for the spectrum of primordial density perturbations, the identity of the physical agent responsible for this accelerated expansion is still an open question. Standard formulations typically posit the existence of one or more scalar fields endowed with suitably tailored potentials. However, no definitive observational evidence has yet confirmed the presence of such fields in nature, and many inflationary constructions require a high degree of parameter tuning, including specific initial conditions and finely adjusted potential shapes, to achieve compatibility with cosmological data \cite{Vazquez:2018qdg}. These considerations have prompted interest in alternative scenarios in which inflation arises not from exotic matter components, but rather from modifications to the underlying gravitational dynamics themselves.

When viewed collectively, these issues—ranging from the challenge of interpreting dark energy, to the presence of singularities in classical GR, to the unresolved origin of inflation—highlight that the standard cosmological model may not capture the full structure of gravitational physics at all scales. This recognition motivates the formulation and systematic study of extended gravitational frameworks capable of introducing new dynamical degrees of freedom, altering the coupling between geometry and matter, or modifying the behavior of curvature at high energies. The guiding expectation is that an appropriately generalized theory could simultaneously address the empirical tensions present in current data, resolve the theoretical pathologies of GR, and provide a unified explanation for both late--time and early--time cosmic acceleration. As a result, the development and analysis of modified theories of gravity has become a central theme in modern cosmology, representing a vigorous and ongoing effort to deepen our understanding of the fundamental laws governing the evolution of spacetime.

In addition to higher--order curvature corrections and new dynamical fields, an alternative approach considers modifying how matter interacts with spacetime geometry. In standard GR, matter couples minimally to gravity, with the matter action depending solely on the metric and energy--momentum conservation following from diffeomorphism invariance. Allowing the matter Lagrangian to couple explicitly to curvature scalars or other geometric quantities gives rise to \emph{non--minimal coupling} (NMC) theories.
In NMC models, the gravitational and matter sectors are not independent prior to variation, leading to modified equations of motion that can introduce new interaction terms. These modifications may result in deviations from geodesic motion, the non--conservation of the stress--energy tensor, and emergent effective interactions, with significant cosmological and astrophysical implications. Early studies showed that such couplings can change the expansion history of the Universe and affect the growth of structure, potentially providing alternatives to $\Lambda$CDM predictions \cite{Koivisto:2005yk}. Later works further developed the formalism and explored its phenomenological consequences \cite{Bertolami:2007gv,Harko:2010mv}.

Recent studies have shown that NMC models can address fundamental cosmological challenges by modifying the effective gravitational interaction. Such theories can induce late--time acceleration without introducing a separate dark energy component, produce self--accelerating solutions, or mimic a dynamical cosmological constant. NMC interactions can also affect matter clustering, potentially alleviating tensions in structure formation, including the $S_8$ discrepancy \cite{BarrosoVarela:2024htf}. 
Comprehensive reviews classify possible coupling functions, identify conditions for theoretical consistency, and examine viability across cosmological and astrophysical scales \cite{Harko:2014gwa,Harko:2018ayt,Harko:2020ibn,Velten:2021xxw}. Depending on the form of the coupling, NMC models may modify conservation laws, alter effective gravitational constants, or introduce new propagating degrees of freedom, and many formulations can be expressed in scalar--tensor or effective fluid language, linking them to broader modified gravity frameworks.

In addition to the modifications of the gravitational action and the introduction of non--minimal couplings, another avenue through which gravity can influence cosmological dynamics involves the phenomenon of particle creation in curved spacetime. In quantum field theory on a non--flat background, the concept of a vacuum state becomes observer--dependent, since the expansion or curvature of spacetime alters the mode decomposition of quantum fields. As first demonstrated by Parker in the late 1960s, a time--varying gravitational field can spontaneously generate particles out of the vacuum, a process now known as \emph{gravitational particle creation} \cite{Parker:1968mv,Parker:1969au}. Within this framework, quantum excitations emerge due to Bogoliubov transformations connecting inequivalent vacuum states at different cosmological epochs, implying that energy can effectively flow from the gravitational field into the matter sector. This mechanism offers a natural pathway for explaining the production of matter and radiation during the early stages of the Universe, and in certain models, can act as an effective source term in the cosmological evolution equations, leading to modifications of the expansion history consistent with late--time acceleration.

A particularly significant manifestation of this effect occurs in the vicinity of black holes, as discovered by Hawking in his seminal works \cite{Hawking:1974rv,Hawking:1975vcx}. By applying quantum field theory to a stationary curved background containing an event horizon, Hawking showed that black holes emit a thermal spectrum of particles with a temperature inversely proportional to their mass. This radiation arises because quantum vacuum fluctuations near the horizon are perceived differently by observers at infinity, resulting in a net flux of particles escaping to spatial infinity. The emission of Hawking radiation endows black holes with thermodynamic properties, introducing a profound connection between quantum mechanics, gravitation, and statistical physics. The achievement that black holes possess entropy proportional to their horizon area and radiate as blackbodies fundamentally altered our understanding of gravitational systems and laid the foundation for black hole thermodynamics \cite{Gibbons:1977mu}. 

{In addition, the analysis of greybody factors, quasinormal modes, and Hawking radiation
has been extensively developed in the context of modified gravity black holes. These
studies provide complementary probes of the spacetime structure beyond purely geometric
observables, allowing one to characterize wave propagation, stability, and emission
spectra in a unified framework. Relevant investigations include, for instance,
Refs.~\cite{f1,f2,f3,f4,f5,f6,f7}, where greybody factors, quasinormal spectra, and quantum emission processes were examined in different extensions of general relativity.}

These results highlight a a fundamental correlation between particle creation in cosmological settings and black hole evaporation. Both phenomena arise from the same quantum--gravitational connection between geometry and field excitations, as established within the framework of quantum field theory in curved spacetime \cite{Birrell:1982ix,Parker:2009uva}. In cosmology, the time dependence of the metric can lead to spontaneous particle production that mimics an effective coupling between matter and curvature, conceptually similar to the interactions introduced in non--minimally coupled theories of gravity. Conversely, in the black hole context, the presence of an event horizon creates causal boundaries that induce particle emission observable at infinity. The unified treatment of these effects reveals that gravitational fields are not passive backgrounds but active participants capable of generating matter and radiation. Consequently, the study of particle creation and Hawking radiation continues to inform theoretical developments in both semiclassical gravity and modified gravity frameworks, reinforcing the notion that the interaction between matter and geometry is richer and more intricate than envisioned in classical GR.

The possibility of gravitationally induced particle production has been extensively explored within the scalar--tensor representation of $f(R,T)$ gravity \cite{Harko:2011kv}. In this framework, the coupling between the scalar curvature and the trace of the stress--energy tensor allows the gravitational field to act as a source for particle production, leading to effective energy transfer from geometry to matter \cite{Pinto:2022tlu,Harko:2014pqa,Harko:2015pma}. In Ref. \cite{Pinto:2022tlu}, the authors demonstrated that the scalar--tensor formulation of $f(R,T)$ gravity gives rise to modified cosmological dynamics, where the rate of particle production is directly determined by the evolution of the scalar field and the expansion rate of the Universe. This approach can be interpreted within the framework of irreversible thermodynamics for open systems, where the creation of particles is associated with entropy generation and a generalized second law \cite{Pinto:2023phl}. More recently, extensions of this mechanism to theories with nonminimal curvature--matter couplings have been proposed, showing that matter creation can be consistently implemented in models such as $f(R,T_{\mu\nu}T^{\mu\nu})$ gravity \cite{Katirci:2013okf}, further extending the phenomenology and offering new possibilities to address the cosmological tensions \cite{Cipriano:2023yhv}.

Motivated by the fruitful connection between quantum fields and modified gravity, this work investigates the influence of particle spin on black hole processes within the framework of modified electrodynamics in $f(R,T)$ gravity \cite{Rois:2024iiu}. We analyze all spin sectors—scalar, vector, tensor, and spinorial—to characterize their effects on particle creation, greybody factors, absorption cross sections, and black hole evaporation. The study examines massless bosonic and fermionic perturbations to determine particle densities, derives analytical expressions for greybody factors (employing suitable approximations for tensor and spinorial modes), and computes the absorption cross sections numerically. Using the Stefan--Boltzmann law, we estimate the black hole evaporation lifetime and discuss the associated energy and particle emission rates. Furthermore, the correspondence between quasinormal modes and greybody factors is explored. This investigation aims to identify general trends across all spin sectors and serves as a basis for extending such analyses, for instance, to more general black hole configurations in higher--order or non--minimally coupled gravity theories.

This article is organized in the following manner: In Section \ref{SectionII}, we introduce the black hole solution within the modified $f(R,T)$ gravity framework and discuss its general properties. Section \ref{SectionIII} is devoted to the study of particle creation, separately analyzing bosonic and fermionic modes. In Section \ref{SectionIV}, we examine the effective potential governing the propagation of these particle modes. Section \ref{sec:GBF} focuses on the derivation of greybody factors for all spin sectors, including scalar, vector, tensor, and spinorial fields. Section \ref{SectionVI} presents a detailed computation of the absorption cross sections, followed by a comparative analysis across different spins. In Section \ref{SectionVII}, we investigate the evaporation process and associated energy and particle emission rates, considering each spin sector and discussing the high--frequency regime. Section \ref{SectionVIII} explores the correspondence between quasinormal modes and greybody factors for all particle types. Finally, Section \ref{Sec:Conclusion} summarizes the main results and conclusions of the work, highlighting the implications of our findings for black hole physics in modified gravity theories.


\section{The black hole solution and the general features}\label{SectionII}

Within $f(R,T)$ gravity, where both the Ricci scalar $R$ and the trace $T$ of the stress–energy tensor modify the Einstein–Hilbert action, one obtains a static and spherically symmetric black hole solution. In this formulation, matter is described through a nonlinear variant of classical electrodynamics rather than by a standard Maxwell field. The scenario analyzed in Ref. \cite{Rois:2024iiu} assumes the functional dependence $f(R,T)=R+\beta T$, introducing a constant coupling $\beta$ that directly links the spacetime curvature with the material sector in a non-minimal way.

The electromagnetic source is modeled by a generalized Lagrangian density,
$\mathcal{L}_{\text{nl}}(F) = f_{0} + F + \alpha F^{p}$,
with $F=\tfrac{1}{4}F_{\mu\nu}F^{\mu\nu}$ representing the usual Maxwell invariant. The constants $\alpha$, $p$, and $f_{0}$ quantify the nonlinear departures from the linear theory: the term $f_{0}$ behaves effectively as a cosmological contribution, whereas $\alpha$ and $p$ control the magnitude and character of the nonlinear corrections to the electromagnetic field.

The electromagnetic field tensor $F_{\mu\nu}$ arises from the antisymmetric combination of derivatives of the gauge potential $A_{\mu}$, expressed as $F_{\mu\nu}=\partial_{\mu}A_{\nu}-\partial_{\nu}A_{\mu}$. When the system is restricted to a magnetic sector, all components vanish except one, $F_{23}=Q\sin(\alpha,\beta)$, which characterizes a monopolar magnetic field associated with a total charge $Q$. In this setup, the electromagnetic invariant reduces to $F=Q^{2}/(2r^{4})$, a relation consistent with the field equations that follow from the variation of the action with respect to the potential $A_{\gamma}$
\ie\label{eq:action}
    S[g_{\mu\nu},A_{\gamma}]=\int \sqrt{-g} \, \mathrm{d}^4x \left[f(R,T)+2\kappa^2{\cal L}_{\nl}(F)\right]\, .
\fe
The constant $\kappa$ represents the coupling coefficient that links the gravitational and matter sectors. Its value is fixed by demanding that the theory reproduces the correct Newtonian limit. From the field equations derived from the action \eqref{eq:action}, one obtains a static, spherically symmetric geometry written in Schwarzschild coordinates $(t, r, \theta, \phi)$ as
\ie
\label{balckkk}
\mathrm{d}s^{2} = - f(r)\,\mathrm{d}t^{2} + \frac{1}{f(r)}\,\mathrm{d}r^{2} + r^{2}\,(\mathrm{d}\theta^{2}+\sin^{2}\theta\,\mathrm{d}\phi^{2}),
\fe
where (we refer the reader to Ref. \cite{Rois:2024iiu} for specfiic details).
\ie
\label{generalfr}
f(r) =  1 - \frac{2M}{r} + \frac{Q^{2}}{r^{2}} - \frac{s_{\eff}}{3}r^2+\frac{2^{1-p}}{3-4p}\alpha[2 \beta(p-1)-1]Q^{2p}r^{2-4p} \,.
\fe

Within this framework, an effective cosmological cosmological contribution appears as $s_{\text{eff}} = 2(2\beta + 1)f_0$, where $M$ corresponds to the mass parameter of the black hole. The presence of the $r^{2-4p}$ term originates from the nonlinear dynamics of the electromagnetic field, controlled by the coupling constant $\alpha$. The parameter $\beta$, stemming from the non-minimal interaction between matter and curvature, simultaneously influences the strength of these nonlinear electromagnetic effects and reshapes the (effective) cosmological cosmological parameter.

The exponent $p$ dictates how strongly the electromagnetic sector deviates from linearity and modifies the radial decay of the magnetic charge term. For $p=1$, the solution reduces to the Reissner--Nordström configuration, with nonlinearities effectively incorporated into a redefined charge $Q_{\text{eff}}^{2}=Q^{2}(1+\alpha)$. Conversely, for $p>1$, the electromagnetic corrections diminish more rapidly with increasing $r$, as expected for higher-order nonlinear terms.

By fixing $p=2$ in Eq. (\ref{generalfr}) and omitting the cosmological constant contribution, one obtains the simplified form reported in Ref. \cite{Rois:2024iiu}
\ie
\label{metricfuntion}
f(r) = 1 - \frac{2M}{r} + \frac{Q^{2}}{r^{2}} - \frac{\alpha(2 \beta -1)Q^{4}}{10 r^{6}},
\fe
which corresponds to the same configuration recently adopted in the literature to investigate quasinormal modes, time-domain evolution, geodesic motion, shadow profiles, gravitational lensing, and topological characteristics \cite{heidari2025gravitational}.

Before discussing the particle creation and evaporation mechanisms, it becomes necessary to determine the location of the event horizon. This requires analyzing the metric function $f(r)$, whose structure reveals six distinct roots. However, only three of these are physically meaningful, corresponding to real and positive values of $r$. Consequently, a single relevant root defines the event horizon, denoted by $r_h$, and its explicit expression is given by \cite{heidari2025gravitational}
\ie
\label{eventhhh}
r_{h} = \left(M + \sqrt{M^2-Q^2}\right) + \frac{\alpha  \left(2 \beta  Q^4-Q^4\right)}{20 \left(M + \sqrt{M^2-Q^2}\right)^3 \left(M^2-Q^2 + M \sqrt{M^2-Q^2}\right)}.
\fe
In the present analysis, the parameters $\alpha$ and $\beta$ are assumed to take small values, as one should naturally expect. When both tend to zero, the metric smoothly reduces to the familiar Reissner--Nordström configuration. More specifically, setting $\alpha \to 0$ restores the standard Maxwell Lagrangian, whose associated stress--energy tensor is traceless. As in the conventional charged black hole scenario, consistency demands a constraint among the parameters $M$, $Q$, $\alpha$, and $\beta$ to guarantee a positive and real event horizon radius, $r_h$. In this manner, we have
\ie
{2 \alpha \beta  Q^4-\alpha\,Q^4 >0,}  \quad \text{and} \quad  M > Q.
\fe


\section{Particle creation}\label{SectionIII}

\subsection{Bosonic particle modes}

The present work explores the influence of the parameters $(\alpha, \beta)$ associated with non–commutative effects on the process of Hawking radiation. The investigation draws its conceptual foundation from Hawking’s pioneering study \cite{hawking1975particle}, which addressed the quantum dynamics of scalar fields in curved backgrounds. In that framework, the field equation was analyzed by solving the corresponding wave equation and formulating the scalar mode functions as
\ie
\frac{1}{\sqrt{-g_{(\alpha,\beta)}}}\partial_{\mu}\Big[g_{(\alpha,\beta)}^{\mu\nu}\sqrt{-g_{(\alpha,\beta)}}\partial_{\nu}\Phi\Big] = 0.
\fe
Throughout this work, the notation $(\alpha,\beta)$ will be employed to refer to the modified electrodynamics formulated within the $f(R,T)$ gravitational framework. In this context, $g_{(\alpha,\beta)}^{\mu\nu}$ designates the inverse metric components, $g_{(\alpha,\beta)}$ stands for the metric determinant, and $\Phi$ denotes the scalar field under consideration. Accordingly, the associated field operator takes the following form:
\se
\Phi &=& \sum_{i} \left (f^{(\alpha,\beta)}_i  a^{(\alpha,\beta)}_{i} + \bar{f}^{(\alpha,\beta)}_{i}  a^{{(\alpha,\beta)}\dagger}_{i} \right) 
	\nonumber \\
&=& \sum_{i} \left( p^{(\alpha,\beta)}_{i}  b^{(\alpha,\beta)}_{i} + \bar{p}^{(\alpha,\beta)}_{i}  b^{{(\alpha,\beta)}\dagger}_{i} + q^{(\alpha,\beta)}_{i}  c^{(\alpha,\beta)}_{i} + \bar{q}^{(\alpha,\beta)}_{i}  c^{{(\alpha,\beta)}\dagger}_{i} \right ) .
\ff

Within this formulation, $f^{(\alpha,\beta)}_{i}$ and $\bar{f}^{(\alpha,\beta)}_{i}$ (the latter being their complex conjugates) represent modes propagating entirely toward the black hole, while $p^{(\alpha,\beta)}_{i}$ and $\bar{p}^{(\alpha,\beta)}_{i}$ describe purely outgoing components. In contrast, $q^{(\alpha,\beta)}_{i}$ and $\bar{q}^{(\alpha,\beta)}_{i}$ correspond to modes without any outward flux. The operators $a^{(\alpha,\beta)}_{i}$, $b^{(\alpha,\beta)}_{i}$, and $c^{(\alpha,\beta)}_{i}$ act as annihilation operators, and their Hermitian conjugates $a^{(\alpha,\beta)\dagger}_{i}$, $b^{(\alpha,\beta)\dagger}_{i}$, and $c^{(\alpha,\beta)\dagger}_{i}$ play the role of creation operators. The purpose of this analysis is to explore how the non–commutative deformation characterized by $(\alpha,\beta)$ modifies these mode structures, thereby revealing the deviations it induces in Hawking’s original radiation framework.

Because the spacetime geometry under consideration possesses spherical symmetry, the scalar perturbations can be conveniently decomposed in terms of spherical harmonics. In the exterior region of the black hole, the corresponding expressions for the ingoing and outgoing wave modes can thus be written as \cite{calmet2023quantum,heidari2024quantum}:
\begin{eqnarray}
f^{(\alpha,\beta)}_{\omega^\prime \ell m} &=& \frac{1}{\sqrt{2 \pi \omega^\prime} r }  \mathcal{F}_{\omega^{\prime}}^{(\alpha,\beta)}(r) e^{i \omega^\prime v^{(\alpha,\beta)}} Y_{\ell m}(\theta,\phi)\ , \\ 
p^{(\alpha,\beta)}_{\omega \ell m} &=& \frac{1}{\sqrt{2 \pi \omega} r }  \mathcal{P}^{(\alpha,\beta)}_\omega(r) e^{i \omega u^{(\alpha,\beta)}} Y_{\ell m}(\theta,\phi) \,. 
\end{eqnarray}
Within this representation, the radial dependence is encoded in the functions $\mathcal{F}_{\omega^{\prime}}^{(\alpha,\beta)}(r)$ and $\mathcal{P}_{\omega}^{(\alpha,\beta)}(r)$, whereas the angular behavior is described by the spherical harmonics $Y_{\ell m}(\theta,\phi)$. The variables $v^{(\alpha,\beta)}$ and $u^{(\alpha,\beta)}$ stand for the advanced and retarded null coordinates, respectively, and are introduced through the relations
$v^{(\alpha,\beta)} = t + r^{}$ and $u^{(\alpha,\beta)} = t - r^{}$,
with $r^{*}$ being the tortoise coordinate associated with the radial geometry.

Having established these definitions, the next step is to determine how non–commutative effects modify the coordinate structure. This can be effectively examined by studying the motion of a test particle following a geodesic in the deformed spacetime, whose path is described in terms of an affine parameter $s$. In this formulation, the corresponding four–momentum of the particle is given by
\ie
p^{(\alpha,\beta)}_{\mu} = g^{(\alpha,\beta)}_{\mu\nu}\frac{\mathrm{d}x}{\mathrm{d}s}^\nu.
\fe
Along the geodesic path, the particle’s momentum is preserved due to the underlying spacetime symmetries. Consequently, its explicit form can be expressed as
\ie
\mathcal{L} = g^{(\alpha,\beta)}_{\mu\nu} \frac{\mathrm{d}x^\mu}{\mathrm{d}s} \frac{\mathrm{d}x^\nu}{\mathrm{d}s}.
\fe
This quantity remains invariant along geodesic motion. For timelike trajectories, one adopts $\mathcal{L} = -1$ and identifies the affine parameter with the proper time, $s = \tau$. In the case of null trajectories—which constitute the central interest here—the condition $\mathcal{L} = 0$ applies, with $s$ serving as a generic affine parameter.

By considering a static, spherically symmetric geometry and restricting attention to purely radial motion in the equatorial plane $(\theta = \pi/2)$, where the angular momentum component $p^{(\alpha,\beta)}_\varphi = L = 0$, the corresponding dynamical relations governing the geodesic evolution can then be established as
\ie
E =  f(r) \dot{t}.
\fe
Within this description, the conserved energy of the particle is introduced as $E = -p^{(\alpha,\beta)}_{t}$, and differentiation along the affine parameter $s$ is indicated by an overdot, meaning \, $\dot{} \equiv \mathrm{d}/\mathrm{d}s$. From these definitions, one can derive an auxiliary relation that takes the form
\ie
\label{1d2r3d4s}
    \left( \frac{\mathrm{d}r}{\mathrm{d}s} \right)^2 = \frac{E^2}{f(r)f(r)^{-1}},
\fe
and after performing a sequence of algebraic manipulations, the resulting expression can be written as
\ie
\label{1c2o3n4ccc}
    \frac{\mathrm{d}}{\mathrm{d}s}\left(t\mp r^{*}\right) = 0,
\fe
where the corresponding tortoise coordinate $r^{*}$ is introduced through the definition
\ie
\begin{split}
\label{1t2o3r4t5o6ise}
\mathrm{d}r^{*} & = \frac{\mathrm{d}r}{f(r)},
\end{split}
\fe
where, it has been explicitly shown in Ref. \cite{heidari2025gravitational}.

It is worth emphasizing that Eq. (\ref{1c2o3n4ccc}) naturally leads to two conserved null coordinates, $v^{(\alpha,\beta)}$ and $u^{(\alpha,\beta)}$. By rearranging the expression associated with the retarded coordinate, one arrives at an equivalent representation given by
\ie
\label{ghjghk}
\frac{\mathrm{d}u^{(\alpha,\beta)}}{\mathrm{d}s}=\frac{2E}{f(r)}.
\fe

Notice that, for an ingoing null trajectory parameterized by the affine variable $s$, the advanced coordinate $u^{(\alpha,\beta)}$ is considered a function of this parameter, expressed as $u^{(\alpha,\beta)}(s)$. Determining its functional form involves two essential procedures: first, rewriting the radial coordinate $r$ in terms of $s$, and then performing the integration prescribed in Eq. \eqref{ghjghk}. The resulting dependence of $u^{(\alpha,\beta)}$ on $s$ directly affects the computation of the Bogoliubov coefficients, which quantify the spectrum of Hawking radiation.

In order to execute this derivation, the metric function $f(r)$ is employed to integrate the square--root term appearing in Eq. \eqref{1d2r3d4s}. The integration extends over $\tilde{r} \in [r_h, r]$, while the affine parameter simultaneously varies within $\tilde{s} \in [0, s]$. Carrying out this procedure ultimately leads to the expression written as
\se
r(Q,\alpha,\beta,s) &=&  \left(M + \sqrt{M^2-Q^2}\right)  - E s 
	\nonumber \\
	 && + \frac{\alpha  \left(2 \beta  Q^4-Q^4\right)}{20 \left(M + \sqrt{M^2-Q^2}\right)^3 \left(M^2-Q^2 + M \sqrt{M^2-Q^2}\right)}\,.
\ff
It should be emphasized that, in obtaining this expression, the negative branch of the square root in Eq. \eqref{1d2r3d4s} was chosen to remain consistent with the direction of an ingoing geodesic.

Advancing the analysis, the relation $r(s,\alpha,\beta)$ is then used to evaluate the integral {and we have expanded $1/f(r)$ from Eq.~(\ref{ghjghk}) in order to obtain a more tractable expression, as we shall be seeing below}. Considering the near--horizon regime \cite{parker2009quantum}, one arrives at the following expression :
\ie
\begin{split}
u_{\text{vicinity}}^{(\alpha,\beta)}(Q,s) & \approx \, -4 M \ln \left[\frac{s}{C}\right] -\frac{Q^4 \ln \left[\frac{s}{C}\right]}{4 M^3} + \frac{\alpha  \beta  Q^4 \ln \left[\frac{s}{C}\right]}{20 M^5}-\frac{\alpha  Q^4 \ln \left[\frac{s}{C}\right]}{40 M^5}.
\end{split}
\fe
Here, $C$ denotes the constant of integration that arises from the previous step. Furthermore, when the analysis is extended to the asymptotic domain, far from the vicinity of the event horizon \cite{parker2009quantum}, the expression assumes the following form:
\ie
\begin{split}
u^{(\alpha,\beta)}_{far}(Q,s)  \approx & \; 2 E s  +\frac{\alpha  (2 \beta -1) Q^4 \left(9 E^2 s ^2+52 M^2-42 E s  M\right)}{240 M^4 (2 M-E s )^3} \\
& +\frac{Q^4 \left(-2 \alpha  \beta +\alpha +10 M^2\right) \ln [2 M-E s]}{40 M^5}.
\end{split}
\fe
In studying the mechanism of particle creation, the analysis is restricted to the near–horizon form $u_{\text{vicinity}}^{(\alpha,\beta)}(s)$, consistent with the conventional treatment found in Refs. \cite{parker2009quantum,calmet2023quantum,araujo2025does,araujo2025particle}. Moreover, the link between the ingoing and outgoing null coordinates emerges naturally from geometric–optics considerations. This correspondence is parameterized by $s$, which obeys
\[
s = \frac{v^{(\alpha,\beta)}_{0} - v^{(\alpha,\beta)}}{D},
\]
where $v^{(\alpha,\beta)}_{0}$ denotes the value of the advanced coordinate at the reflection point on the horizon ($s=0$), and $D$ is a proportionality constant \cite{calmet2023quantum}.

With these preliminary elements established, the analysis now turns to the derivation of the outgoing modes that satisfy the modified Klein–Gordon equation, incorporating the influence of the parameters $(\alpha,\beta)$. The resulting expressions can therefore be written as
\ie
p^{(\alpha,\beta)}_{\omega} =\int_0^\infty \left ( \alpha^{(\alpha,\beta)}_{\omega\omega^\prime} f^{(\alpha,\beta)}_{\omega^\prime} + \beta^{(\alpha,\beta)}_{\omega\omega^\prime} \bar{ f}^{(\alpha,\beta)}_{\omega^\prime}  \right)\mathrm{d} \omega^\prime,
\fe
where the quantities $\alpha^{(\alpha,\beta)}_{\omega\omega^\prime}$ and $\beta^{(\alpha,\beta)}_{\omega\omega^\prime}$ denote the Bogoliubov coefficients \cite{parker2009quantum,hollands2015quantum,wald1994quantum,fulling1989aspects}, responsible for describing the correlation between mode conversion and particle production arising from the effects of nonlinear electrodynamics
\begin{equation}
\begin{split}
\alpha^{(\alpha,\beta)}_{\omega\omega^\prime} = & -i N e^{i\omega^\prime v^{(\alpha,\beta)}_{0}}e^{\pi \left\{ 2 M  +\frac{Q^4 }{2 M^3} - \frac{\alpha  \beta  Q^4 }{10 M^5}+\frac{\alpha  Q^4 }{20 M^5} \right\} \omega} \\
& \times \int_{-\infty}^{0} \,\mathrm{d}x\,\Big(\frac{\omega^\prime}{\omega}\Big)^{1/2}e^{\omega^\prime x}  \, e^{i\omega\left\{ 4 M \ln \left[\frac{|x|}{C D}\right] +\frac{Q^4 \ln \left[\frac{|x|}{CD}\right]}{4 M^3} - \frac{\alpha  \beta  Q^4 \ln \left[\frac{|x|}{CD}\right]}{20 M^5} + \frac{\alpha  Q^4 \ln \left[\frac{|x|}{CD}\right]}{40 M^5}\right\}},
    \end{split}
\end{equation}
and
\begin{equation}
\begin{split}
\beta^{(\alpha,\beta)}_{\omega\omega^\prime} = & \, \, i N e^{-i\omega^\prime v^{(\alpha,\beta)}_{0}}e^{-\pi \left\{ 2 M  +\frac{Q^4 }{2 M^3} - \frac{\alpha  \beta  Q^4 }{10 M^5}+\frac{\alpha  Q^4 }{20 M^5} \right\} \omega} \\
& \times \int_{-\infty}^{0} \,\mathrm{d}x\,\Big(\frac{\omega^\prime}{\omega}\Big)^{1/2}e^{\omega^\prime x}  \, e^{i\omega\left\{ 4 M \ln \left[\frac{|x|}{C D}\right] +\frac{Q^4 \ln \left[\frac{|x|}{CD}\right]}{4 M^3} - \frac{\alpha  \beta  Q^4 \ln \left[\frac{|x|}{CD}\right]}{20 M^5} + \frac{\alpha  Q^4 \ln \left[\frac{|x|}{CD}\right]}{40 M^5}\right\}}.
    \end{split}
\end{equation}
In these expressions, $N$ denotes the normalization constant. The obtained relation reveals that the amplitude associated with particle emission is affected by the non--linear electrodynamics, since the parameters $(\alpha,\beta)$ alter the spacetime geometry in a manner consistent with earlier discussions.

Remarkably, even though such non--linear corrections (due to $\alpha$ and $\beta$) modify the amplitude, the resulting spectrum at this level continues to exhibit a blackbody profile. To verify this property, one proceeds to evaluate the following relation:
\ie
    \big|\alpha^{(\alpha,\beta)}_{\omega\omega'}\big|^2 = e^{  \left( 8 M  +\frac{Q^4 }{2 M^3} - \frac{\alpha  \beta  Q^4 }{10 M^5} + \frac{\alpha  Q^4 }{20 M^5}\right) \omega}\big|\beta^{(\alpha,\beta)}_{\omega\omega'}\big|^2\,.
\fe
By examining the particle emission rate in the frequency range between $\omega$ and $\omega + \mathrm{d}\omega$ \cite{o10}, one arrives at the following formulation:
\ie
    \mathcal{P}(\omega, \alpha,\beta)=\frac{\mathrm{d}\omega}{2\pi}\frac{1}{\left \lvert\frac{\alpha^{(\alpha,\beta)}_{\omega\omega^\prime}}{\beta^{(\alpha,\beta)}_{\omega\omega^\prime}}\right \rvert^2-1}\, ,
\fe
or, equivalently, the expression can be written as
\ie
    \mathcal{P}(\omega, \alpha,\beta) = \frac{\mathrm{d}\omega}{2\pi}\frac{1}{e^{ \left( 8 M  +\frac{Q^4 }{2 M^3} - \frac{\alpha  \beta  Q^4 }{10 M^5} + \frac{\alpha  Q^4 }{20 M^5}\right) \omega}-1}\,.
\fe

A noteworthy observation arises when the derived relation is contrasted with the Planck distribution: a clear correspondence emerges, enabling the radiation spectrum to be interpreted as that of a thermal emitter
\ie
    \mathcal{P}(\omega, \alpha,\beta)=\frac{\mathrm{d}\omega}{2\pi}\frac{1}{e^{\frac{\omega}{T^{(\alpha,\beta)}}}-1},
\fe
so that that Hawking temperature can be matched
\ie
\begin{split}
\label{hawtemp1}
T^{(\alpha,\beta)} & =  \, \frac{1}{8 M} -\frac{Q^4}{128 M^5} + \frac{\alpha  \beta  Q^4}{640 M^7}-\frac{\alpha  Q^4}{1280 M^7}.
\end{split}
\fe
It is important to notice that this outcome coincides precisely with the result recently reported in the literature through the surface–gravity approach \cite{heidari2025gravitational}, thereby reinforcing the internal consistency of the theoretical framework.

It should also be noted that, up to this point, the conservation of energy for the complete system has not been explicitly incorporated. As the black hole emits radiation, its total mass gradually decreases, leading to a continuous shrinkage of the horizon. To account for this backreaction effect, the tunneling framework developed by Parikh and Wilczek \cite{011} is employed in the following analysis. The procedure follows the formulation presented in Refs. \cite{011,vanzo2011tunnelling,parikh2004energy,calmet2023quantum}. When the metric is rewritten in Painlevé--Gullstrand coordinates, it assumes the form:
\ie
\mathrm{d}s^2 = - f(r)\mathrm{d}t^2 + 2 h(r) \mathrm{d}t \mathrm{d}r + \mathrm{d}r^2 + r^2\mathrm{d}\Omega^2,
\fe
where the auxiliary function is defined as $h(r) = \sqrt{f(r)\,[f(r)^{-1} - 1]}$. In explicit form, it can be written as
\ie
h(r) = \sqrt{\frac{2 M}{r}+\frac{\alpha  (2 \beta -1) Q^4}{10 r^6}-\frac{Q^2}{r^2}}.
\fe
Furthermore, the quantum tunneling probability is determined by the imaginary part of the action \cite{araujo2025non,parikh2004energy,calmet2023quantum,heidari2025nonas,vanzo2011tunnelling,araujo2025particle,araujo2025does,araujo2025particle2}.

The motion of a test particle along a geodesic in a curved spacetime background is described through the corresponding action functional
\ie
\mathcal{S}_{(\alpha,\beta)} = \int p^{(\alpha,\beta)}_\mu \, \mathrm{d}x^\mu.
\fe
To isolate the imaginary part of the action, one examines the following relation
\ie 
p^{(\alpha,\beta)}_\mu \mathrm{d}x^\mu = p^{(\alpha,\beta)}_t \mathrm{d}t + p^{(\alpha,\beta)}_r \mathrm{d}r.
\fe
Because the temporal term $p^{(\alpha,\beta)}_t\,\mathrm{d}t = -\omega\,\mathrm{d}t$ is purely real, it has no effect on the imaginary contribution to the action. Consequently, the radial term provides the sole contribution, yielding
\ie
\text{Im}\,\mathcal{S}_{(\alpha,\beta)} = \text{Im}\,\int_{r_i}^{r_f} \,p^{(\alpha,\beta)}_r\,\mathrm{d}r = \text{Im}\,\int_{r_i}^{r_f}\int_{0}^{p^{(\alpha,\beta)}_r} \,\mathrm{d}p^{(\alpha,\beta) '}_r\,\mathrm{d}r.
\fe

Using Hamilton’s equations for a system whose Hamiltonian is defined as $H = M - \omega'$, one obtains the differential relation $\mathrm{d}H = -\mathrm{d}\omega'$. Here, the variable $\omega'$ denotes the energy carried by the emitted particle, restricted to the interval $0 \leq \omega' \leq \omega$. From these considerations, the subsequent expression can be established as
\ie
\begin{split}
\text{Im}\, \mathcal{S}_{(\alpha,\beta)} & = \text{Im}\,\int_{r_i}^{r_f}\int_{M}^{M-\omega} \,\frac{\mathrm{d}H}{\mathrm{d}r/\mathrm{d}t}\,\mathrm{d} r  =\text{Im}\,\int_{r_i}^{r_f}\,\mathrm{d}r\int_{0}^{\omega} \,-\frac{\mathrm{d}\omega'}{\mathrm{d}r/\mathrm{d}t}\,.
\end{split}
\fe
By changing the order of integration and performing a suitable variable substitution, the relation can be obtained in the form
\ie
    \frac{\mathrm{d}r}{\mathrm{d}t} = - h(r)+\sqrt{f(r) + h(r)^2}=1-\sqrt{\frac{\Delta(r,\alpha,\beta,\omega^\prime)}{r}}, 
    \fe
so that we may write 
\ie
\Delta(r,\alpha,\beta,\omega^\prime) = 2 M+\frac{\alpha  (2 \beta -1) Q^4}{10 r^5}-\frac{Q^2}{r}.
\fe  
Such a definition is taken into account for the sake of simplifying the evaluation of the integral. Within this framework, the resulting expression takes the form
\ie
\label{ims}
\text{Im}\, \mathcal{S}^{(\alpha,\beta)} =\text{Im}\,\int_{0}^{\omega} -\mathrm{d}\omega'\int_{r_i}^{r_f}\,\frac{\mathrm{d}r}{1-\sqrt{\frac{\Delta(r,\,\alpha,\beta, \,\omega^\prime)}{r}}}.
\fe

Considering that $\alpha$ and $\beta$ are treated as very small quantities, the integrand from the preceding relation can be rewritten as
\se
	\label{ims22}
	\text{Im}\, \mathcal{S}^{(\alpha,\beta)} & \approx & \,  \text{Im}\,\int_{0}^{\omega} -\mathrm{d}\omega' 
	 \int_{r_i}^{r_f}\,\mathrm{d}r \left\{ \frac{1}{1-\sqrt{2} \sqrt{\frac{M-\omega '}{r}}} -\frac{Q^2 \sqrt{\frac{M-\omega'}{r}}}{2 \sqrt{2} r (M-\omega') \left(\sqrt{2} \sqrt{\frac{M-\omega'}{r}}-1\right)^2} \right. 
		\nonumber	\\
	&& \hspace{-1.75cm} \left. +\frac{Q^4 \left(\sqrt{2} r \sqrt{\frac{M-\omega'}{r}}-6 M+6 \omega'\right)}{32 r^3 (M-\omega')^2 \left(\sqrt{2} \sqrt{\frac{M-\omega'}{r}}-1\right)^3}  +  \left[  \frac{(2 \beta -1) Q^4 \sqrt{\frac{M-\omega'}{r}}}{20 \sqrt{2} r^5 (M-\omega') \left(\sqrt{2} \sqrt{\frac{M-\omega'}{r}}-1\right)^2}   \right] \alpha   \right\}.
\ff

A remarkable aspect arises when $M$ is replaced by $(M - \omega')$, as this substitution relocates the pole of the integrand to $2(M - \omega')$. Evaluating the contour integral about this pole in the counterclockwise sense then gives the following outcome
\ie
    \text{Im}\, \mathcal{S}^{(\alpha,\beta)}  = \, 4 \pi  \omega  \left( M - \frac{\omega}{2} \right) -\frac{\pi  Q^4}{8 M^2}+ \frac{\pi  \alpha  (2 \beta -1) Q^4}{160 M^4}+\frac{\pi  Q^4 \left[-2 \alpha  \beta +\alpha +20 (M-\omega )^2\right]}{160 (M-\omega )^4}  .
\fe
As shown in Ref. \cite{vanzo2011tunnelling}, the probability associated with particle emission—now incorporating the influence of non--linear electrodynamics —can be expressed as
\ie
\Gamma \sim e^{-2 \, \text{Im}\, S^{(\alpha,\beta)}} = e^{- 8 \pi  \omega  \left( M - \frac{\omega}{2} \right) +\frac{\pi  Q^4}{4 M^2} - \frac{\pi  \alpha  (2 \beta -1) Q^4}{80 M^4} - \frac{\pi  Q^4 \left[-2 \alpha  \beta +\alpha +20 (M-\omega )^2\right]}{80 (M-\omega )^4} } .
\fe

{It is worth mentioning that above equation is meaningful only within the semiclassical non--extremal regime, where $M-\omega$ remains sufficiently large compared with the charge scale and possible quantum--gravity corrections. Therefore, the apparent divergence as $\omega\to M$ should be understood as a breakdown of the approximation rather than as a physical singularity of the emission probability.}

When the energy of the emitted quanta approaches zero $(\omega \to 0)$, the spectrum smoothly reduces to the conventional Planck distribution obtained in Hawking’s seminal analysis. Accordingly, the radiation spectrum can be written as
\ie
    \mathcal{P}(\omega,\alpha,\beta,Q)=\frac{\mathrm{d}\omega}{2\pi}\frac{1}{e^{8 \pi  \omega  \left( M - \frac{\omega}{2} \right) - \frac{\pi  Q^4}{4 M^2} + \frac{\pi  \alpha  (2 \beta -1) Q^4}{80 M^4} + \frac{\pi  Q^4 \left[-2 \alpha  \beta +\alpha +20 (M-\omega )^2\right]}{80 (M-\omega )^4} }-1}.
\fe

The parameters $\omega$ and $(\alpha,\beta)$ generate measurable departures from the conventional blackbody radiation profile, as becomes evident upon inspection of the emission spectrum. For sufficiently small $\omega$, the distribution preserves a Planck–type structure, though characterized by a corrected Hawking temperature. Furthermore, the particle number density may be rewritten in terms of the tunneling probability as
\ie
\begin{split}
& n(\omega,\alpha,\beta,Q)  = \frac{\Gamma}{1 - \Gamma} \\
& = \frac{1}{\exp \left\{\frac{1}{80} \pi  \left[\frac{\alpha  (2 \beta -1) Q^4}{M^4}-\frac{20 Q^4}{M^2}+\frac{Q^4 \left(-2 \alpha  \beta +\alpha +20 (M-\omega )^2\right)}{(M-\omega )^4}+320 \omega  (2 M-\omega )\right]\right\}-1}.
\end{split}
\fe

To analyze the behavior of $n(\omega,\alpha,\beta,Q)$, the results are summarized in Table~\ref{tab1} rather than through graphical representations, unlike what will be done for the fermionic sector. This approach is adopted because, in the present configuration, the changes in $n$ are too subtle to be visually distinguished in a plot, whereas in the fermionic case such variations are considerably more evident. Even so, the table highlights two main tendencies: as the charge $Q$ increases, $n(\omega,\alpha,\beta,Q)$ decreases, whereas reducing the parameters $\alpha = \beta$ enhances the particle density.

\begin{table}[!h]
\begin{center}
\begin{tabular}{c c c|| c c c } 
 \hline\hline \hline
 $Q$ & $\alpha=\beta$ &  $n(\omega,\alpha,\beta,Q)$ & $Q$ & $\alpha=\beta$ &  $n(\omega,\alpha,\beta,Q)$  \\ [0.2ex] 
 \hline
 0.60  & -0.01 & 0.0985200 & 0.99  & -0.01 & 0.0833849  \\ 

 0.70  & -0.01 & 0.0963459 & 0.99  & -0.02 & 0.0834038   \\
 
 0.80  & -0.01 & 0.0931116 & 0.99  & -0.03 & 0.0834235  \\
 
 0.90  & -0.01 & 0.0886164  & 0.99  & -0.04 & 0.0834438  \\
 
 0.99  & -0.01 & 0.0833849 & 0.99  & -0.05 & 0.0834649   \\ 
 [0.2ex] 
 \hline \hline \hline
\end{tabular}
\caption{\label{tab1} The numerical values corresponding to $n(\omega,\alpha,\beta,Q)$ are presented. On the left panel, the magnetic charge $Q$ is varied with fixed parameters $\alpha = \beta$, while on the right panel, $Q$ remains constant and the values of $\alpha = \beta$ are modified. Here, we consider $M=1$ and $\omega =0.1$.
}
\end{center}
\end{table}


\subsection{Fermionic sector}

Black holes emit thermal radiation due to their intrinsic temperature, producing spectra analogous to blackbody emission though typically without accounting for greybody corrections. Such radiation encompasses particles of different spins, including fermionic ones. The pioneering investigation of Kerner and Mann \cite{o69}, complemented by several later analyses \cite{o70,o72,o73,o75,o71,o74}, demonstrated that massless bosons and fermions radiate at a common temperature. Further research on spin--1 bosons confirmed that the Hawking temperature remains unchanged even after higher–order quantum effects are incorporated \cite{o76,o77}.

For fermionic fields, the associated action is linked to the phase of the spinor solution that satisfies the Hamilton--Jacobi equation. Alternative formulations for this action have been explored in several works \cite{o84,o83,vanzo2011tunnelling}. The interaction between the spin of the particle and the spacetime spin connection does not generate divergences at the horizon; instead, it introduces only mild corrections that primarily affect spin precession, which are negligible in the present framework. The spin contribution to the black hole’s total angular momentum is also exceedingly small—especially for static, nonrotating black holes with masses far exceeding the Planck scale \cite{vanzo2011tunnelling}.

Because emission occurs symmetrically for particles of opposite spin, the average angular momentum of the system remains unchanged.
Building upon this theoretical foundation, the analysis now turns to the tunneling process of fermionic modes across the event horizon in the black hole background considered. The emission rate is obtained in Schwarzschild--like coordinates, which, although singular at the horizon, are convenient for the derivation. Equivalent treatments employing generalized Painlevé--Gullstrand or Kruskal--Szekeres coordinates have been discussed in previous studies \cite{o69}. To begin, the general form of the spacetime metric is expressed as follows:
\ie
\mathrm{d}s^{2} = - f(r) \mathrm{d}t^{2} + \frac{1}{f(r)}\,\mathrm{d}r^{2} + r^{2}\left(\mathrm{d}\theta^{2} + \sin^{2}\theta \mathrm{d}\varphi^{2} \right).
\fe

In a curved spacetime background, the Dirac equation takes the generalized form
\ie
\left(\gamma^\mu \nabla_\mu + \frac{m}{\hbar}\right) \Psi^{\uparrow \downarrow}_{\pm}(t,r,\theta,\varphi) = 0
\fe
where we have
\ie
\nabla_\mu = \partial_\mu + \frac{i}{2} {\Gamma^\alpha_{\;\mu}}^{\;\beta} \,\Sigma_{\alpha\beta}
\fe 
and 
\ie
\Sigma_{\alpha\beta} = \frac{i}{4} [\gamma_\alpha,  \gamma_\beta].
\fe
The matrices $\gamma^\mu$ obey the Clifford algebra relations, which are expressed as
\ie
\{\gamma_\alpha,\gamma_\beta\} = 2 g_{\alpha\beta} \mathbbm{1}.
\fe 
Here, $\mathbbm{1}$ denotes the $4 \times 4$ identity operator. In this framework, the specific representation adopted for the $\gamma$ matrices is given by
\begin{eqnarray*}
 \gamma ^{t} &=&\frac{i}{\sqrt{f(r)}}\left( \begin{array}{cc}
\vec{1}& \vec{ 0} \\ 
\vec{ 0} & -\vec{ 1}%
\end{array}%
\right), \qquad
\gamma ^{r} =\sqrt{f(r)}\left( 
\begin{array}{cc}
\vec{0} &  \vec{\sigma}_{3} \\ 
 \vec{\sigma}_{3} & \vec{0}%
\end{array}%
\right), \\
\gamma ^{\theta } &=&\frac{1}{r}\left( 
\begin{array}{cc}
\vec{0} &  \vec{\sigma}_{1} \\ 
 \vec{\sigma}_{1} & \vec{0}%
\end{array}%
\right), \qquad
\gamma ^{\varphi } =\frac{1}{r\sin \theta }\left( 
\begin{array}{cc}
\vec{0} &  \vec{\sigma}_{2} \\ 
 \vec{\sigma}_{2} & \vec{0}%
\end{array}%
\right),
\end{eqnarray*}%
in which $\vec{\sigma}$ denotes the set of Pauli matrices, obeying the usual commutation relations given by
\ie
 \sigma_i  \sigma_j = \vec{1} \delta_{ij} + i \varepsilon_{ijk} \sigma_k, \quad \text{ with} \quad i,j,k =1,2,3. 
\fe 
Conversely, the $\gamma^5$ matrix is defined in the form
\begin{equation*}
\gamma ^{5}=i\gamma ^{t}\gamma ^{r}\gamma ^{\theta }\gamma ^{\varphi } =  \frac{i}{r^{2}\sin \theta }\left( 
\begin{array}{cc}
\vec{ 0} & - \vec{ 1} \\ 
\vec{ 1} & \vec{ 0}%
\end{array}%
\right)\:.
\end{equation*}
To describe a Dirac field whose spin is aligned in the upward direction along the positive $r$–axis, the following ansatz is employed \cite{vagnozzi2022horizon}:
\begin{equation}
\Psi^{\uparrow}_{+}(t,r,\theta ,\varphi ) = \left( \begin{array}{c}
\mathrm{H}(t,r,\theta ,\varphi ) \\ 
0 \\ 
\mathrm{Y}(t,r,\theta ,\varphi ) \\ 
0%
\end{array}%
\right) \exp \left[ \frac{i}{\hbar }\Psi^{\uparrow}_{+}(t,r,\theta ,\varphi )\right]\;.
\label{spinupbh} 
\end{equation} 
The discussion here focuses on the spin--up configuration ($+$ and $\uparrow$), whereas the spin--down case ($-$ and $\downarrow$), corresponding to alignment along the negative $r$--axis, can be addressed in a similar manner. Substituting the ansatz (\ref{spinupbh}) into the Dirac equation yields
\ie
\begin{split}
-\left( \frac{i \,\mathrm{H}}{\sqrt{f(r)}}\,\partial _{t} \psi^{\uparrow}_{+} + \mathrm{Y} \sqrt{f(r)} \,\partial_{r} \psi^{\uparrow}_{+}\right) + \mathrm{H}\, i\, m &=0, \\
-\frac{\mathrm{Y}}{r}\left( \partial _{\theta }\psi^{\uparrow}_{+} +\frac{i}{\sin \theta } \, \partial _{\varphi }\psi^{\uparrow}_{+}\right) &= 0, \\
\left( \frac{i \,\mathrm{Y}}{\sqrt{f(r)}}\,\partial _{t}\psi^{\uparrow}_{+} - \mathrm{H} \sqrt{f(r)}\,\partial_{r}\psi^{\uparrow}_{+}\right) + \mathrm{Y} \, i\, m & = 0, \\
-\frac{\mathrm{H}}{r}\left(\partial _{\theta }\psi^{\uparrow}_{+} + \frac{i}{\sin \theta }\,\partial _{\varphi }\psi^{\uparrow}_{+}\right) &= 0,
\end{split}
\fe%
To leading order in $\hbar$, the action can be expressed as
$
\psi^{\uparrow}_{+} = - \omega\, t + \vartheta(r) + L(\theta ,\varphi )  $
so that
\cite{vanzo2011tunnelling}
\begin{eqnarray}
\left( \frac{i\, \omega\, \mathrm{H}}{\sqrt{f(r)}} - \mathrm{Y} \sqrt{f(r)}\,  \vartheta^{\prime }(r)\right) +m\, i\, \mathrm{H} &=&0,
\label{abahasapin5} \\
-\frac{\mathrm{Y}}{r}\left( L_{\theta }+\frac{i}{\sin \theta }L_{\varphi }\right) &=&0,
\label{abahsapain6} \\
-\left( \frac{i\,\omega\, \mathrm{Y}}{\sqrt{f(r)}} + \mathrm{H}\sqrt{f(r)}\, \mathcal \vartheta^{\prime }(r)\right) +\mathrm{Y}\,i\,m &=&0,
\label{abahaspaian7} \\
-\frac{\mathrm{H}}{r}\left( L_{\theta } + \frac{i}{\sin \theta }L_{\varphi }\right) &=& 0.
\label{bshssspsin8}
\end{eqnarray}

The explicit forms of $\mathrm{H}$ and $\mathrm{Y}$ are not essential for the conclusion that Eqs. (\ref{abahsapain6}) and (\ref{bshssspsin8}) lead to the constraint $L_{\theta} + i(\sin \theta)^{-1} L_{\varphi} = 0$, implying that $L(\theta,\varphi)$ must be a complex function. This relation holds for both outgoing and ingoing modes. Consequently, when computing the ratio between the corresponding probabilities, all terms containing $L$ cancel out, permitting its omission from the subsequent analysis. For the case of a massless fermion, Eqs. (\ref{abahasapin5}) and (\ref{abahaspaian7}) yield two independent solutions:
\ie
\mathrm{H} = -i \mathrm{Y}, \qquad \vartheta^{\prime}(r) = \vartheta_{\text{out}}' = \frac{\omega}{f(r)},
\fe
\ie
\mathrm{H} = i \mathrm{Y}, \qquad \vartheta^{\prime }(r) = \vartheta_{\text{in}}' = - \frac{\omega}{f(r)}.
\fe

Here, $\vartheta_{\text{out}}$ and $\vartheta_{\text{in}}$ correspond to the solutions describing, respectively, the outgoing and ingoing modes \cite{vanzo2011tunnelling}. The associated tunneling rate is expressed as $\Gamma_{\psi_{\text{metric}}} \sim e^{-2,\text{Im}(\vartheta_{\text{out}} - \vartheta_{\text{in}})}$. Consequently,
\ie
\mathcal \vartheta_{ \text{out}}(r)= - \mathcal \vartheta_{ \text{in}} (r) = \int \mathrm{d} r \,\frac{\omega}{f(r)}\,.
\fe
In the vicinity of $r = r_h$, the relevant functions can be expanded to first order and approximated as
\ie
f(r) = f'(r_{h})(r - r_{h}) + \dots \, .
\fe
This expansion exposes a simple pole characterized by a specific residue. Employing Feynman’s prescription for evaluating the integral, one obtains
\ie
2\mbox{ Im}\;\left( \mathcal \vartheta_{ \text{out}} -  \vartheta_{ \text{in}} \right) = \mbox{Im}\int \mathrm{d} r \,\frac{4\omega}{f(r)}=\frac{2\pi\omega}{\kappa},
\fe
with the surface gravity being introduced through the definition
\ie
\kappa = \frac{1}{2} f'(r_{h})  .
\fe

Within this formulation, the particle number density $n_{\psi}$ associated with the black hole configuration is expressed as $\Gamma_{\psi} \sim e^{-\frac{2\pi\omega}{\kappa}}$
\ie
n_{\psi} = \frac{\Gamma_{\psi}}{1+\Gamma_{\psi}}  =  \frac{1}{\exp \left(\frac{2 \pi  \omega }{-\frac{Q^2}{\left(\lambda +\sqrt{M^2-Q^2}+M\right)^3}+\frac{M}{\left(\lambda +\sqrt{M^2-Q^2}+M\right)^2}+\frac{3 \alpha  (2 \beta -1) Q^4}{10 \left(\lambda +\sqrt{M^2-Q^2}+M\right)^7}}\right)+1},
\fe
with 
\ie
\lambda = \frac{\alpha  (2 \beta -1) Q^4}{20 \left(\sqrt{M^2-Q^2}+M\right)^3 \left(M \left(\sqrt{M^2-Q^2}+M\right)-Q^2\right)}.
\fe

\begin{figure}
    \centering
      \includegraphics[scale=0.43]{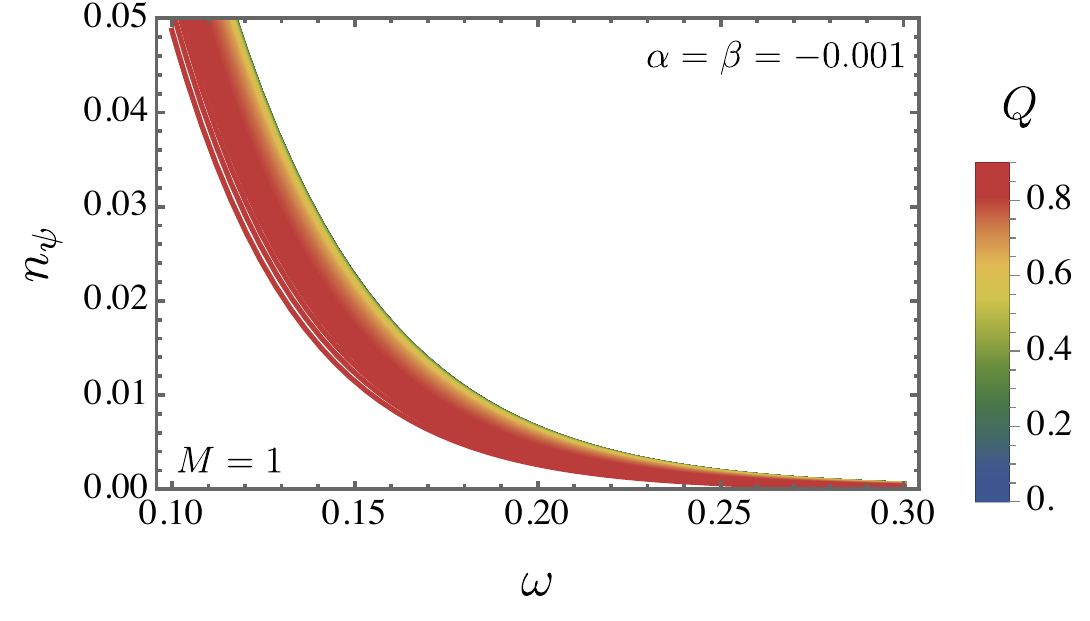}
       \includegraphics[scale=0.43]{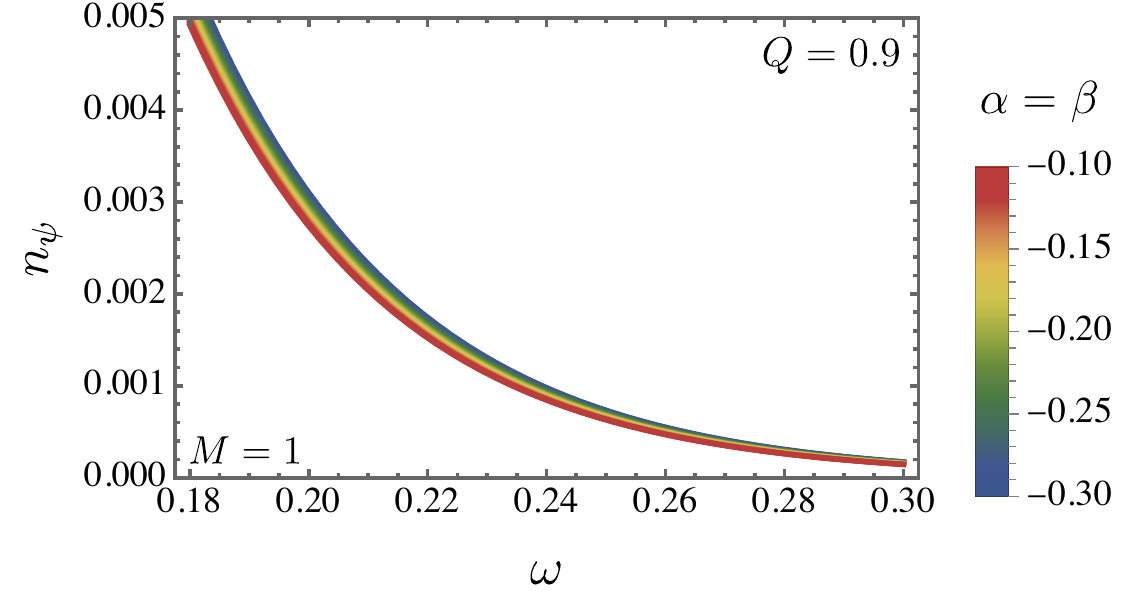}
    \caption{The particle density for fermions $n_{\psi}$ is shown against the frequency $\omega$. On the left panel, we consider the variation of $Q$ for fixed values of $\alpha = \beta = -0.001$. On the right panel, we vary $\alpha = \beta$ for a fixed value of $Q=0.9$.  }
    \label{particlefermions}
\end{figure}

To clarify the physical interpretation of the results, Fig.~\ref{particlefermions} illustrates the fermionic particle density $n_{\psi}$ as a function of the frequency $\omega$. The left panel depicts the influence of varying the electric charge $Q$, keeping $\alpha = \beta = -0.001$ fixed, whereas the right panel shows the dependence on $\alpha = \beta$ for a constant value of $Q = 0.9$. Complementarily, Table~\ref{tab2} provides a quantitative summary of these results. Overall, the analysis indicates that increasing $Q$ (with $\alpha = \beta = -0.01$ fixed) leads to a decrease in $n_{\psi}$, while reducing $\alpha = \beta$ enhances the corresponding particle density.

\begin{table}[!h]
\begin{center}
\begin{tabular}{c c c|| c c c } 
 \hline\hline \hline
 $Q$ & $\alpha=\beta$ &  $n_{\psi}$ & $Q$ & $\alpha=\beta$ &  $n_{\psi}$  \\ [0.2ex] 
 \hline
 0.60  & -0.01 & 0.07278720 & 0.99  & -0.01 & 0.00334425  \\ 

 0.70  & -0.01 & 0.07010770 & 0.99  & -0.02 & 0.00369590   \\
 
 0.80  & -0.01 & 0.06413920 & 0.99  & -0.03 & 0.00407625  \\
 
 0.90  & -0.01 & 0.04879970  & 0.99  & -0.04 & 0.00448593  \\
 
 0.99  & -0.01 & 0.00334425 & 0.99  & -0.05 & 0.00492539   \\ 
 [0.2ex] 
 \hline \hline \hline
\end{tabular}
\caption{\label{tab2} The numerical values corresponding to $n_{\psi}$ are presented. On the left panel, the magnetic charge $Q$ is varied with fixed parameters $\alpha = \beta$, while on the right panel, $Q$ remains constant and the values of $\alpha = \beta$ are modified. Here, we consider $M=1$ and $\omega =0.1$.
}
\end{center}
\end{table}


\section{Effective potential for the wave function}\label{SectionIV}

 A fundamental methodology for analyzing the behavior of quantum fields in curved spacetime involves reformulating the dynamics in terms of an effective one-dimensional scattering problem. For bosonic and fermionic particles, described by the Klein-Gordon and Dirac equations, respectively, this is achieved by separating the angular variables to obtain a radial wave equation. This radial equation can be recast into the standard form of a Schr\"{o}dinger equation, characterized by an effective potential $V_\text{eff}$
 . This potential is paramount, as it governs the transmission and reflection of waves, thereby directly influencing physical observables such as greybody factors, quasinormal modes, and absorption cross sections. Consequently, our initial objective is to derive and examine this $V_\text{eff}$  for both classes of particles, which will serve as the basis for our later exploration of scattering phenomena.

\subsection{Bosonic particles}

The dynamics of bosonic perturbations with spins $s = 0, 1,$ and $2$ can all be reduced to a one–dimensional wave equation of Schr\"{o}dinger form. 
To explore the analyses of these fields, we begin from their respective covariant field equations and derive the corresponding effective potentials. For the scalar field ($s=0$), the starting point is the Klein--Gordon equation
\begin{equation}\label{klein}
    \frac{1}{\sqrt{-g}} \partial_\mu \!\left( \sqrt{-g}\, g^{\mu\nu} \partial_\nu \Psi \right) = 0.
\end{equation}
Applying the separation of variables ansatz
\begin{equation}\label{ansatz}
    \Psi_{\omega \ell m}(\mathbf{r},t) = \frac{\psi_{\omega \ell}(r)}{r} Y_{\ell m}(\theta, \varphi) e^{-i\omega t},
\end{equation}
and introducing the tortoise coordinate $r^{*}$
\begin{equation}\label{rstar}
    \mathrm{d}r^{*} = \frac{\mathrm{d}r}{f(r)},
\end{equation}
the corresponding Klein--Gordon equation reduces to a Schr\"{o}dinger--like radial equation
\begin{equation}\label{waves}
    \left[\frac{\mathrm{d}^2}{\mathrm{d}r^{*2}} + \bigl(\omega^2 - V_{\mathrm{eff}}\bigr)\right] \psi_{\omega \ell}(r) = 0,
\end{equation}
where $V_{\text{eff}}$ corresponds to the scalar effective potential $\mathcal{V}_{S}$.

Similarly, we have a vector field ($s=1$) governed by the Proca equation, which reads
\begin{equation}\label{proca}
    \nabla_\nu F^{\mu\nu} + m^2 A^\mu = 0.
\end{equation}
After separation of variables and elimination of angular components, a single master variable also satisfies a Schr\"{o}dinger--type equation of the form Eq.~\eqref{waves}.  
Here the potential $V_{\mathrm{eff}} \!\to\! \mathcal{V}_V$ incorporates both the geometric term $f(r)\ell(\ell+1)/r^2$ and possible mass or coupling contributions from the background.

For tensor perturbations ($s = 2$), the linearized Einstein equations lead to the well--known Regge--Wheeler (axial) and Zerilli (polar) master equations, both of which reduce to the canonical wave form given in Eq.~\eqref{waves}, with a spin-dependent potential $\mathcal{V}_T$. It should be noted that the present analysis focuses exclusively on the odd (axial) perturbations.

Consequently, all bosonic degrees of freedom in a static and spherically symmetric background can be represented by a unified Schrödinger--like equation, each characterized by its own effective potential, as shown below
\begin{eqnarray}
    \mathcal{V}_{S} &=& f(r) \!\left( \frac{\ell(\ell + 1)}{r^2} + \frac{1}{r}\frac{\mathrm{d}f}{\mathrm{d}r} \right), \label{Veff:Sc}\\[3pt]
    \mathcal{V}_{V} &=& f(r) \frac{\ell(\ell + 1)}{r^2}, \label{Veff:V}\\[3pt]
    \mathcal{V}_{T} &=& f(r) \!\left[\frac{2}{r^2}\!\left(f(r)-1\right) + \frac{\ell(\ell + 1)}{r^2} - \frac{1}{r}\frac{\mathrm{d}f}{\mathrm{d}r}\right]. \label{Veff:T}
\end{eqnarray}
Notice that each potential carries essential information about how the spacetime geometry reacts to perturbations of distinct spin configurations. It expresses the coupling between the field and the background curvature while describing how such interactions modify the propagation and stability of the perturbative modes. Through these potentials, we can follow how the curvature distorts the wave dynamics, influences the scattering amplitude, affects the absorption probability, and determines the characteristic quasinormal spectra associated with each massless bosonic degree of freedom.


\subsection{Fermionic particles}

In addition to the bosonic sectors, a spin-$\tfrac{1}{2}$ Dirac field satisfies the covariant Dirac equation,
\begin{equation}
\gamma^{\alpha}\!\left( \partial_{\alpha} - \omega_{\alpha} \right)\!\Psi = 0,
\end{equation}
where $\gamma^\alpha$ denote the Dirac matrices and $\omega_\alpha$ represents the spin connection.
By applying the separation of variables, the angular and radial dependencies can be decoupled, as we presented for the bosonic perturbations, and the radial components $\Psi^{\pm}$ satisfy a Schrödinger--like equation that gives rise to the corresponding effective potential
\begin{equation}
\left[\frac{\mathrm{d}^2}{\mathrm{d}{r^*}^2} + \bigl(\omega^2 - V_{\text{eff}}^{\pm}\bigr)\right]\Psi^\pm = 0,
\label{eq:schrodinger}
\end{equation}
where the $\pm$ signs correspond to the two chiralities. The effective potentials are ~\cite{albuquerque2023massless, al2024massless,arbey2021hawking,devi2020quasinormal}
\begin{equation}
V_{\text{eff}}^{\pm} = \frac{(\ell + 1/2)^2}{r^2} f(r) \pm (\ell + 1/2) f(r) \frac{\mathrm{d}}{\mathrm{d}r}\!\left(\frac{\sqrt{f(r)}}{r}\right),
\label{eq:Veffpm}
\end{equation}
which form a supersymmetric pair. We take $V_{\text{eff}}^{+}$ as the representative Dirac potential,
\begin{equation}
\mathcal{V}_{\psi} = \frac{(\ell + 1/2)^2}{r^2} f(r) + (\ell + 1/2) f(r) \frac{\mathrm{d}}{\mathrm{d}r}\!\left(\frac{\sqrt{f(r)}}{r}\right).
\label{eq:Veffpsi}
\end{equation}

Having established all the preliminary aspects concerning the effective potentials $\mathcal{V}_S$, $\mathcal{V}_V$, $\mathcal{V}_T$, and $\mathcal{V}_\psi$, the subsequent analyses can now be carried out—namely, the computation of the greybody factors, absorption cross section, emission rate, evaporation lifetime, and the correspondence between the quasinormal modes and the greybody factors.


\section{Greybody factors}
\label{sec:GBF}


It is widely recognized that as Hawking radiation travels from the event horizon toward spatial infinity, spacetime curvature alters its spectrum. These deviations from an ideal blackbody distribution are encapsulated by the so–called greybody factors. In this framework, the following section focuses on investigating such factors, emphasizing how the particle spin influences the radiation process. The analysis encompasses scalar, vector, tensor, and spinor fields. To carry out this study, we build upon recent advances in the literature—particularly the work presented in Ref.~\cite{heidari2025gravitational}—where the authors implemented the separation of variables necessary to obtain the effective potentials associated with each spin sector.

The lower limit for the greybody factor, denoted by $|T_b|$, can be expressed as
\ie
\label{tgmetric}
|T_{b}| \ge {\mathop{\rm sech}\nolimits} ^2 \left(\int_\infty^ {+\infty} {\mathcal{G} \,\rm{d}}r^{*}\right),
\fe
where 
\ie
\mathcal{G} = \frac{{\sqrt {{{(\xi')}^2} + {{({\omega ^2} - \mathcal{V} - {\xi^2})}^2}} }}{{2\xi}}.
\fe
It is essential to realize that the function $\xi$ remains strictly positive and is constrained by the asymptotic limits $\xi(-\infty) = \xi(+\infty) = \omega$. When $\xi$ is fixed to this constant value, Eq.~\eqref{tgmetric} takes the simplified form
\ie
\begin{split}
& |T_{b}|  \ge {\mathop{\rm sech}\nolimits} ^2 \left[\int_{-\infty}^ {+\infty} \frac{\mathcal{V}} {2\omega}\mathrm{d}r^{*}\right] \ge {\mathop{\rm sech}\nolimits} ^2 \left[\int_{r_{ h}}^ {+\infty} \frac{\mathcal{V}} {2\omega f(r)}\mathrm{d} r\right].
\label{tbbbb}
\end{split}
\fe

In the subsequent subsections, the discussion will address the greybody factors corresponding to particles with spins $0$, $1$, $1/2$, and $2$.


\subsection{Spin--$0$ particle modes}

This subsection is devoted to investigating the scalar spin--$0$ emission. To do so, we write down the corresponding effective potential \cite{heidari2025gravitational}
\ie
\begin{split}
\mathcal{V}_{\text{S}}(r,\alpha,\beta,Q) = f(r) \left(\frac{\ell (\ell+1)}{r^2} + \frac{2 M}{r^3} -\frac{2 Q^2}{r^4} + \frac{3 \alpha  (2 \beta -1) Q^4}{5 r^8} \right)
\label{Vss}
\end{split},
\fe
so that substituting Eq. (\ref{Vss}) into Eq. (\ref{tbbbb}), we obtain
\begin{eqnarray}
\label{greybodyscalar}
	|T_{b}^{\text{S}}|  & \ge & {\mathop{\rm sech}\nolimits} ^2 \left[\int_{r_{ h}}^ {+\infty} \frac{\mathcal{V}_{\text{S}}(r,\alpha,\beta,Q)} {2\omega f(r)}\mathrm{d} r\right] 
		\nonumber \\
	&=&  {\mathop{\rm sech}\nolimits} ^2 \Bigg\{\frac{10}{21 \omega \zeta ^7}  \, \left(\sqrt{(M-Q) (M+Q)}+M\right)^3 \left(M \left(\sqrt{(M-Q) (M+Q)}+M\right)-Q^2\right) 
		\nonumber \\
	&& \qquad  \times \, \Big( (-1+2 \beta) 115200000 \alpha  Q^4 \left(\sqrt{(M-Q) (M+Q)}+M\right)^{18} \times 
		\nonumber \\
	&& \qquad \qquad   \times \left(Q^2-M \left(\sqrt{(M-Q) (M+Q)}+M\right)\right)^6 
		\nonumber \\
	&&  -5600 \zeta ^4 \left(\sqrt{(M-Q) (M+Q)}+M\right)^6 \left(Q^3-M Q \left(\sqrt{(M-Q) (M+Q)}+M\right)\right)^2 
		\nonumber \\
	&&   +21 \zeta ^6 \ell(\ell +1) +420 \zeta ^5 M \left(\sqrt{(M-Q) (M+Q)}+M\right)^3 \times
	\nonumber \\
	&&  
	\qquad \qquad \times \left(M \left(\sqrt{(M-Q) (M+Q)}+M\right)-Q^2\right)\bigg)   \Bigg\},
\end{eqnarray}
where
\ie
\begin{split}
\zeta   = & \,\,320 M^6+320 M^5 \sqrt{(M-Q) (M+Q)}-560 M^4 Q^2-400 M^3 Q^2 \sqrt{(M-Q) (M+Q)} \\
& +260 M^2 Q^4+100 M Q^4 \sqrt{(M-Q) (M+Q)}-Q^4 \left(-2 \alpha  \beta +\alpha +20 Q^2\right).
\end{split}
\fe

\begin{figure}
    \centering
      \includegraphics[scale=0.44]{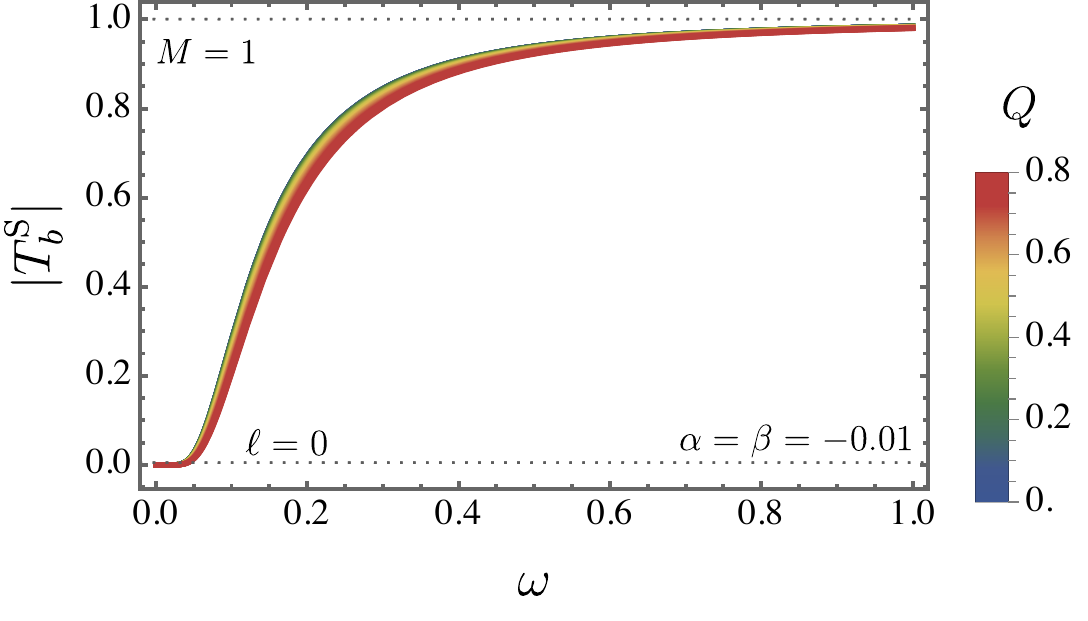}
       \includegraphics[scale=0.44]{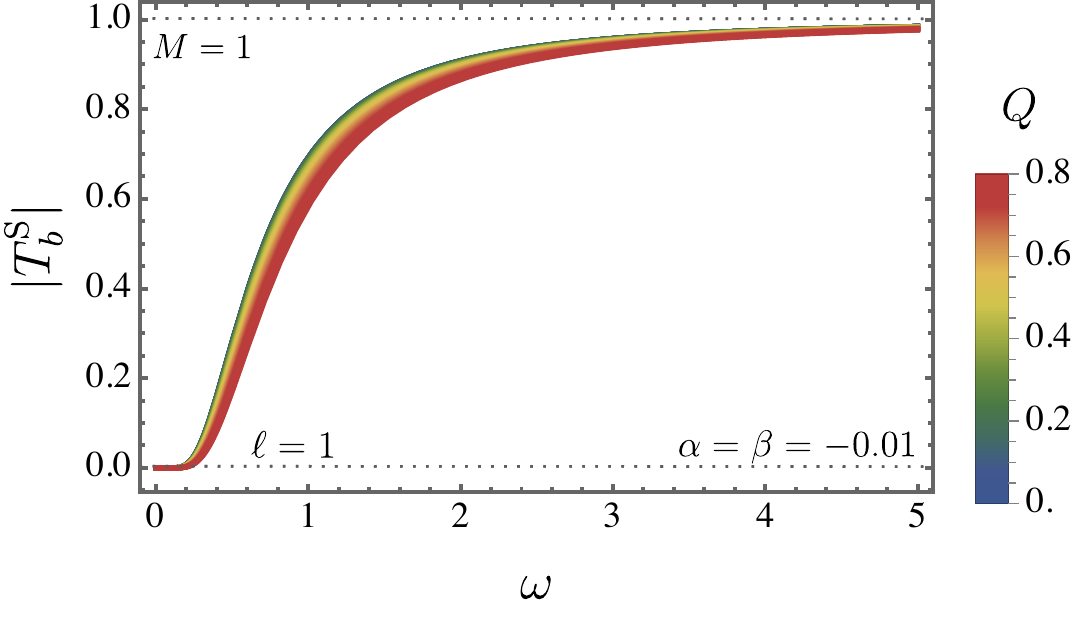}
          \includegraphics[scale=0.44]{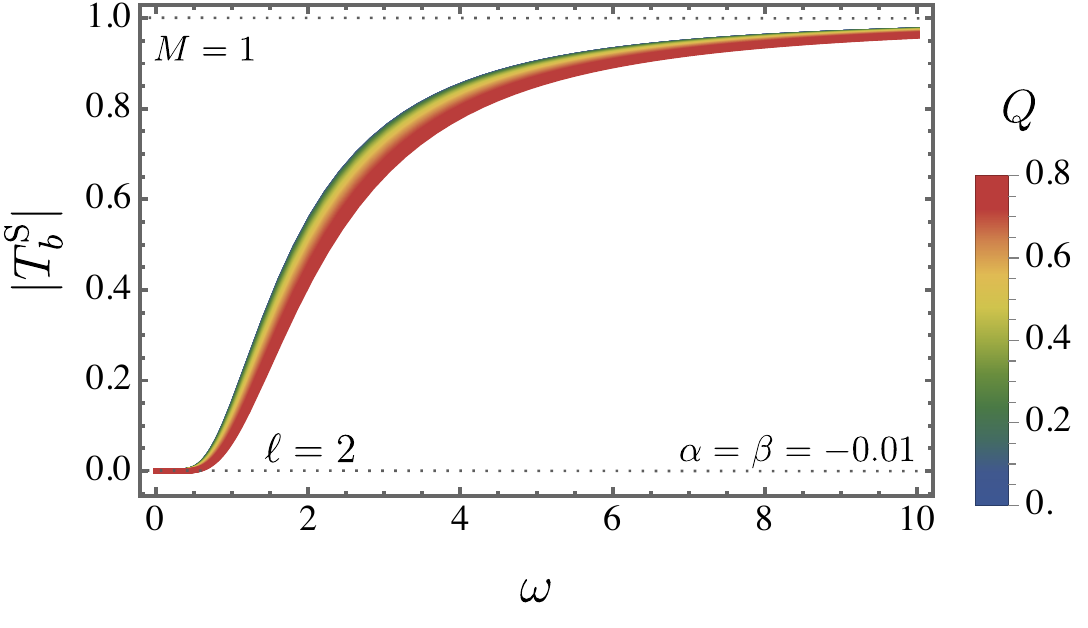}
    \caption{The greybody factors for spin--0 particle modes are presented for fixed values of $\alpha = \beta = -0.01$, considering various values of $Q$ and angular momentum number $\ell$. The top left panel corresponds to $\ell = 0$, the top right panel displays the case $\ell = 1$, and the bottom panel shows the results for $\ell = 2$.  }
    \label{spin0greybody}
\end{figure}

Now, let us provide the interpretation of the above result in Eq. (\ref{greybodyscalar}) through Fig. \ref{spin0greybody}. Thereby, the greybody factors for spin--$0$ particle modes are displayed with $\alpha = \beta = -0.01$ fixed, while varying $Q$ and the angular momentum number $\ell$. The top left panel illustrates the case $\ell = 0$, the top right one corresponds to $\ell = 1$, and the bottom panel depicts the scenario for $\ell = 2$.
In general terms, the charge $Q$ tends to suppress the tunneling process, thereby reducing the flux of scalar particles emitted to infinity.

\begin{figure}
    \centering
      \includegraphics[scale=0.43]{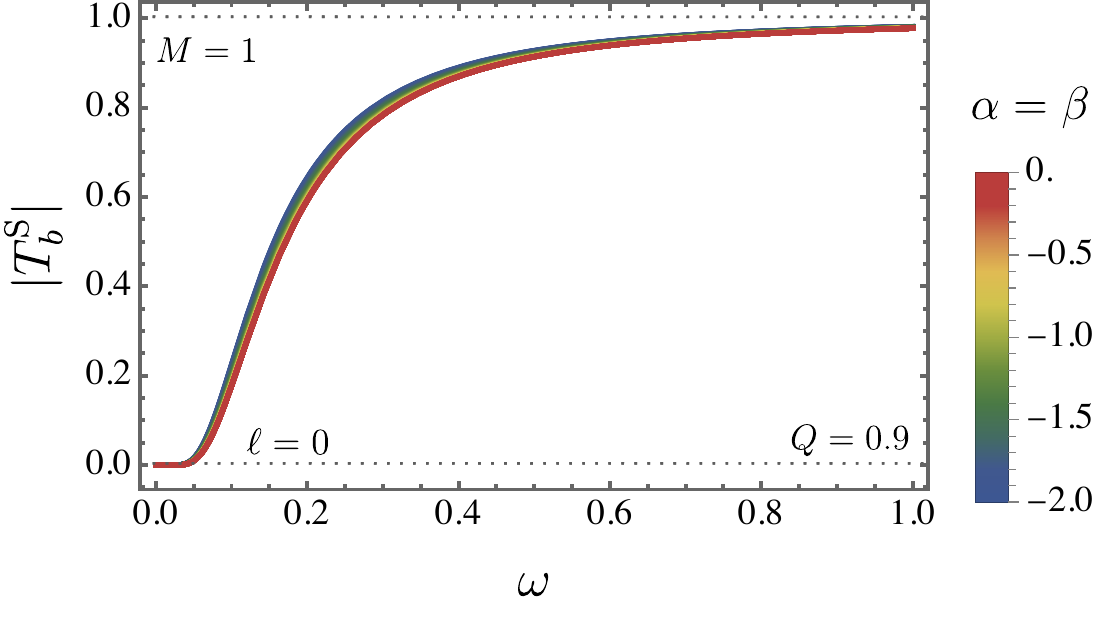}
       \includegraphics[scale=0.43]{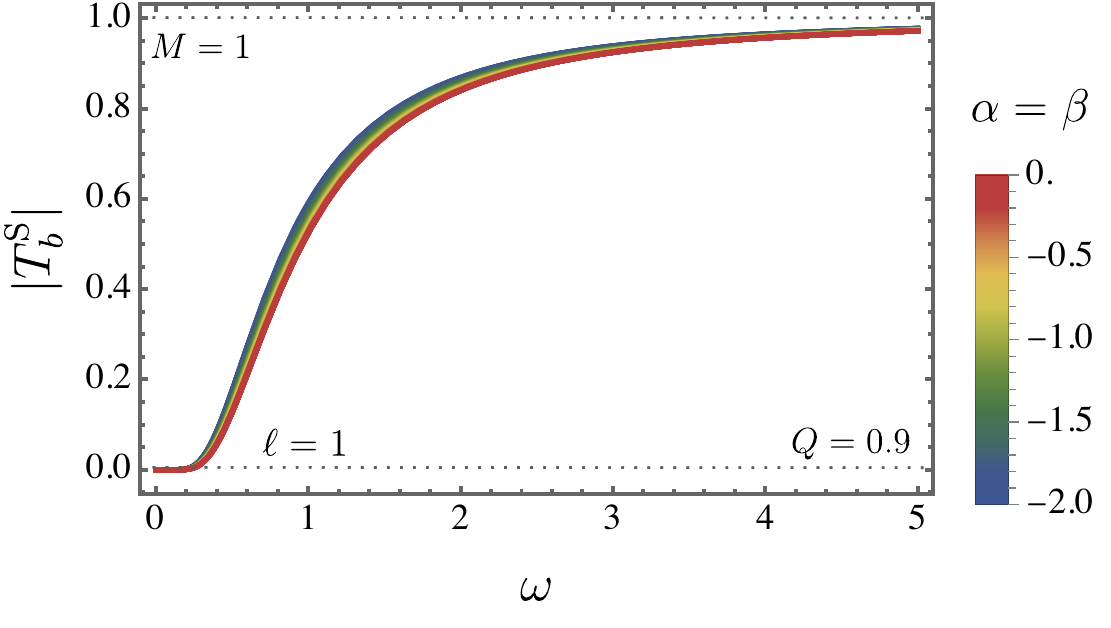}
          \includegraphics[scale=0.43]{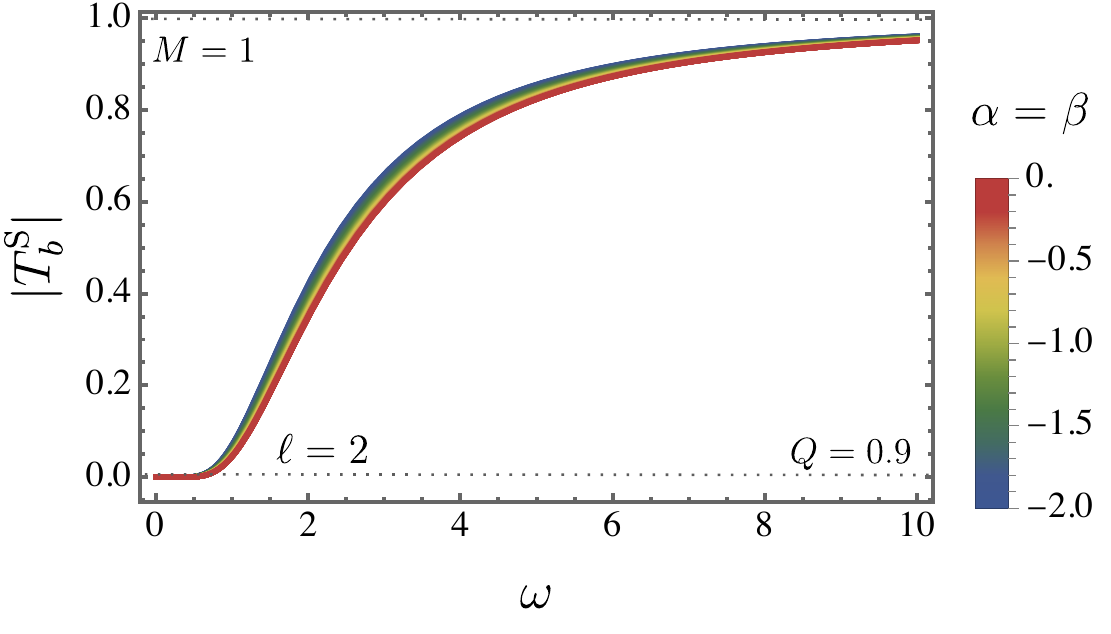}
    \caption{The greybody factors for spin--0 particle modes are presented for fixed values of $Q = 0.9$, considering various values of $\alpha = \beta$ and angular momentum number $\ell$. The top left panel corresponds to $\ell = 0$, the top right panel displays the case $\ell = 1$, and the bottom panel shows the results for $\ell = 2$.  }
    \label{spin0greybodyd}
\end{figure}

In addition to this analysis, we present Fig. \ref{spin0greybodyd}, which address the greybody factors for spin--$0$ particle modes are displayed with $Q = 0.9$ fixed, while varying $\alpha = \beta$ and the angular momentum number $\ell$. The top left panel illustrates the case $\ell = 0$, the top right one corresponds to $\ell = 1$, and the bottom panel depicts the scenario for $\ell = 2$. Notice that the reduction ascribed to $\alpha = \beta$ leads to an increment in the magnitude of $|T^{\text{S}}_{b}|$.

\subsection{Spin--$1$ particle modes}

Now, this subsection focuses complementary on the analysis of the spin--$1$ (vector) emission. For this purpose, the associated effective potential is expressed as \cite{heidari2025gravitational}
\ie
\begin{split}
\mathcal{V}_{\text{V}}(r,\alpha,\beta,Q) = f(r) \left(\frac{\ell (\ell+1)}{r^2} \right).
\label{Vvvv}
\end{split}
\fe
Substituting Eq.~(\ref{Vvvv}) into Eq.~(\ref{tbbbb}) yields
\begin{eqnarray}
\label{greybodyvector}
& |T_{b}^{\text{V}}|   \ge  {\mathop{\rm sech}\nolimits} ^2 \left[\int_{r_{ h}}^ {+\infty} \frac{\mathcal{V}_{\text{V}}(r,\alpha,\beta,Q)} {2\omega f(r)}\mathrm{d} r\right] 
	\nonumber \\
& =  {\mathop{\rm sech}\nolimits}^2 \left\{ {\frac{1}{2\omega}} \frac{20 \ell (\ell+1) \left(\sqrt{(M-Q) (M+Q)}+M\right)^3 \left(M \left(\sqrt{(M-Q) (M+Q)}+M\right)-Q^2\right)}{320 M^6+320 M^5 \sqrt{(M-Q) (M+Q)}-560 M^4 Q^2 \Tilde{\zeta}}  \right\},
\end{eqnarray}
where $\tilde{\zeta} = -400 M^3 Q^2 \sqrt{(M-Q) (M+Q)}+260 M^2 Q^4+100 M Q^4 \sqrt{(M-Q) (M+Q)}-Q^4 \left(-2 \alpha  \beta +\alpha +20 Q^2\right)$.

\begin{figure}
    \centering
      \includegraphics[scale=0.44]{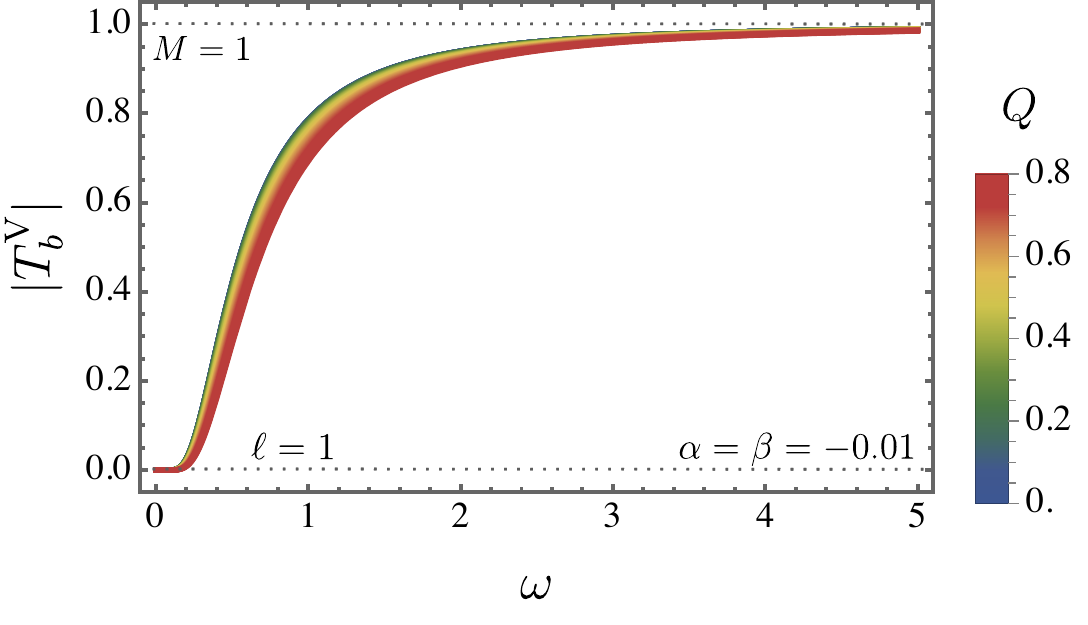}
       \includegraphics[scale=0.44]{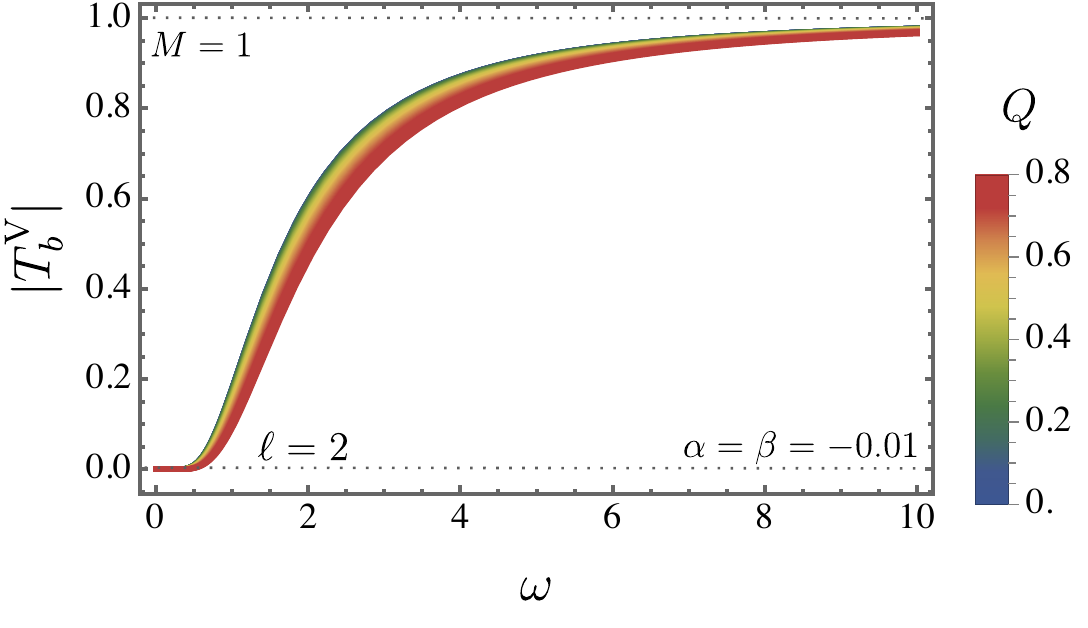}
    \caption{The greybody factors corresponding to spin--1 particle modes are shown for $\alpha = \beta = -0.01$, with different values of the charge $Q$ and the angular momentum quantum number $\ell$. The left and right panels represent the cases $\ell = 1$, and $\ell = 2$, respectively. }
    \label{spin1greybody}
\end{figure}

To interpret the result expressed in Eq.~(\ref{greybodyvector}), Fig.~\ref{spin1greybody} illustrates the behavior of the greybody factors for vector (spin–1) modes. In this figure, the parameters are fixed at $\alpha = \beta = -0.01$, while both the charge $Q$ and the angular momentum number $\ell$ are varied. The left panel corresponds to $\ell = 1$, and the right one to $\ell = 2$. Overall, increasing the charge $Q$ tends to weaken the tunneling probability, which in turn diminishes the amount of radiation that escapes to infinity.

\begin{figure}
    \centering
      \includegraphics[scale=0.43]{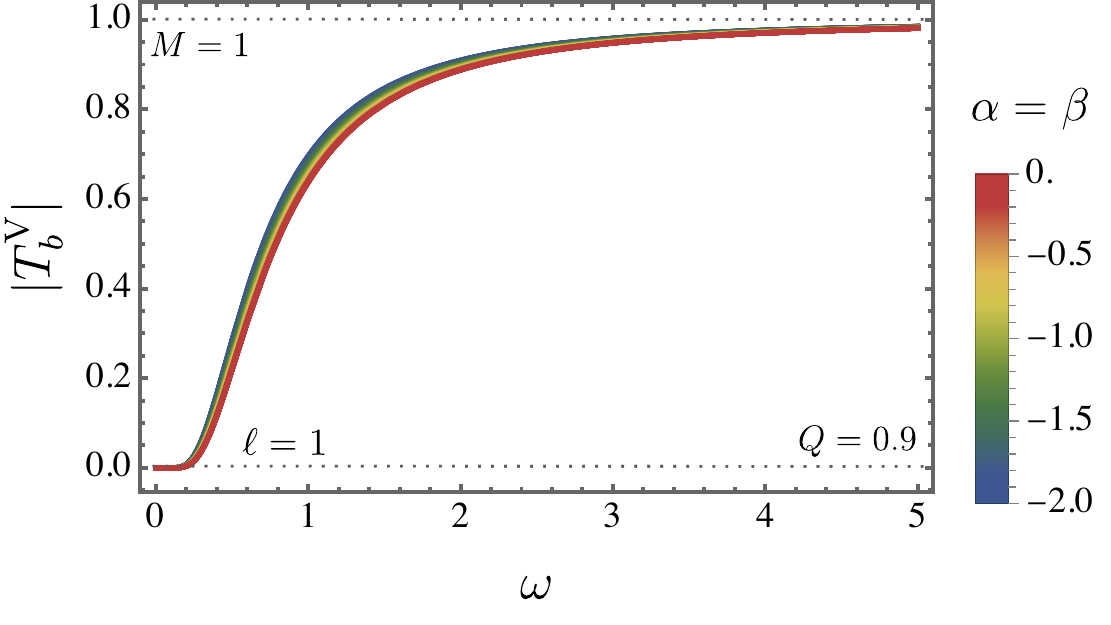}
       \includegraphics[scale=0.43]{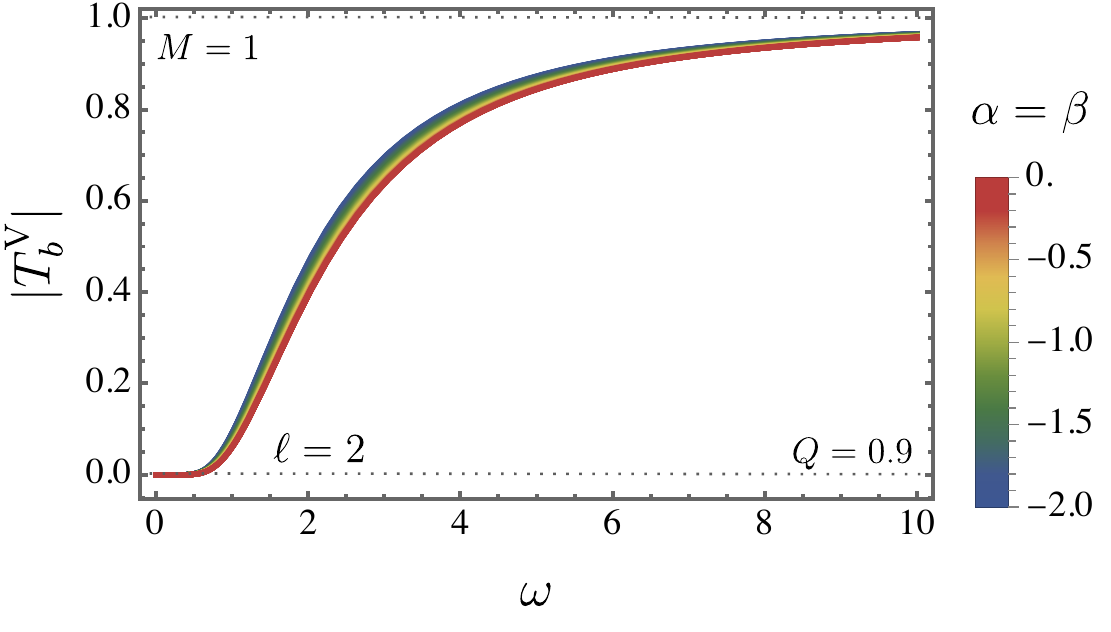}
    \caption{The greybody factors for spin--1 particle modes are displayed for a fixed charge $Q = 0.9$, while varying the parameters $\alpha = \beta$ and the angular momentum number $\ell$. The left panel depicts the case where $\ell = 1$, and the right panel presents the results for $\ell = 2$.  }
    \label{spin1greybodyd}
\end{figure}

Complementarily, Fig.~\ref{spin1greybodyd} shows the corresponding greybody factors for scalar (spin–0) modes, this time keeping $Q = 0.9$ constant while varying $\alpha = \beta$ and $\ell$. The panels follow a similar ordering as before: $\ell = 1$ (left), and $\ell = 2$ (right). It can be observed that lowering the values of $\alpha = \beta$ enhances the amplitude of $|T^{\text{V}}_{b}|$, reflecting an increase in transmission through the potential barrier.

\subsection{Spin--$2$ particle modes}

In this section, we present the computation of the greybody factors associated with tensor perturbations. Unlike the previous analyses, the evaluation here involves an expansion of the integrand in Eq.~(\ref{tbbbb}). Specifically, the term $\mathcal{V}/f(r)$ is expanded up to first order in $\alpha$ (or equivalently in $\beta$) and up to the fourth order in the charge $Q$ ($Q^{4}$). This truncation is adopted to maintain analytical tractability within computational limits when determining $|T_{b}^{\text{T}}|$. Accordingly, we obtain 
\se
	|T_{b}^{\text{T}}| &  \ge & {\mathop{\rm sech}\nolimits} ^2 \left[\int_{r_{h}}^ {+\infty} \frac{\mathcal{V}_{\text{T}}(r,\alpha,\beta,Q)} {2\omega f(r)}\mathrm{d} r\right] 
		\nonumber \\
	& = & {\rm{sech}}^{2} \Bigg\{ \frac{1}{2\omega} \Bigg(  \frac{20 \left(\sqrt{M^2-Q^2}+M\right)^3 \left(M \sqrt{M^2-Q^2}+M^2-Q^2\right)}{7 \Xi^{7}}  \,   
		\nonumber \\
	&& \quad \times \bigg[ 25600000\left(2 \beta - 1 \right)
	\alpha  Q^4   \left(\sqrt{M^2-Q^2}+M\right)^{18} \left(M \sqrt{M^2-Q^2}+M^2-Q^2\right)^6 
		\nonumber \\
	&& \hspace{-1cm} + 7 \ell^{2} \Xi ^6+7 \ell \,\Xi ^6-140 M \Xi ^5 \left(\sqrt{M^2-Q^2}+M\right)^3 \left(M \sqrt{M^2-Q^2}+M^2-Q^2\right)   \Bigg) \Bigg]\Bigg\} ,
	\label{tensorbgreybody}
\ff
where 
\ie
\begin{split}
\Xi = & \, \, 320 M^6-560 M^4 Q^2+260 M^2 Q^4+100 M Q^4 \sqrt{M^2-Q^2}+320 M^5 \sqrt{M^2-Q^2}\\
& -400 M^3 Q^2 \sqrt{M^2-Q^2}-Q^4 \left(-2 \alpha  \beta +\alpha +20 Q^2\right).
\end{split}
\fe

\begin{figure}
    \centering
          \includegraphics[scale=0.51]{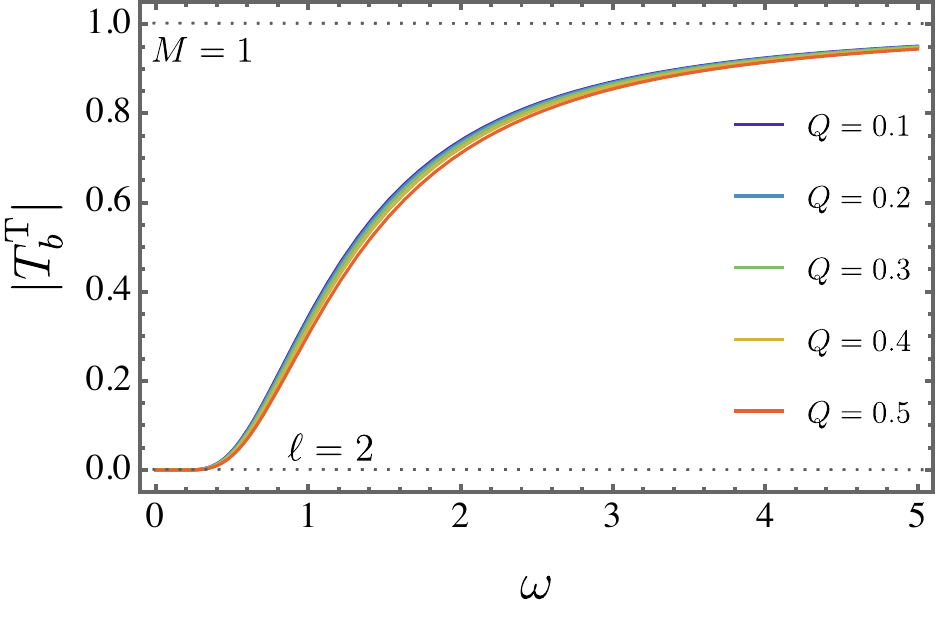}
           \includegraphics[scale=0.51]{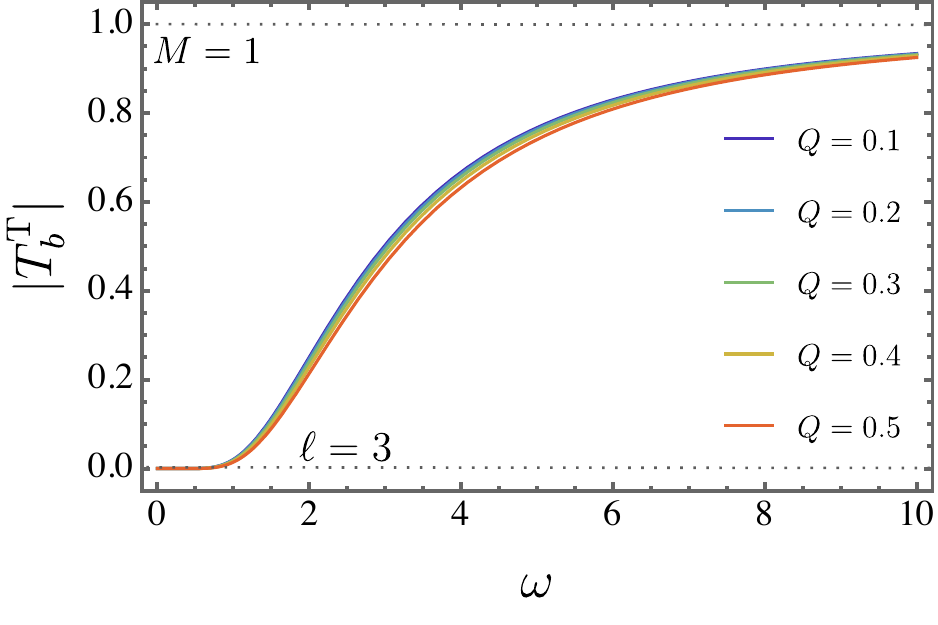}
           \includegraphics[scale=0.51]{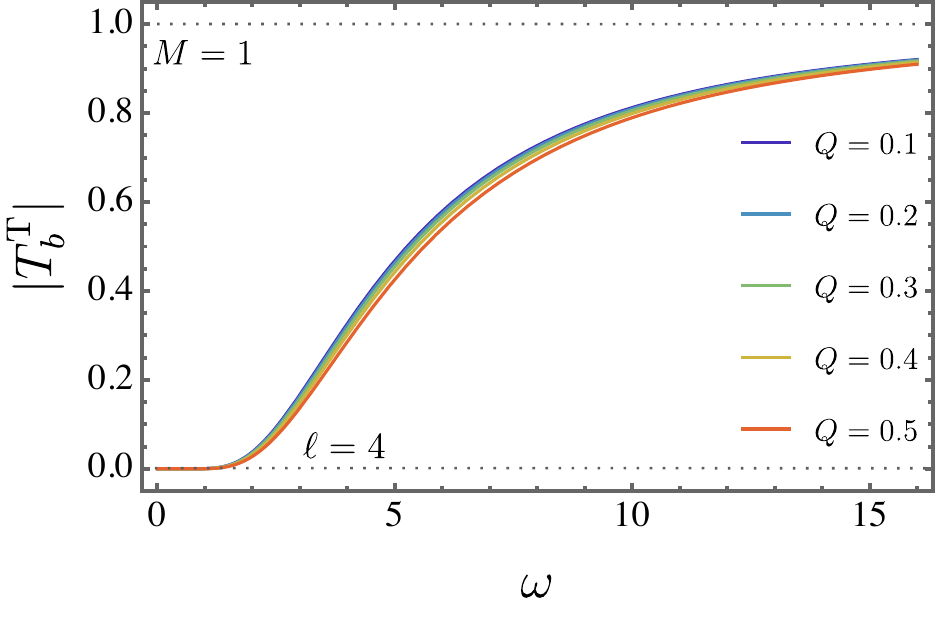}
    \caption{The greybody factors associated with spin–2 particle modes are shown for $\alpha = \beta = -0.01$, while the parameters $Q$ and the angular momentum number $\ell$ are varied. The top left panel illustrates the configuration with $\ell = 2$, the top right one depicts $\ell = 3$, and the lower panel presents the corresponding results for $\ell = 4$.}
    \label{spin2greybody}
\end{figure}

Accordingly, Fig. \ref{spin2greybody} is presented.
It is worth emphasizing that, although an analytical expression for $|T_{b}^{\text{T}}|$ was obtained as shown above, achieving it required an additional approximation: due to computational limitations, the potential $\mathcal{V}_{\text{T}}(r,\alpha,\beta,Q)/f(r)$ was expanded up to first order in $\alpha$ and up to fourth order in $Q$. Moreover, the plots are generated using its full, unexpanded form\footnote{To do so, we have accomplish it via numerical approach due to computational limitations.}. This choice allows us to clearly distinguish the curves corresponding to variations in $Q$ and $\alpha = \beta$. For comparison, recall that in the previous cases of $|T_{b}^{\text{S}}|$ and $|T_{b}^{\text{V}}|$, the parameter--induced changes in the greybody factors were relatively small. Here, performing expansions in $Q$ and $\alpha$ (or $\beta$) would make the resulting curves nearly indistinguishable, justifying the use of the exact expression therefore.

As observed for the scalar (spin–0) and vector (spin–1) cases, the greybody factors for spin–2 modes also decrease in magnitude as the charge $Q$ increases. Conversely, varying $\alpha = \beta$ reveals the opposite trend: lowering these parameters leads to an increase in the corresponding greybody factor, consistent with the behavior found for the other spin configurations analyzed. A plot is not presented here since the variation is exceedingly small, rendering the curves practically indistinguishable from one another.

\subsection{Spin--$1/2$ particle modes}

Now, the investigation of the greybody factors ends by addressing them for the spin--$1/2$. Differently what happened for the previous calculations for scalar and vector perturbations, here, we shall provide an expansion in the integrand of Eq. (\ref{tbbbb}) -- analogously to what we did for the tensor perturbations. In other words, to obtain an analytical expression for $|T_{b}^{\psi}|$, we expand $\mathcal{V}_{\psi}(r,\alpha,\beta,Q)/f(r)$, since the computational limitations prevent solving it in closed form otherwise. In line with the tensor perturbation analysis, the expansion is carried out up to first order in $\alpha$ and fourth order in $Q$. However, in this case, an additional expansion in the black hole mass $M$ is also performed to derive tractable analytical expressions. Thereby, we can write 
\se
	|T_{b}^{\psi}| &  \ge & {\mathop{\rm sech}\nolimits} ^2 \left[\int_{r_{h}}^ {+\infty} \frac{\mathcal{V}_{\psi}(r,\alpha,\beta,Q)} {2\omega f(r)}\mathrm{d} r\right] 
		\nonumber \\
	& = & {\rm{sech}}^{2} \Bigg\{ \frac{5}{2\omega\,\Xi^{8}} \left(\sqrt{M^2-Q^2}+M\right)^3 \left(M \sqrt{M^2-Q^2}+M^2-Q^2\right) \times  
		\nonumber \\
	&& \Bigg[ 128000000 \left( 1 + 2 \ell  - 2 \beta  - 4 \ell \beta \right)\alpha    M Q^4  \times
	 	\nonumber \\
	&& \qquad \qquad \times
	 \left(\sqrt{M^2-Q^2}+M\right)^{21} \left(M \sqrt{M^2-Q^2}+M^2-Q^2\right)^7 
	 	\nonumber \\
	&& + 6400000 \left(1  + 2 \ell  - 2  \beta  - 4 \beta  \ell \right) \alpha  \Xi  Q^4 \times
	 	\nonumber \\
	&& \qquad \qquad \times \left(\sqrt{M^2-Q^2}+M\right)^{18} \left(M \sqrt{M^2-Q^2}+M^2-Q^2\right)^6  
		\nonumber \\
	&& -  2400000 \left( 1 + 2 \ell \right) M \Xi ^2 Q^4 \left(\sqrt{M^2-Q^2}+M\right)^{15} \left(M \sqrt{M^2-Q^2}+M^2-Q^2\right)^5    
		\nonumber \\
	&& - 40000  \left( 1 + 2 \ell \right) \Xi ^3 Q^4
	\left(\sqrt{M^2-Q^2}+M\right)^{12} \left(M \sqrt{M^2-Q^2}+M^2-Q^2\right)^4 
		\nonumber \\
	&& + 8000 \left( 1 + 2 \ell  \right)M \Xi ^4 Q^2 
	\left(\sqrt{M^2-Q^2}+M\right)^9 \left(M \sqrt{M^2-Q^2}+M^2-Q^2\right)^3 
		\nonumber \\
	&& + 400 \left( 1  + 2 \ell   \right)\Xi ^5 Q^2
	\left(\sqrt{M^2-Q^2}+M\right)^6 \left(M \sqrt{M^2-Q^2}+M^2-Q^2\right)^2 
		\nonumber \\
	&& - 40  \left( 1 + 2 \ell  \right)M \Xi ^6
	\left(\sqrt{M^2-Q^2}+M\right)^3 \left(M \sqrt{M^2-Q^2}+M^2-Q^2\right) 
			\nonumber \\
	&& + 4 \ell^2 \Xi ^7+8 \ell \Xi ^7+3 \Xi ^7  \Bigg] \Bigg\}.
	\label{psigrey}
\ff

As in the preceding analysis, the greybody factors are initially obtained through an analytical expansion in the relevant parameters, following the procedure outlined earlier. Owing to computational constraints, however, Eq.~(\ref{psigrey}) is further expanded to first order in the mass parameter $M$. As will be discussed in the section on black hole evaporation, this approximation allows one to derive an analytical expression for the evaporation lifetime.

However, for improved visualization, we also compute them numerically. It is worth noting that even with the full numerical treatment, the curves corresponding to different values of $Q$ are nearly indistinguishable. Had we relied solely on the expanded expressions, these differences would have been even smaller, making it considerably more difficult to distinguish the curves in the plots.

\begin{figure}
    \centering
      \includegraphics[scale=0.51]{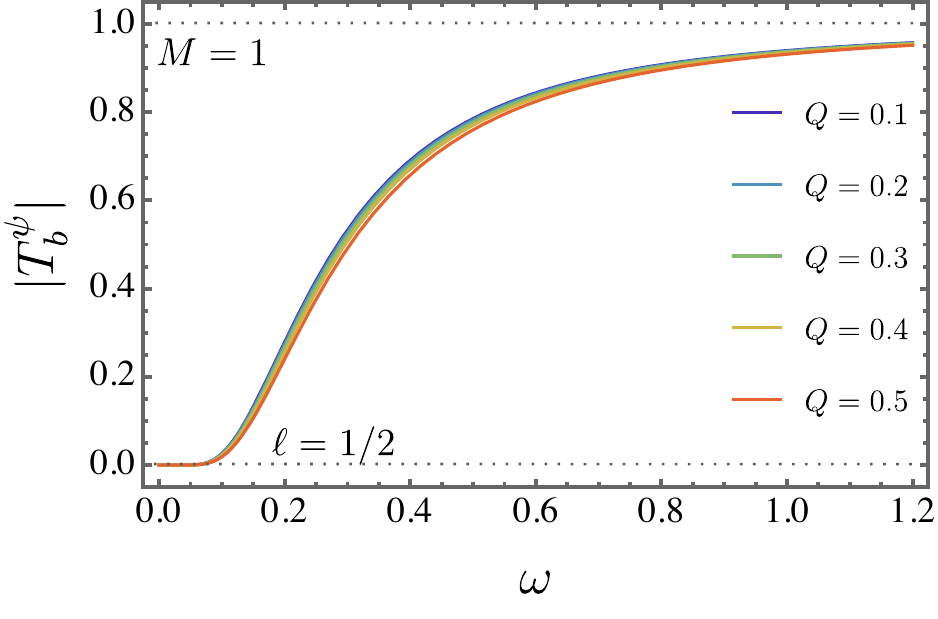}
       \includegraphics[scale=0.51]{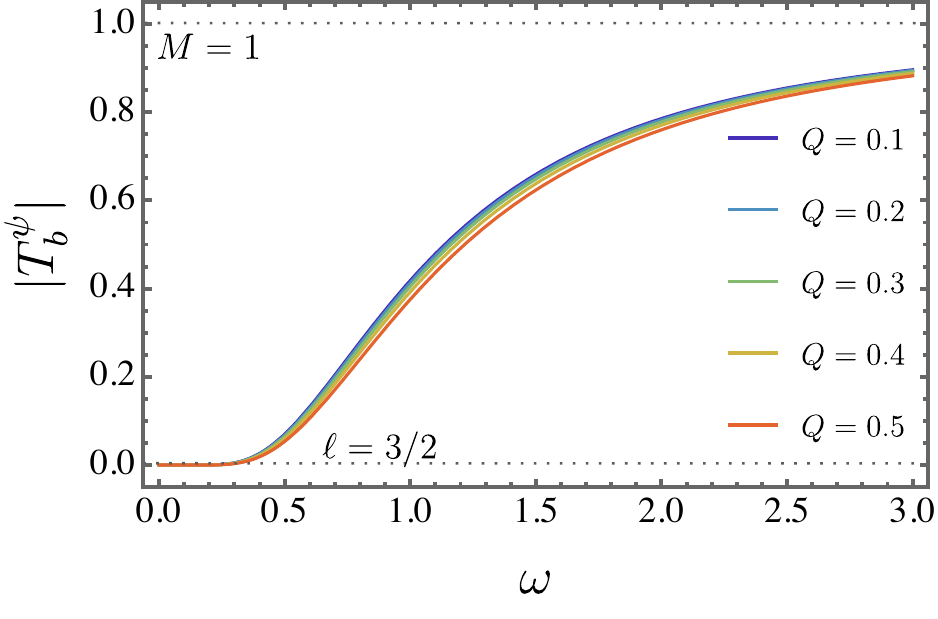}
          \includegraphics[scale=0.51]{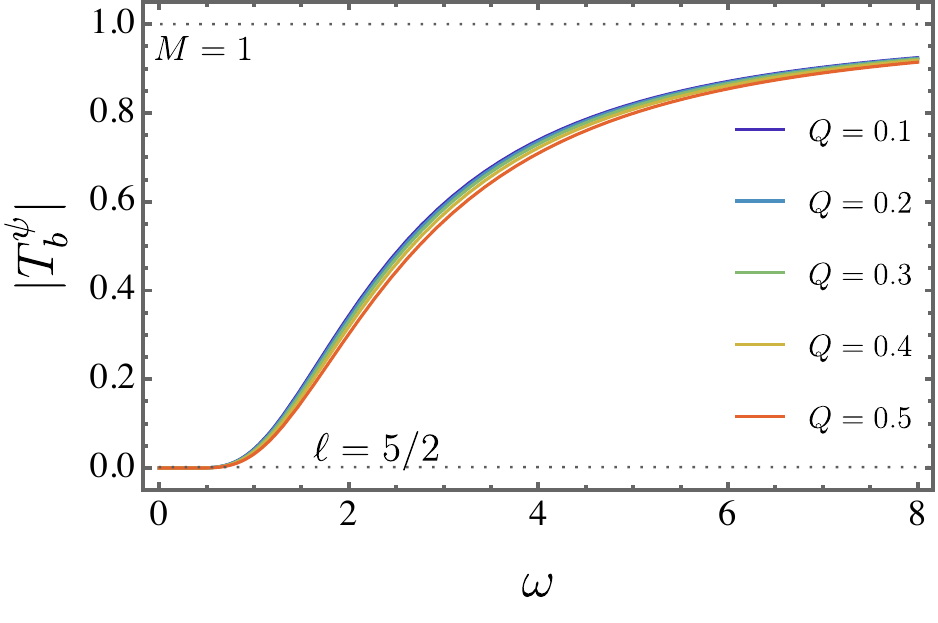}
    \caption{ The greybody factors for spin--$1/2$ modes are displayed for $\alpha = \beta = -0.01$, while varying the charge $Q$ and the angular momentum quantum number $\ell$. The upper left, upper right, and lower panels correspond to $\ell = 1/2$, $\ell = 3/2$, and $\ell = 5/2$, respectively.  }
    \label{spin1-2greybody}
\end{figure}

In Fig. \ref{spin1-2greybody}, we
display them $|T_{b}^{\psi}|$ as a function of $\omega$ for several values of $Q$. Similarly to what happened to the other spin configurations, as $Q$ increased, $|T_{b}^{\psi}|$ had their corresponding values diminished. Also, we shall ommit the corresposing graphical analysis for varying $\alpha = \beta$ (for a fixed value of $Q$) due to dificult in distinghising the different lines. Nevertheless, as $\alpha =\beta$ decreases, the corresponding magnitude of the associded greybody factors increases, being in agreement with the other spin configurations studied here.

\begin{figure}
    \centering
      \includegraphics[scale=0.65]{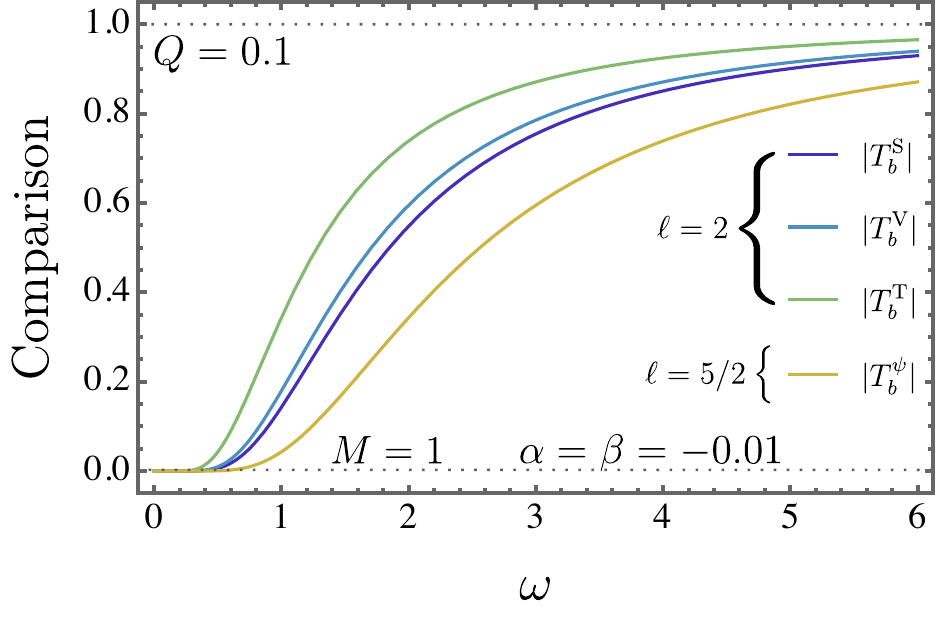}
    \caption{ The comparison of the greybody factors as functions of the frequency $\omega$ is carried out for all perturbations considered—scalar ($|T_{b}^{\text{S}}|$), vector ($|T_{b}^{\text{V}}|$), tensor ($|T_{b}^{\text{T}}|$), and spinorial ($|T_{b}^{\psi}|$)—with the parameters fixed at $M = 1$, $\ell =2$ (for bosonic case), $\ell=5/2$ (for the spinor case),$Q = 0.1$, and $\alpha = \beta = -0.01$.  }
    \label{compgreyall}
\end{figure}

Finally, in Fig.~\ref{compgreyall}, we present a comparison of the greybody factor intensities for a fixed configuration, namely $\alpha = \beta = -0.01$, $M = 1$, $Q = 0.1$, and $\ell = 2$ (for the bosonic case), $\ell = 5/2$ (for fermionic case). Overall, we observe that
\ie
|T_{b}^{\text{T}}| > |T_{b}^{\text{V}}| >  |T_{b}^{\text{S}}| > |T_{b}^{\psi}|. 
\label{comparisongreybodyall}
\fe
At least for the values of parameters employed here, this indicates that the greybody factors are most pronounced for tensor perturbations. Among the remaining cases, vector perturbations dominate over spinorial and scalar ones, with the spinorial contribution exceeding the scalar. Interestingly, the magnitude of the greybody bounds scaled directly with the spin of the perturbative mode—higher--spin configurations exhibited stronger greybody intensities.

\section{Absorption cross section}\label{SectionVI}

In this section, we analyze the scattering behavior of the perturbative fields. The absorption cross section represents a fundamental quantity in describing wave propagation and gravitational perturbations around black holes, as it measures the probability that the incoming radiation is absorbed by the geometry rather than transmitted to infinity. The probability that an outgoing mode escapes—or, equivalently, that an ingoing one is absorbed—is governed by the greybody factor, which connects the near-horizon and asymptotic regions by describing the tunneling of the perturbative field through the effective potential barrier inherent to the spacetime geometry. This quantity, examined in detail in Sec.~\ref{sec:GBF}, reveals the influence of the parameters $\alpha$, $\beta$, and $Q$ on the transmission properties of the wavefunction. The present analysis focuses on how these parameters modify the absorption cross section and, consequently, the scattering features of the system.

Within this framework, the scattering problem is analyzed using the semi--analytical Wentzel--Kramers--Brillouin (WKB) approximation, which requires imposing suitable boundary conditions that characterize the asymptotic behavior of the wavefunction $\mathcal{R}_{\omega \ell}$ both near the event horizon and at spatial infinity~\cite{cardoso2001quasinormal,macedo2015scattering,gogoi2024quasinormal}, i.e.,
\begin{equation}
\mathcal{R}_{\omega \ell} =
\begin{cases}
e^{-i\omega r^*} + \mathcal{A}_R\, e^{i\omega r^*}, & r^* \to -\infty \quad (r \to r_h),\\[6pt]
\mathcal{A}_T\, e^{-i\omega r^*}, & r^* \to +\infty \quad (r \to \infty),
\end{cases}
\end{equation}
where $\mathcal{A}_R$ and $\mathcal{A}_T$ represent the reflection and transmission coefficients, respectively. These coefficients are obtained from the WKB analysis as \cite{iyer1987black1,iyer1987black2,Konoplya:2011qq}
\begin{equation}
|\mathcal{A}_R|^{2} = \frac{1}{1 + e^{-2i\pi \Sigma}}, \qquad
|\mathcal{A}_T|^{2} = \frac{1}{1 + e^{+2i\pi \Sigma}} = 1 - |\mathcal{A}_R|^{2},
\end{equation}
where the complex parameter $\Sigma$ is defined as~\cite{konoplya2003quasinormal,Konoplya:2011qq}
\begin{equation}
\Sigma = \frac{i(\omega^{2} - \mathcal{V}_{\text{eff}})}{\sqrt{-2\mathcal{V}''_{\text{eff}}}} - \sum_{j=2}^{6}\Lambda_{j}.
\end{equation}
Here, $\mathcal{V}_{\text{eff}}$ denotes the peak value of the effective potential, $\mathcal{V}''_{\text{eff}}$ is its second derivative with respect to the tortoise coordinate $r^*$ evaluated at the maximum, and $\Lambda_{j}$ are higher--order WKB correction terms that depend on successive derivatives of the potential.

The transmission coefficient derived above enables the computation of the \emph{partial} absorption cross section~\cite{crispino2009scattering,anacleto2020absorption,macedo2016absorption}, which is expressed as
\begin{equation}
\sigma_{\mathrm{abs}} = \frac{\pi(2\ell + 1)}{\omega^{2}} |\mathcal{A}_T|^{2},
\end{equation}
where $\omega$ represents the frequency of the mode and $\ell$ is the angular quantum number.

In what follows, we systematically explore how variations in the physical parameters of the spacetime as charge $Q$, nonlinear correction parameter $\alpha$, coupling parameter $\beta$, and the spin of the perturbing field, influence the absorption cross section. Specifically, we investigate and compare the absorption characteristics for scalar (spin--$0$), electromagnetic (spin--$1$), gravitational (spin--$2$) fields, and fermionic (spin--$1/2$), emphasizing the distinctive behavior of each case and its dependence on the background parameters.


\subsection{Spin--$0$ particle modes}


\begin{figure}
    \centering
      \includegraphics[scale=0.43]{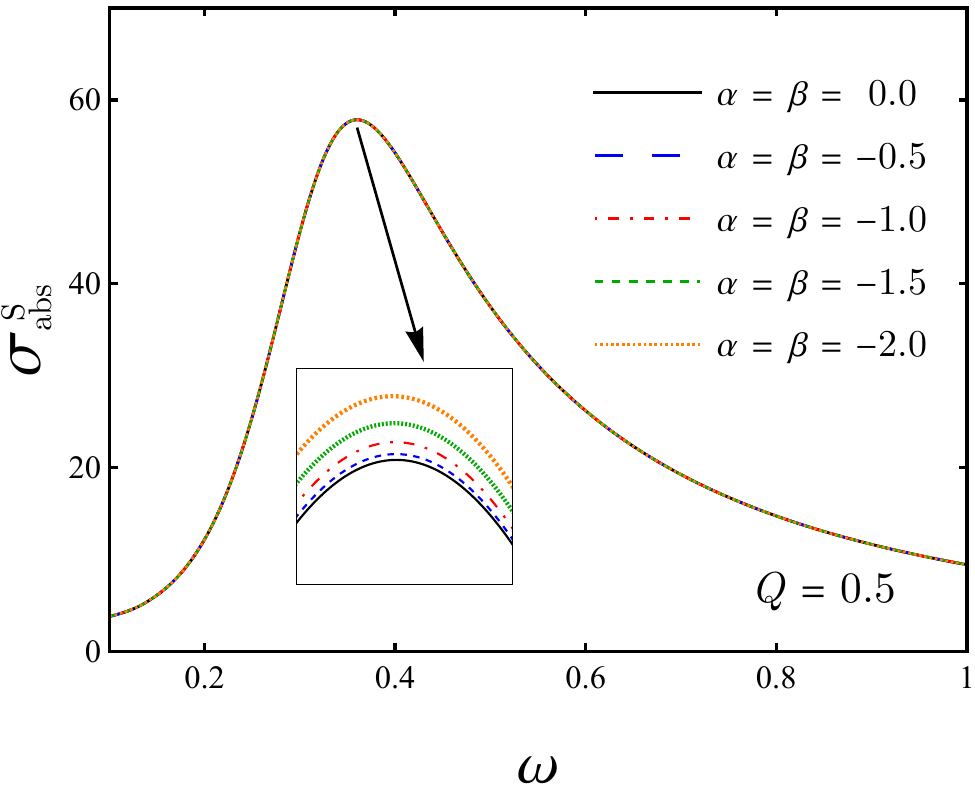}\quad\quad      
       \includegraphics[scale=0.43]{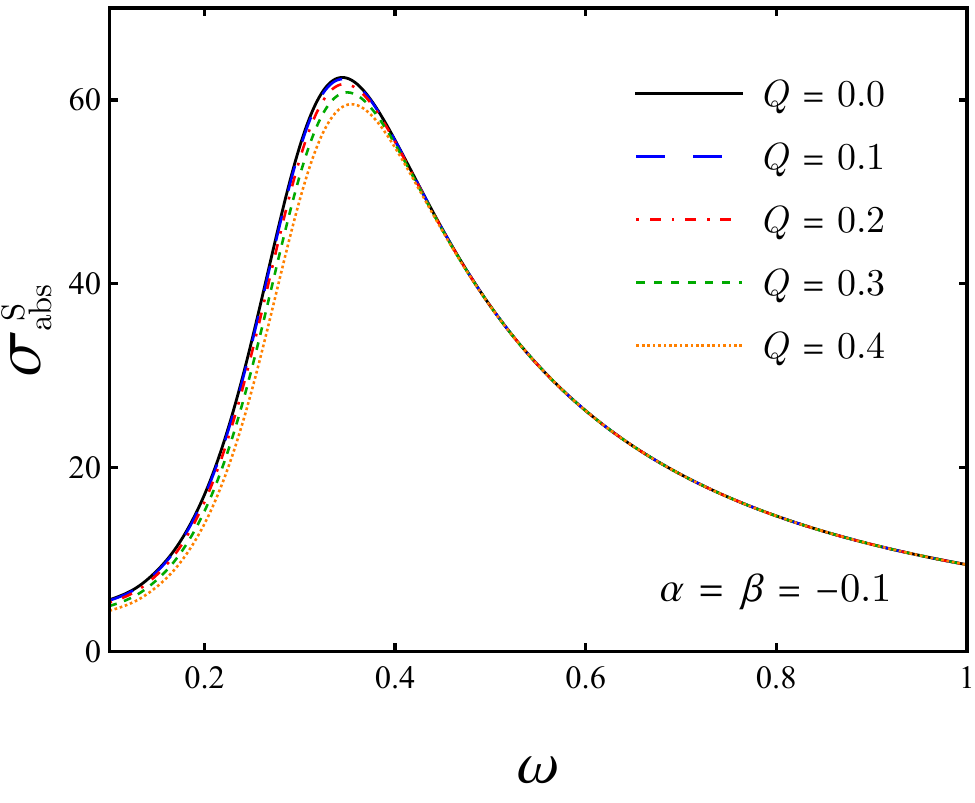}
          \includegraphics[scale=0.43]{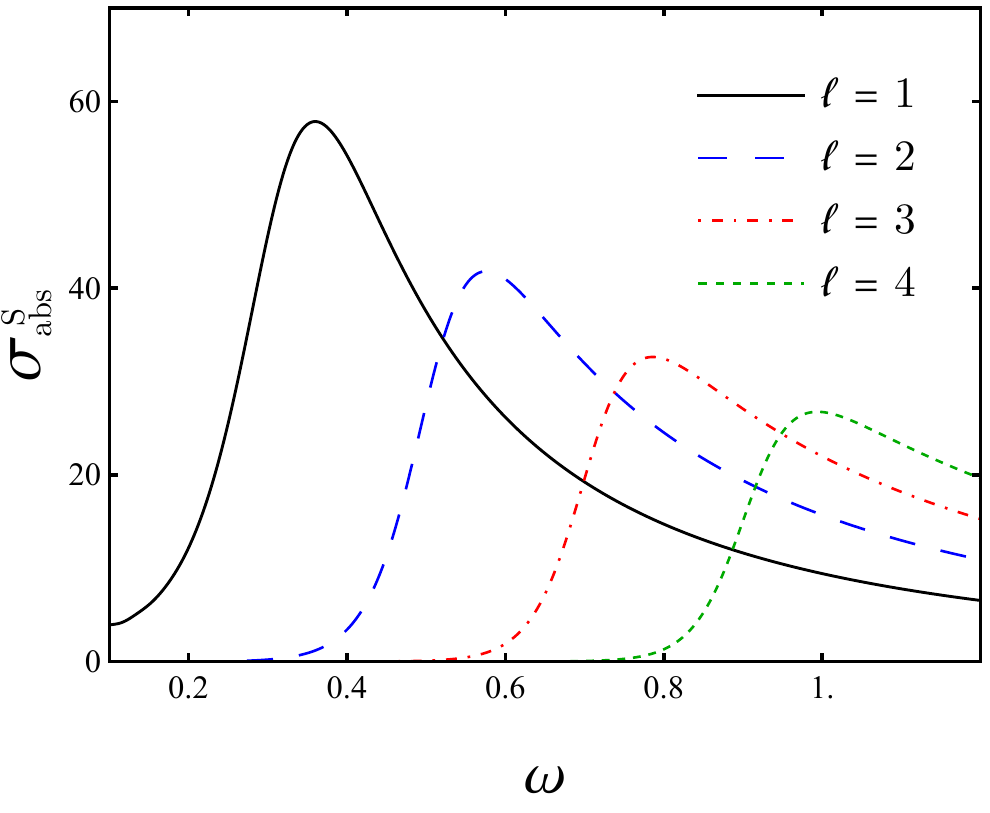}
    \caption{The absorption cross sections for spin--$0$ particle modes are shown for a fixed mass parameter $M = 1$, evaluated across different values of $\alpha$, $\beta$, $Q$, and the angular momentum number $\ell$. The top left and right panels correspond to $\ell = 1$, while the bottom panel presents the results for $Q = 0.5$ with $\alpha = \beta = -0.1$.}
    \label{fig:spin0abs}
\end{figure}
The absorption cross sections for spin--$0$ particles, computed for a fixed mass parameter $M = 1$, reveal a dependence on the charge $Q$, angular momentum number $\ell$, and the deformation parameters $\alpha$ and $\beta$ in Fig. \ref{fig:spin0abs}. For a fixed charge $Q = 0.5$, more negative values of the deformation parameters $\alpha = \beta$ (from $0.0$ down to $-2.0$) lead to a substantial increase in the absorption cross section, indicating that the $f(R, T)$ framework creates a shorter or narrower potential barrier. Meanwhile, for a fixed $\ell = 1$ and $\alpha = \beta = -0.1$, increasing the charge $Q$ from $0.0$ to $0.4$ slightly suppresses the peak of the cross section and induces a slight shift of the absorption peak toward higher energies. Furthermore, varying the angular momentum number shows that higher multipoles contribute significantly only at higher frequencies, adding characteristic oscillatory structure to the total cross section, which is dominated by lower modes at low frequencies.   


\subsection{Spin--$1$ particle modes}
\begin{figure}
    \centering
      \includegraphics[scale=0.43]{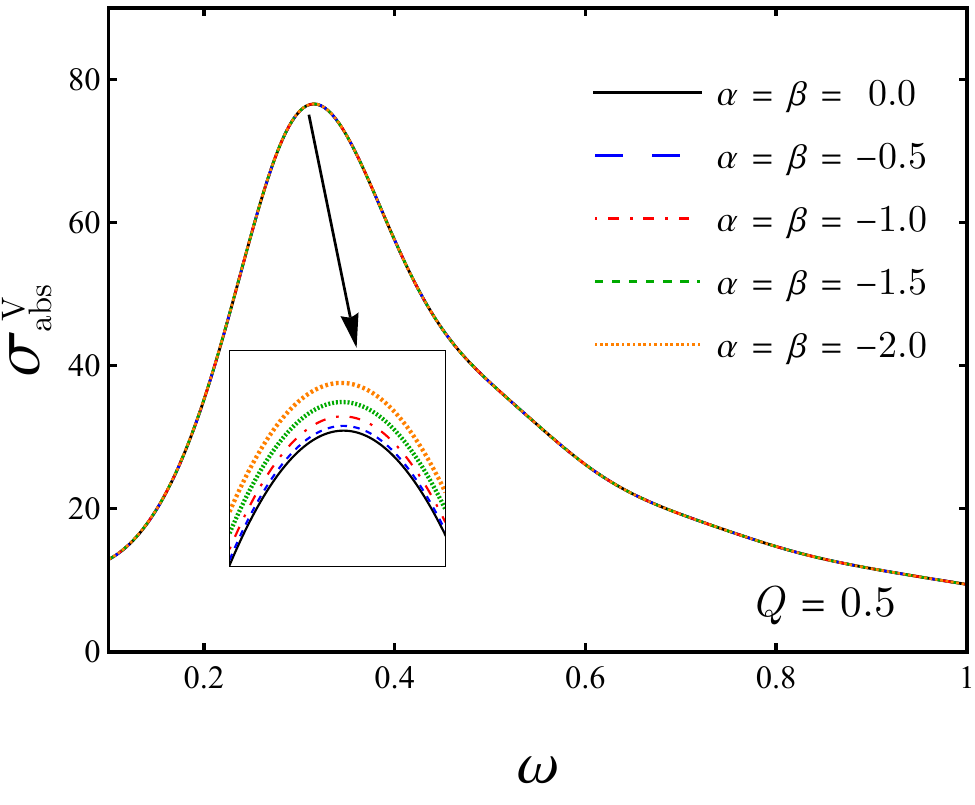}\quad\quad      
       \includegraphics[scale=0.43]{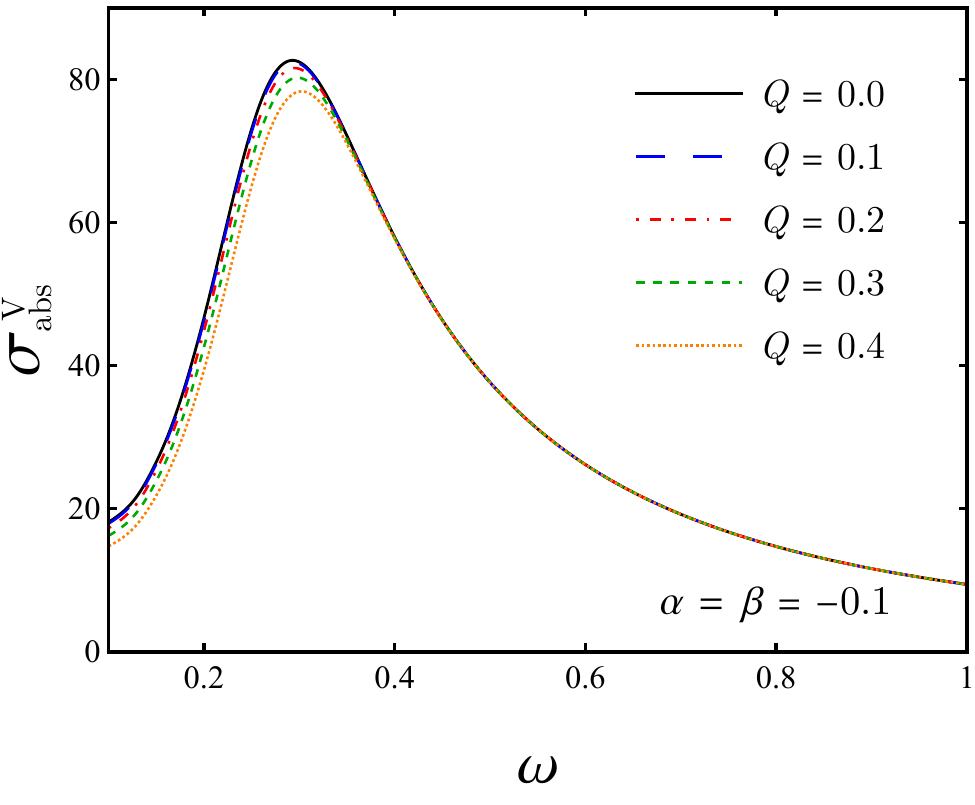}          \includegraphics[scale=0.43]{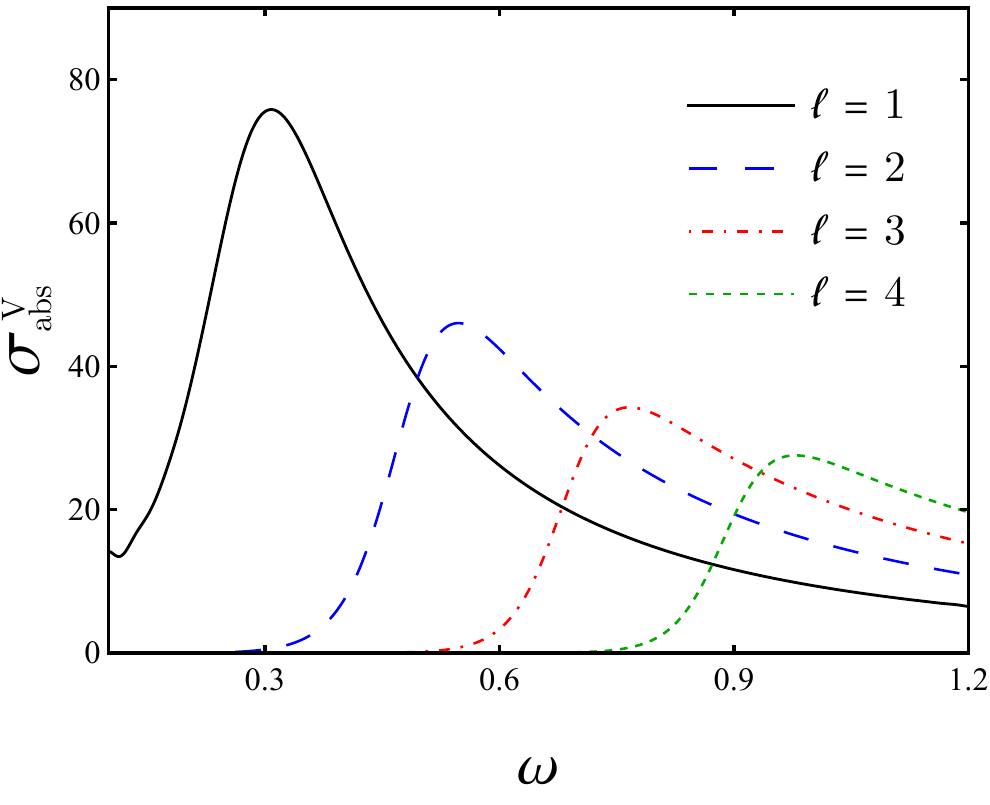}
    \caption{Absorption cross sections for the vector field are plotted for a fixed mass parameter $M = 1$, considering different values of $\alpha$, $\beta$, $Q$, and the angular momentum $\ell$. The top panels correspond to $\ell = 1$, and the bottom panel shows the case $Q = 0.5$ with $\alpha = \beta = -0.1$.}

    \label{fig:spin1abs}
\end{figure}

The absorption cross sections for spin--$1$ fields are presented in Fig.~\ref{fig:spin1abs},  revealing a clear dependence on the parameters $\alpha$, $\beta$, $Q$, and $\ell$. As $\alpha$ and $\beta$ become more negative, the absorption peak slightly goes higher, indicating a modification of the effective potential that enhances the absorption. In contrast, increasing the black hole charge $Q$ suppresses the absorption magnitude and causes a slight high--frequency shift, indicative of heightened electromagnetic repulsion. Furthermore, the behaviour of the absorption cross section for various values of multipole number $\ell$ shows that the dominant low--frequency absorption is governed by the dipole ($\ell=1$) mode. As the frequency increases, higher multipoles ($\ell \geq 2$) begin to contribute substantially, imprinting a characteristic oscillatory pattern onto the total cross section. 


\subsection{Spin--$2$ particle modes}
\begin{figure}
    \centering
      \includegraphics[scale=0.43]{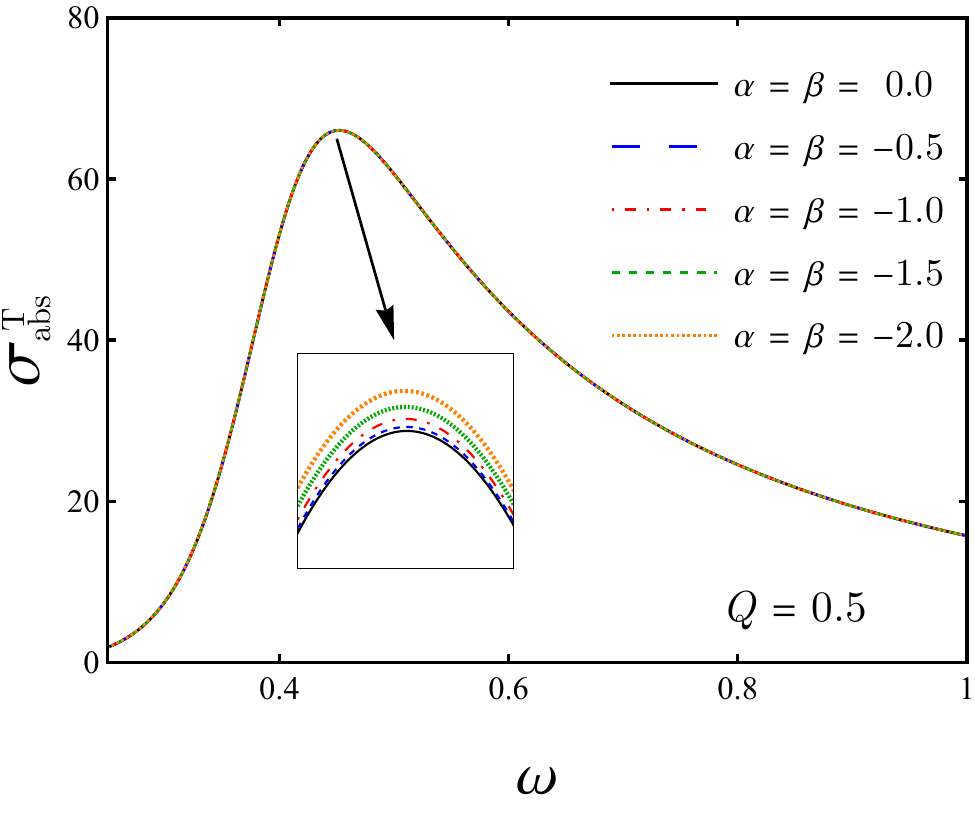}\quad\quad      
       \includegraphics[scale=0.43]{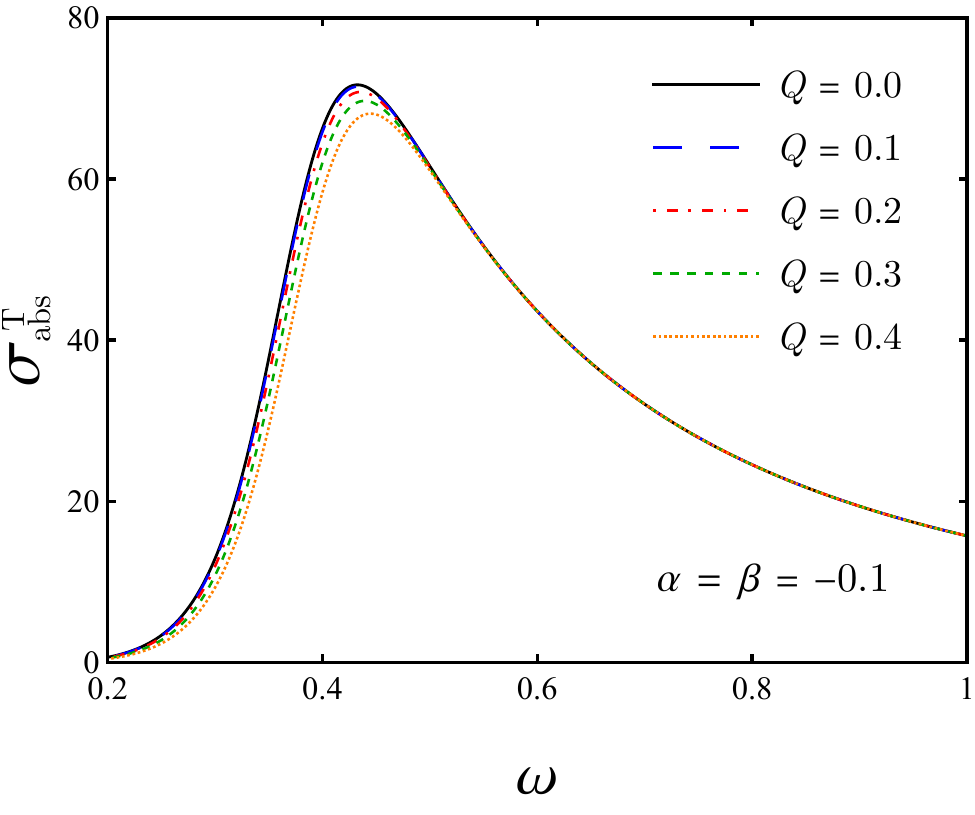}
          \includegraphics[scale=0.43]{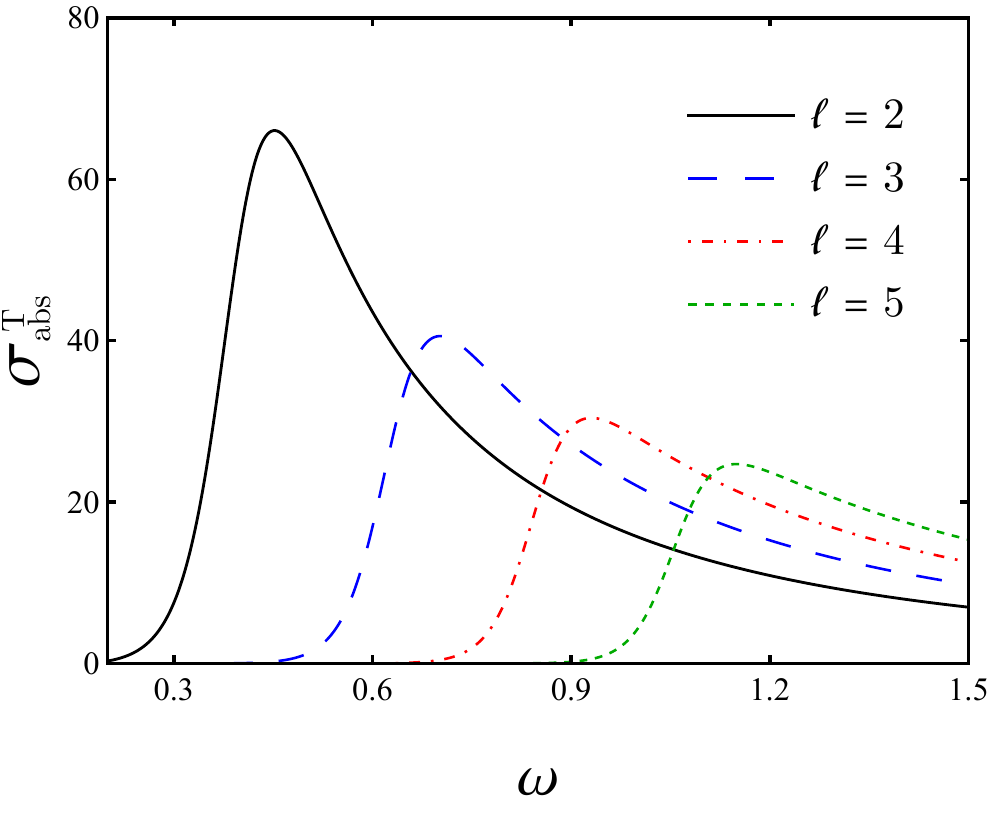}
    \caption{The absorption cross sections for tensor (spin--2) field modes are presented for a fixed black hole mass parameter $M = 1$. The results are computed for different values of the parameters $\alpha$, $\beta$, $Q$, and the angular momentum number $\ell$. The top panels correspond to $\ell = 2$, while the bottom panel illustrates the case $Q = 0.5$ with $\alpha = \beta = -0.1$.}

    \label{fig:spin2abs}
\end{figure}

The absorption cross section for spin--$2$ field perturbations is analyzed as a function of frequency for a fixed mass $M=1$ in Fig.~\ref{fig:spin2abs}. Like the other bosonic perturbations described previously, $s = 0$ and $s = 1$, the cross section exhibits a characteristic profile that varies with the parameters of $f(R, T)$ and the number of angular momentum $\ell$. The influence of the modification parameters $\alpha$ and $\beta$ on the cross section is subtle but observable. For larger magnitudes of $|\alpha|=|\beta|$, the primary effect is a slight increase in the amplitude of the resonant peak. At the same time, the overall functional form and frequency--dependence of the cross section remain largely unaltered.  In contrast, the electric charge $Q$ has a more observable role: an increase in $Q$ systematically suppresses the overall magnitude of the cross section. Finally, the contribution of various angular momentum modes is examined. The lowest allowed multipole governs the low--frequency behavior, whereas higher multipoles ($\ell > 2$) become progressively relevant at higher frequencies with lower amplitudes.


\subsection{Spin--$1/2$ particle modes}

\begin{figure}
    \centering
      \includegraphics[scale=0.53]{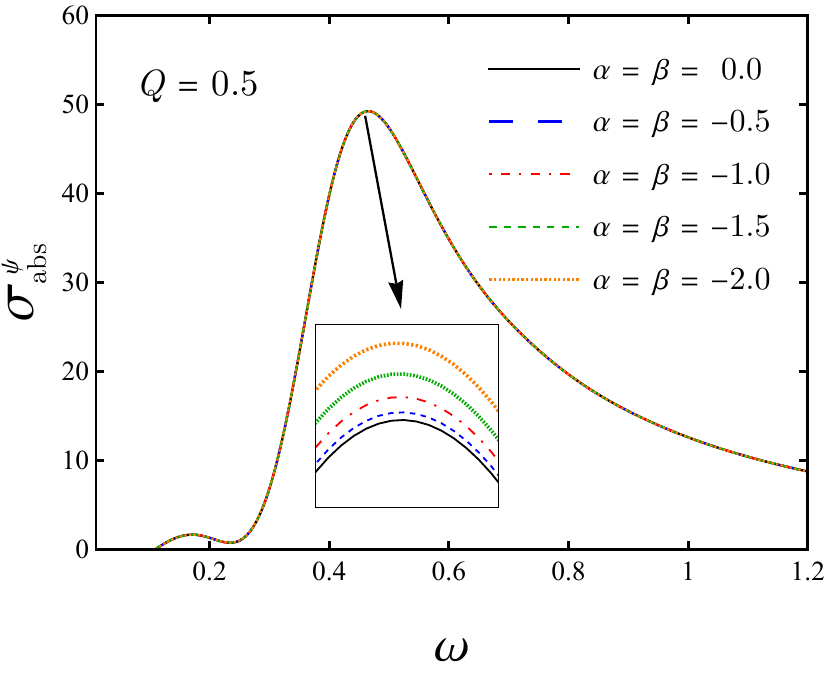}\quad\quad      
       \includegraphics[scale=0.53]{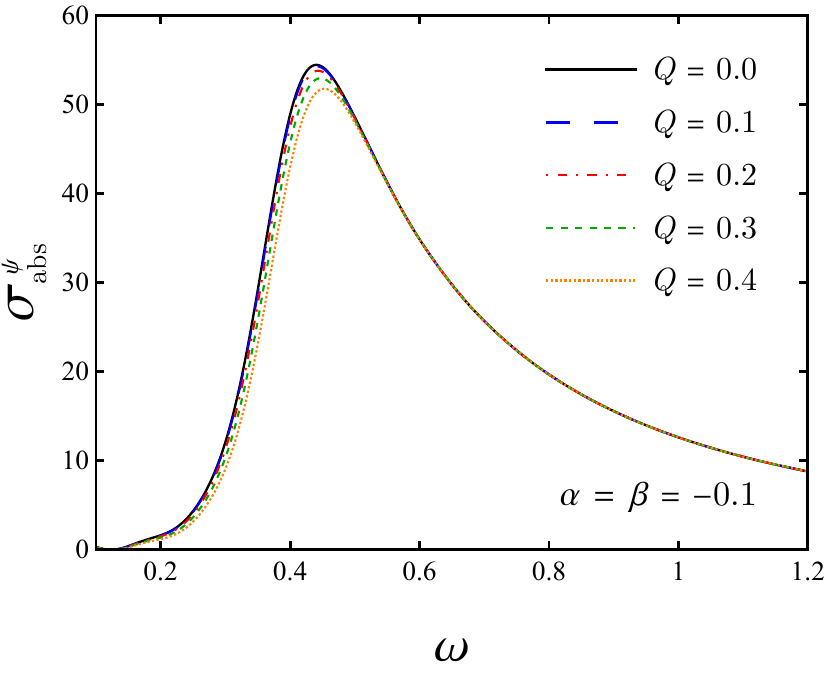}
          \includegraphics[scale=0.53]{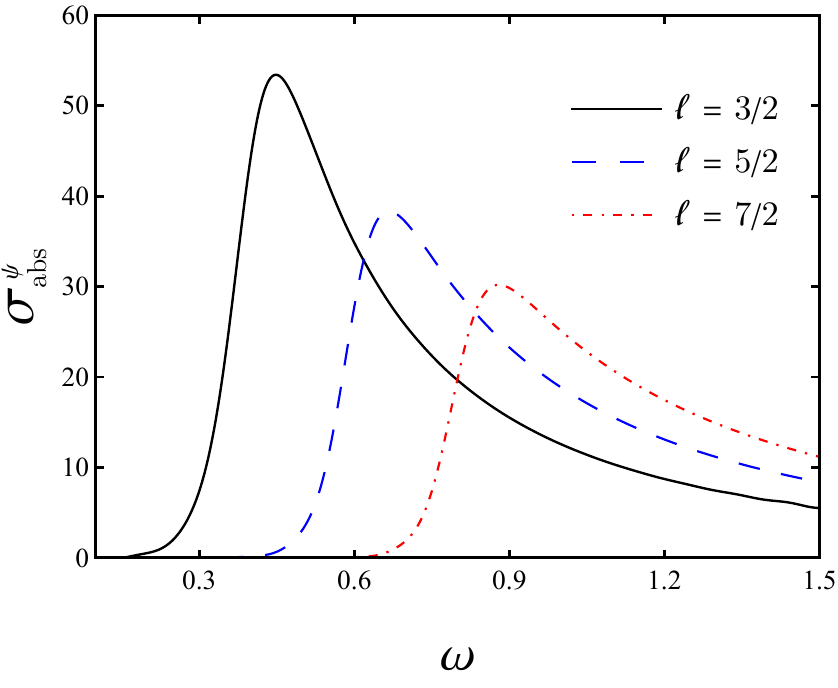}
    \caption{Fermionic absorption cross sections for the case $M=1$. Top Left: Variation with $\alpha$ and $\beta$ for $\ell=1$ and different $Q$. Top panels: A complementary view of the parameter space for $\ell=1$. Bottom panel: Dependence on the multipole number $\ell$ for a fixed configuration with $Q=0.5$ and $\alpha = \beta = -0.1$.}
    \label{fig:spin12abs}
\end{figure}

The absorption cross section for the spin--$1$/2 particles, $\sigma_{\text{abs}}^{\psi}$, is displayed in Fig.~\ref{fig:spin12abs}. The overall spectral profile exhibits the typical pattern of fermionic wave propagation in curved backgrounds: a gradual rise at low frequencies, a dominant resonance peak, and a sequence of damped oscillations at higher frequencies. The top left panel in Fig.~\ref{fig:spin12abs} shows the dependence of $\sigma_{\text{abs}}^{\psi}$ on the $f(R,T)$ coupling parameters $\alpha$ and $\beta$. Variations of these parameters produce only mild changes in the height of the absorption peak, leaving the global shape of the spectrum largely unaffected. This behavior suggests that, within the explored range, the $f(R, T)$ framework introduces only small corrections to the effective potential experienced by the Dirac field, which yield a small increase in the absorption cross section by increasing the absolute value of $\alpha$ and $\beta$.
The influence of the charge parameter $Q$, shown in the top right panel of Fig.~\ref{fig:spin12abs}, is more pronounced: increasing $Q$ leads to a systematic reduction in the peak amplitude and a slight shift of the resonance toward higher frequencies. Finally, the bottom panel in Fig.~\ref{fig:spin12abs} presents the variation with the angular momentum number $\ell$. As expected, larger $\ell$ values suppress the overall magnitude of the absorption cross section and increase the oscillation pattern in frequency space, due to the modification of the effective potential at higher multipole orders.

\subsection{Comparative Analysis of Absorption cross sections}
\label{sec:comparative}

\begin{figure}
    \centering
      \includegraphics[scale=0.5]{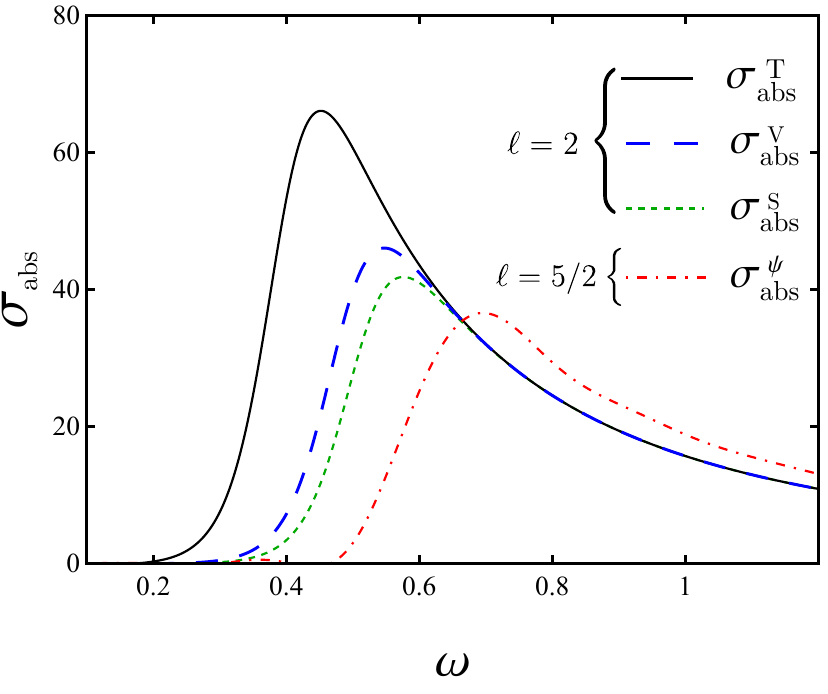}
    \caption{The comparison of the absorption cross section as functions of the frequency $\omega$ is presented for all spin types $0,1/2,1,2$. The $\sigma^{\text{S}}_\text{abs}$, $\sigma^{\text{T}}_\text{abs}$, $\sigma^{\text{V}}_\text{abs}$, $\sigma^{\psi}_\text{abs}$ are denoted to scalar, vector, tensor, and spinorial perturbations, respectively. The values are set to $M = 1$, $Q = 0.1$ and $\alpha = \beta = -0.01$. The angular momentum for bosonic and fermionic cases is fixed at $\ell = 2$ and $\ell = 5/2$, respectively.}
    \label{fig:compareabs}
\end{figure}

The preceding sections have detailed the absorption cross section for individual field perturbations with spin $s = 0, 1/2, 1, 2$, revealing a universal sensitivity to the parameters $\alpha$, $\beta$, $Q$, and $\ell$. A consistent pattern emerges: the cross section is suppressed by increasing electric charge $Q$, experiences subtle modulations from the parameters $\alpha$ and $\beta$, and is progressively constructed from the contributions of higher angular momentum modes.
We now present a unified comparison of the absorption cross section $\sigma_{\text{abs}}(\omega)$ across all spins. This analysis is performed for a fixed background with $M = 1$, $Q = 0.5$, $\alpha = \beta = -0.1$, and for the angular momentum mode $\ell = 2$, contrasting the scalar ($\sigma^{S}_{\text{abs}}$), vector ($\sigma^{V}_{\text{abs}}$), tensor ($\sigma^{T}_{\text{abs}}$) contributions, and $\ell = 5/2$ for spinorial ($\sigma^{\psi}_{\text{abs}}$) case, as illustrated in Fig.~\ref{fig:compareabs}.

The important feature is the strong dependence of the cross section's magnitude and resonant peak on the spin of the perturbing field. The tensor perturbations yield the most pronounced absorption peak, followed sequentially by vector, Dirac, and finally scalar fields, which exhibit the weakest absorption in the middle range of frequency ($0 \leq \omega \leq 1$). In the bosonic cases, the tensor perturbations ($s=2$) not only yield the most pronounced absorption peak but also exhibit this maximum at the lowest frequency. The vector ($s=1$) and scalar ($s=0$) peaks follow in both intensity and spectral position, with the scalar field requiring the highest frequency to reach its maximum absorption. This systematic trend—where higher--spin bosonic fields achieve peak absorption at progressively lower frequencies—suggests that spin increases the coupling to the gravitational potential. The fermionic ($s=1/2$) mode, on the other hand, presents a distinctive case. While its peak amplitude lies after the tensor, vector, and scalar cases, its resonance occurs at a frequency higher that all the bosonic sequences, highlighting the unique nature of the effective potential governing Dirac fields. Thereby, the full analysis of all spins reveals the following pattern
\begin{equation}
\sigma^{T}_{\text{abs}}>\sigma^{V}_{\text{abs}}>\sigma^{S}_{\text{abs}}>\sigma^{\psi}_{\text{abs}}.
\end{equation}

It is worth mentioning that this result is consistent with the comparison of the greybody factor in the previous section and the results demonstrated in Eq. \eqref{comparisongreybodyall}. \\
Moreover, at high frequencies, the absorption cross sections for all spin types converge toward the geometric optics limit, becoming nearly independent of the field's spin. However, the distinct variations in peak magnitude and spectral position at intermediate energies unequivocally demonstrate the critical role of spin in shaping the scattering and absorption dynamics of fields in curved spacetime.

\section{Evaporation and emission rate}\label{SectionVII}

In this section, we qualitatively examine the black hole’s evaporation process described by Eq.~(\ref{generalfr}). The analysis is based primarily on the \textit{Stefan--Boltzmann} law, which provides the thermodynamic framework for evaluating the energy loss due to radiation \cite{ong2018effective}
\ie
\label{slawbotz}
\frac{\mathrm{d}M}{\mathrm{d}t}  =  - a  \Bar{\Gamma}_{\ell \omega}  \sigma_{\ell \omega} \, T^{4},
\fe
where
\ie
\sigma_{\ell\,\omega} = \frac{\pi(2\ell + 1)}{\omega^{2}} |T_{b}^{\text{S,V,T},\psi}|,
\fe
with $\sigma_{\ell \omega}$ denotes the \textit{partial} cross sectional area, $a$ is the radiation constant, $\Bar{\Gamma}_{\ell \omega}$ represents the greybody factors and $T$ represents the \textit{Hawking} temperature. In the following subsections, we analyze the quantities $\Bar{\Gamma}^{\text{S}}_{\ell \omega}$ (scalar field), $\Bar{\Gamma}^{\text{V}}_{\ell \omega}$ (vector field), $\Bar{\Gamma}^{\text{T}}_{\ell \omega}$ (tensor field), and $\Bar{\Gamma}^{\psi}_{\ell \omega}$ (spinor field). As in the earlier discussions, all spin configurations—$0$, $1$, $2$, and $1/2$—are examined. Consistent with the analysis presented in the absorption section, analytical solutions could not be obtained (except for the spinorial case, as will be shown). Consequently, numerical techniques are employed to carry out this part of the investigation.


\subsection{Spin--$0$ particle modes}

For the subsequent calculations in this section, aimed at estimating the black hole evaporation lifetime for the spin--$0$ configuration, we employ Eq. (\ref{slawbotz}). For scalar particle emission, we take
$\bar{\Gamma}^{\text{S}}_{\ell \omega} = |T_{b}^{\text{S}}|$ and
$\sigma^{\text{Spin}\,0}_{\ell \omega}$. Since the expression for $\mathrm{d}M/\mathrm{d}t$ is too much lengthy for the scalar case, it will not be explicitly presented here. Moreover, after applying Eq.~(\ref{slawbotz}) to estimate the lifetime, we obtain
\ie
\int \mathrm{d}t = \int_{M_{f}}^{M_{i}} \frac{\mathrm{d}M}{a  \Bar{\Gamma}_{\ell \omega}  \sigma_{\ell \omega} \, T^{4}}.
\label{tgggg}
\fe
To carry out the calculations, we expand $1/(a\, \bar{\Gamma}_{\ell \omega}\, \sigma_{\ell \omega}\, T^{4})$ up to fourth order in $Q$ and first order in $\alpha$. Unless stated otherwise, this procedure is applied to all perturbations considered throughout this work. In addition, we have considered $\ell = 2$, which will facilitate to compare all effects ascribed to the perturbations investigated in this manuscript.

As already pointed out in the previous section, no closed--form solution exists for the evaporation lifetime, so a purely numerical approach is used here as well. For the computations, we set the initial mass to $M_{i}=2$, and take the final mass $M_{f}$ as the remnant reported in Ref. \cite{heidari2025gravitational}: $ M_{f} \approx \, \frac{Q}{2} + \frac{\alpha  (1-2 \beta )}{20 Q}$.
The resulting values for the evaporation lifetime are displayed in Tab. \ref{evapS}.
Inspection of the table shows that $t^{\text{S}}_{\text{evap-final}}$ decreases rapidly as $Q$ grows. Conversely, more negative values of $\alpha = \beta$ lead to a longer lifetime. Finally, increasing the frequency range $\omega$ produces a shorter overall evaporation time.

\begin{figure}
    \centering
      \includegraphics[scale=0.435]{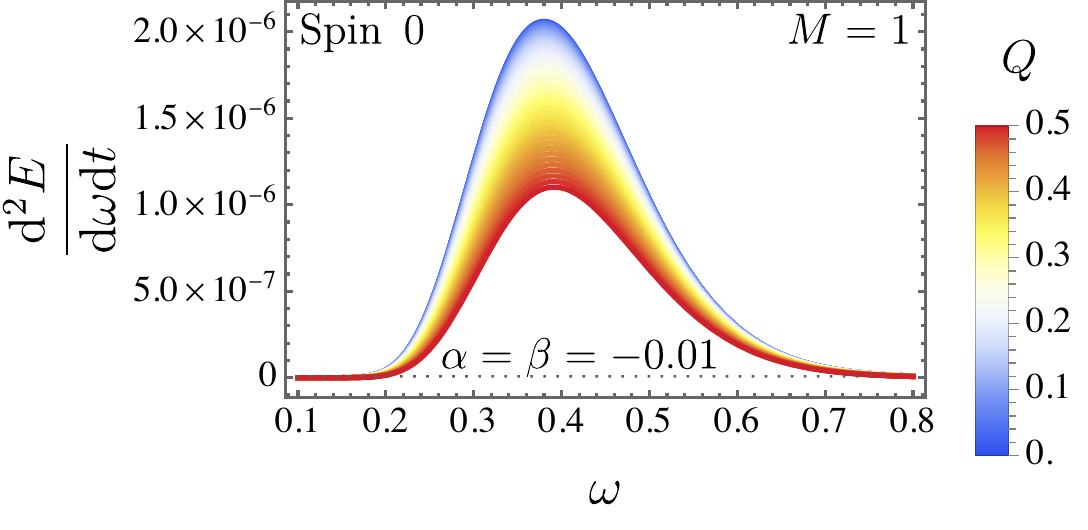}
       \includegraphics[scale=0.435]{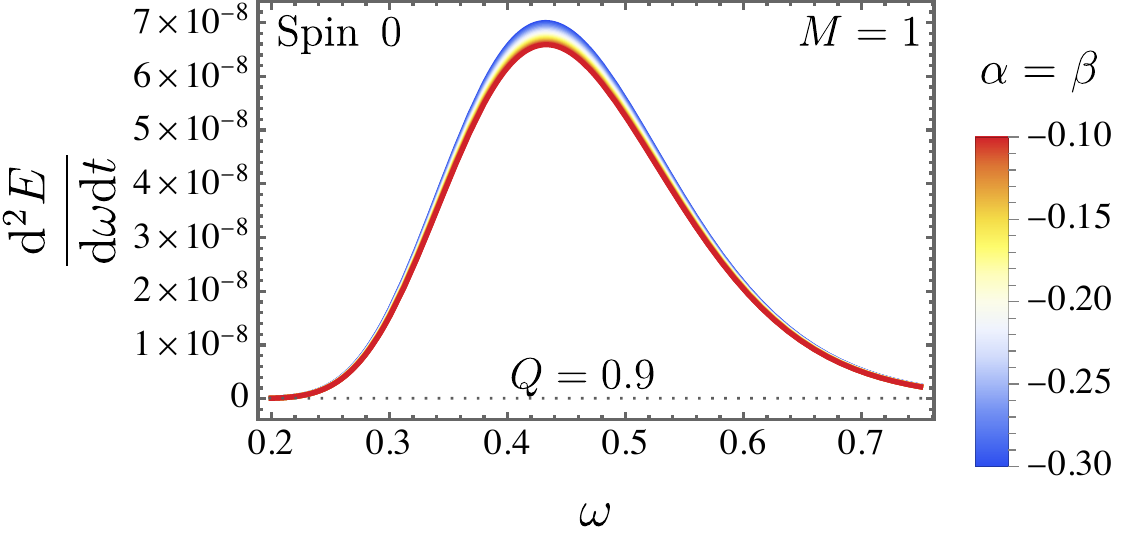}
    \caption{The energy emission rate for spin-$0$ particle modes is displayed. The left panel shows the dependence on the charge $Q$ with $M = 1$ and $\alpha = \beta = -0.01$, while the right panel displays the variation with $\alpha = \beta$ for fixed $M = 1$ and $Q = 0.9$. Here, we consider $\ell =2$.  }
    \label{spin0emission}
\end{figure}

We now turn to the analysis of the energy emission rate
\ie
\label{energyemission}
	\frac{{{\mathrm{d}^2}E}}{{\mathrm{d}\omega \mathrm{d}t}} = \frac{{2{\pi ^2}{\sigma}^{\text{Spin}\,0}_{\ell \omega}}}{{{e^{\frac{\omega }{T}}} - 1}} {\omega ^3}.
\fe
With these considerations, the results are shown in Fig. \ref{spin0emission}. From this figure, we observe that increasing $Q$ (for $\alpha = \beta = -0.01$) enhances the intensity of the energy emission rate. In contrast, for a fixed value of $Q = 0.9$, increasing $\alpha = \beta$ produces the opposite behavior, resulting in a reduction of the emission rate. The outcomes are obtained by considering $\ell =2$.

\begin{figure}
    \centering
      \includegraphics[scale=0.435]{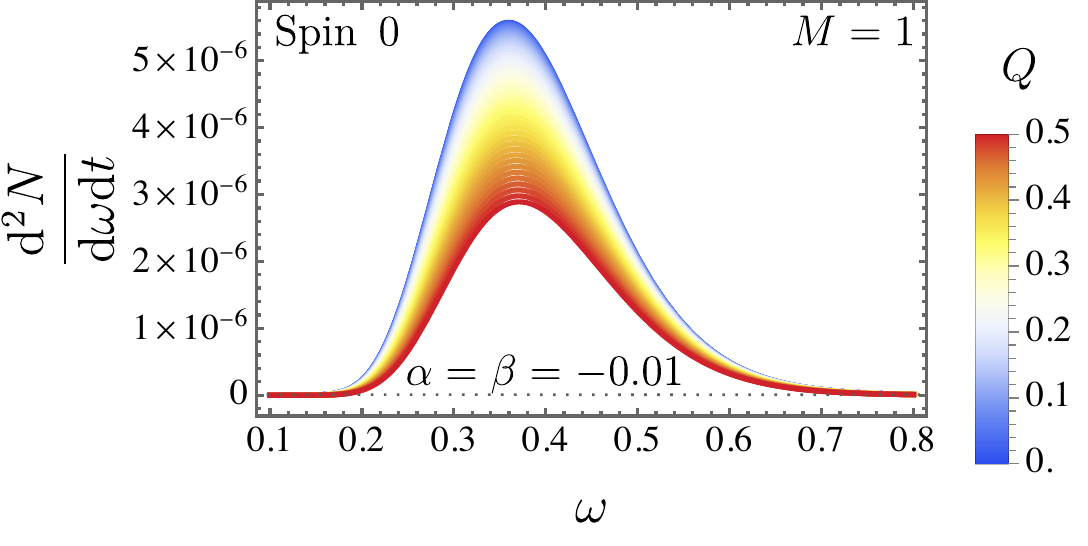}
       \includegraphics[scale=0.435]{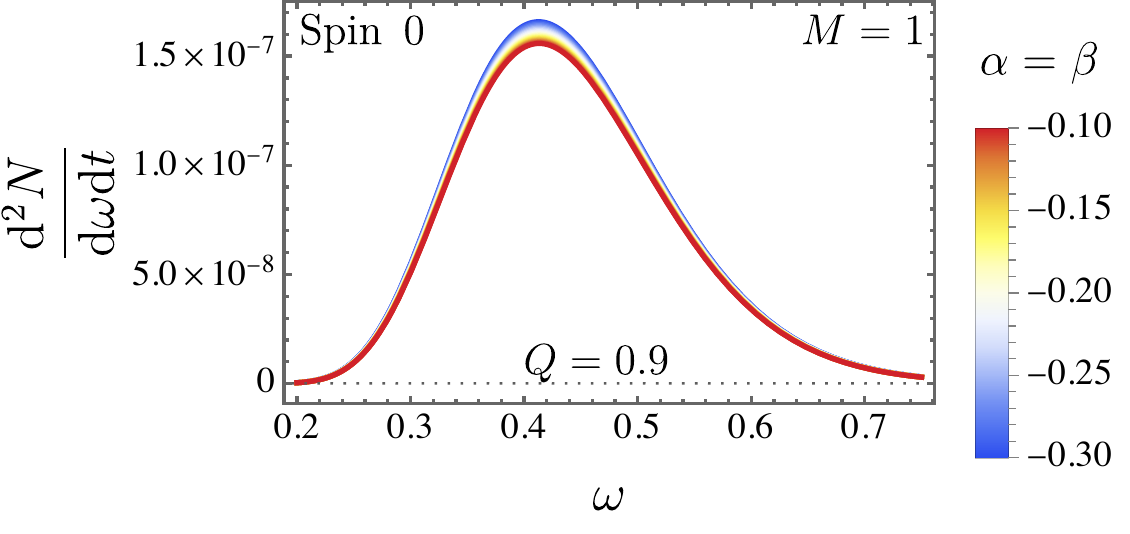}
    \caption{The particle emission rate for spin--$0$ modes is shown. The left panel illustrates how the spectrum varies with $Q$ when $M = 1$ and $\alpha = \beta = -0.01$, whereas the right panel depicts the effect of changing $\alpha = \beta$ while keeping $M = 1$ and $Q = 0.9$ constant. Here, we consider $\ell=2$.}
    \label{spin0emissionparticle}
\end{figure}

Moreover, the particle emission rate can be expressed as follows
\ie
\frac{\mathrm{d}^{2}N}{\mathrm{d}\omega \mathrm{d}t}
= \frac{2\pi^{2}\,\sigma^{\text{Spin}\,0}_{\ell \omega}\,\omega^{2}}
       {{{e^{\frac{\omega }{T}}} - 1}}.
       \label{particlesseee}
\fe
Based on the previous assumptions, we now present Fig. \ref{spin0emissionparticle}. The same qualitative behavior observed for the energy emission rate is confirmed here: for $\alpha = \beta = -0.01$, increasing $Q$ enhances the emission intensity; conversely, when $Q$ is kept fixed at $0.9$, increasing $\alpha = \beta$ produces the opposite effect, leading to a reduction in the emission rate. Again, the results are generated by regarding $\ell =2$.

\begin{table}[!h]
\begin{center}
\begin{tabular}{c c c c || c c c c || c c c c } 
 \hline\hline \hline
 $Q$ & $\alpha=\beta$ & $\omega$ & $t^{\text{S}}_{\text{evap-final}}$ & $Q$ & $\alpha=\beta$ & $\omega$ &  $t^{\text{S}}_{\text{evap-final}}$ & $Q$ & $\alpha=\beta$ & $\omega$ &  $t^{\text{S}}_{\text{evap-final}}$  \\ [0.2ex] 
 \hline
  0.10  & -0.01 & 0.10 & $4.96\times10^{626}$ & 0.5  & -0.01 & 0.10 & $1.47\times 10^{114}$ & 0.5  & -0.01 & 0.10 & $1.47\times 10^{114}$  \\
  
  0.20  & -0.01 & 0.10 & $7.40\times10^{288}$ & 0.5  & -0.02 & 0.10 & $4.49\times 10^{114}$ & 0.5  & -0.01 & 0.20 & $6.16\times 10^{57}$  \\
  
  0.30  & -0.01 & 0.10 & $1.77\times 10^{190}$ & 0.5  & -0.03 & 0.10 & $1.44\times10^{115}$ & 0.5  & -0.01 & 0.30 & $1.20\times 10^{39}$  \\
  
  0.40  & -0.01 & 0.10 & $2.82\times 10^{142}$ & 0.5  & -0.04 & 0.10 & $4.90\times 10^{115}$ & 0.5  & -0.01 & 0.40 & $5.84\times 10^{29}$  \\
  
   0.50  & -0.01 & 0.10 & $1.47\times 10^{114}$ & 0.5  & -0.05 & 0.10 & $1.75\times 10^{116}$ & 0.5  & -0.01 & 0.50 & $1.60\times 10^{24}$  \\
   
 0.60  & -0.01 & 0.10 & $2.96\times10^{95}$ & 0.5  & -0.06 & 0.10 & $6.67 \times 10^{116}$ & 0.5  & -0.01 & 0.60 & $3.27\times 10^{20}$  \\ 

 0.70  & -0.01 & 0.10 & $1.58\times 10^{82}$ & 0.5  & -0.07 & 0.10 & $2.68\times 10^{117}$ & 0.5  & -0.01 & 0.70 & $7.80\times 10^{17}$  \\
 
 0.80  & -0.01 & 0.10 & $1.98\times 10^{72}$ & 0.5  & -0.08 & 0.10 & $1.14\times 10^{118}$ & 0.5  & -0.01 & 0.80 & $8.63\times 10^{15}$ \\
 
 0.90  & -0.01 & 0.10 & $4.25\times 10^{64}$  & 0.5  & -0.09 & 0.10 & $5.18\times 10^{118}$ & 0.5  & -0.01 & 0.90 & $2.65\times 10^{14}$ \\
 
 0.99  & -0.01 & 0.10 & $1.18\times 10^{59}$ & 0.5  & -0.10 & 0.10 & $2.50\times 10^{119}$ & 0.5  & -0.01 & 0.99 & $2.13\times 10^{13}$  \\ 
 [0.2ex] 
 \hline \hline \hline
\end{tabular}
\caption{\label{evapS} Quantitative values of the evaporation lifetime driven by the emission of spin--0 particle modes, $t^{\text{S}}_{\text{evap-final}}$, for different configurations of the parameters $Q$, $\alpha = \beta$, and $\omega$. Here, it is considered $\ell =2$ and $M_{i} =2$.}
\end{center}
\end{table}


\subsection{Spin--$1$ particle modes}

Using the same procedure described in the previous section, Eq. (\ref{slawbotz}) is now employed to estimate the evaporation lifetime in the case of vector perturbations. In this setting, we adopt the greybody factors $\bar{\Gamma}^{\text{V}}_{\ell \omega}$ and the corresponding \textit{partial} cross sections $\sigma^{\text{Spin 1}}_{\ell \omega}$, which account exclusively for spin--1 modes. As in the scalar case, the results concerning to the evaporation process are obtained analytically and are presented in Tab. \ref{evapV}.

In general lines, three trends can be observed. First, increasing the electric charge $Q$ leads to a shorter evaporation lifetime for spin--1 modes, $t^{\text{V}}_{\text{evap-final}}$, while keeping $\alpha=\beta$ and $\omega$ fixed. Second, decreasing $\alpha=\beta$ with $Q$ and $\omega$ fixed also shortens the lifetime. Third, higher frequencies $\omega$ enhance the emission power, causing the black hole to evaporate faster when $Q$ and $\alpha=\beta$ are kept constant.

Since the previous subsection modeled the emission of spin--0 quanta only, while the present analysis focuses solely on spin--1 emission, a natural question arises: which channel leads to faster evaporation? To answer this, one must compare Tables \ref{evapS} (spin--0) and \ref{evapV} (spin--1). The comparison shows that $t^{\text{V}}_{\text{evap-final}} < t^{\text{S}}_{\text{evap-final}}$,
meaning that a black hole emitting only spin--1 particles evaporates more rapidly than one emitting only spin--0 particles.

We now examine the energy emission rate using Eq. (\ref{energyemission}), replacing the \textit{partial} cross section with $\sigma_{\ell \omega}^{\text{Spin}\, 1}$.
Unlike the scalar perturbation case, for vector perturbations the effective potential $\mathcal{V}_{\text{V}}(r,\alpha,\beta,Q)$ becomes trivial in the limit $\ell \to 0$.

\begin{figure}
    \centering
      \includegraphics[scale=0.435]{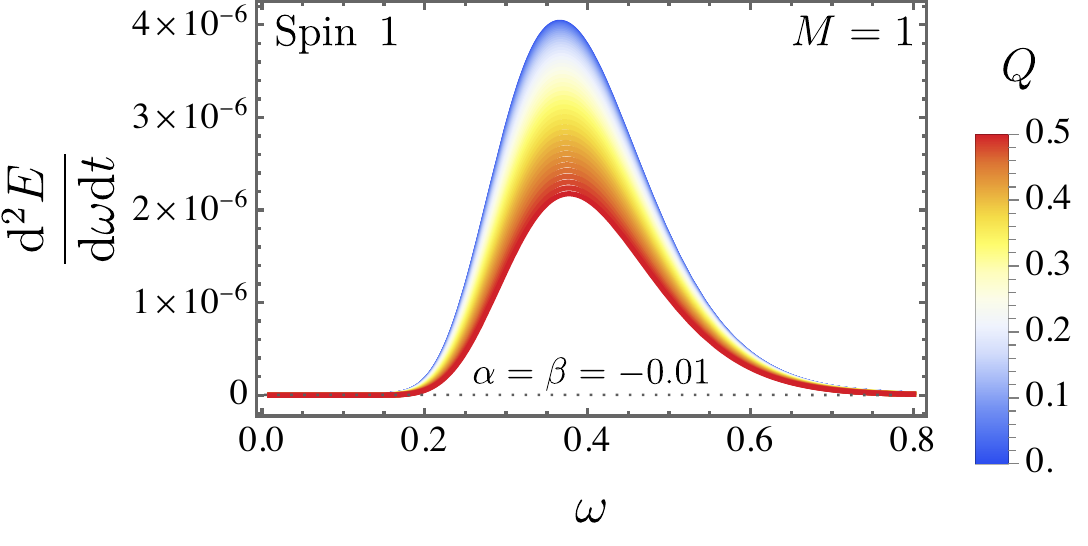}
       \includegraphics[scale=0.435]{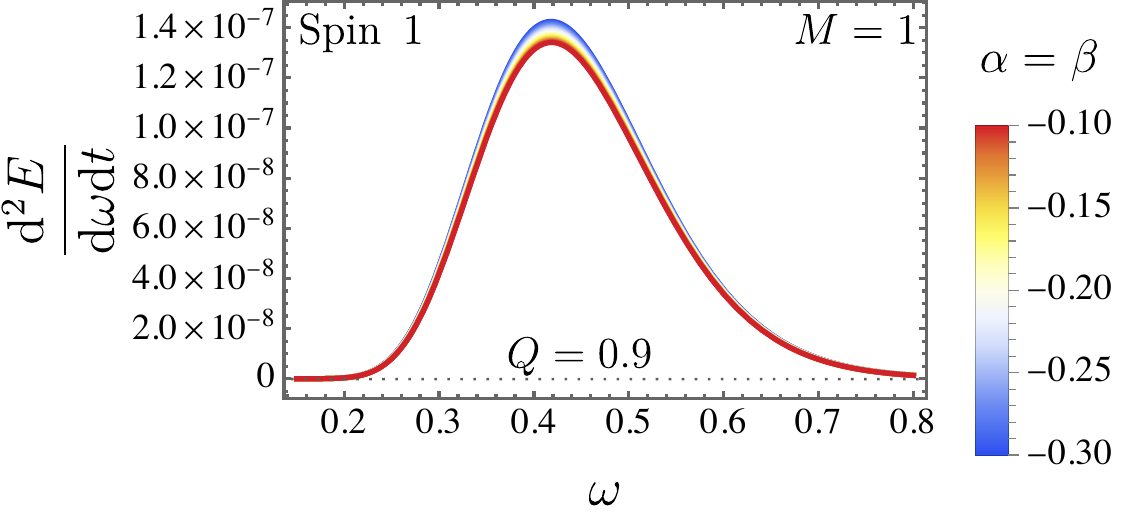}
    \caption{The energy emission rate for spin--$1$ particle modes is displayed. The left panel shows the dependence on the charge $Q$ with $M = 1$ and $\alpha = \beta = -0.01$, while the right panel displays the variation with $\alpha = \beta$ for fixed $M = 1$ and $Q = 0.9$.  }
    \label{spin1emission}
\end{figure}

Figure \ref{spin1emission} shows the resulting energy emission rate:
the upper panel illustrates its dependence on $Q$, while the lower panel shows the effect of varying $\alpha = \beta$. All results are conducted by considering $\ell=2$.

\begin{figure}
    \centering
      \includegraphics[scale=0.55]{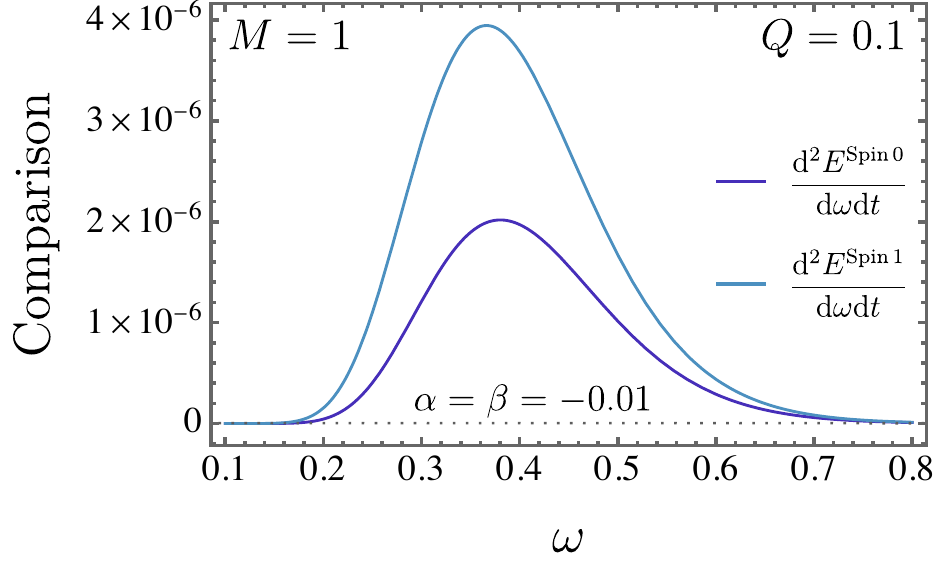}
    \caption{The comparison of the energy emission rates is shown for spin--0 and spin--1 cases with $\alpha = \beta = -0.01$, $M = 1$, $Q = 0.1$, and $\ell = 2$.}
    \label{emissioncompsv}
\end{figure}

We observe that increasing $Q$ suppresses the emission rate, whereas decreasing $\alpha = \beta$ enhances it—mirroring the qualitative behavior found in the scalar (spin--$0$) case.
Furthermore, by comparing the results for spin--$0$ and spin--$1$ modes, we observe that the energy emission is higher for the vector case, as illustrated in Fig.~\ref{emissioncompsv}. Therefore, we can express that
\[\frac{{{\mathrm{d}^2}E^{\text{Spin} \,1}}}{{\mathrm{d}\omega \mathrm{d}t}} > \frac{{{\mathrm{d}^2}E^{\text{Spin} \,0}}}{{\mathrm{d}\omega \mathrm{d}t}}.\]
Notice that the above result is fully consistent with our previous findings. In particular, for spin--1 particle modes, the black hole exhibited a shorter evaporation lifetime before reaching its final state. In other words, more energy is released through spin--1 particle modes than through spin--0 ones, as confirmed by the energy emission rate.

\begin{figure}
    \centering
      \includegraphics[scale=0.435]{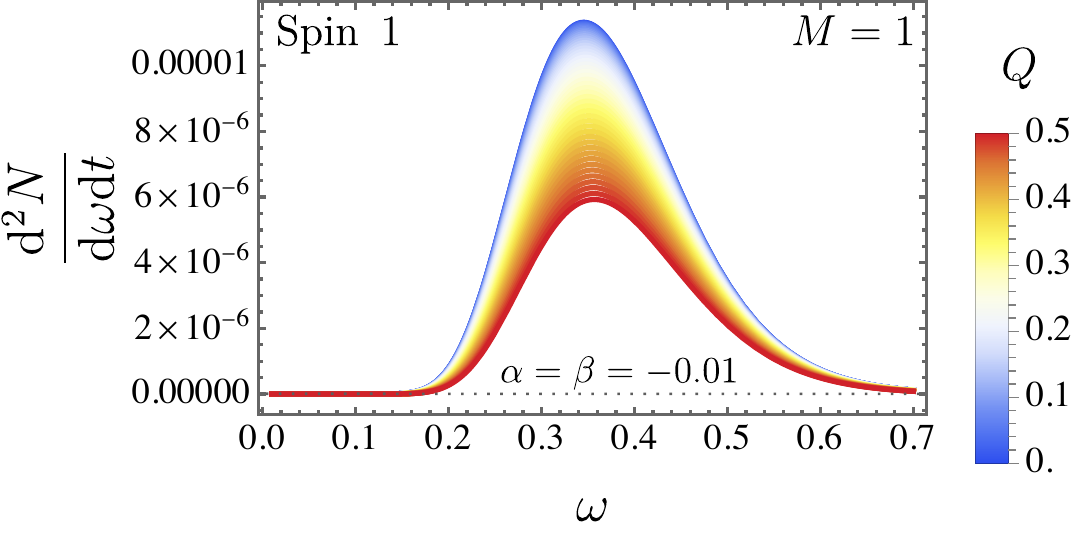}
       \includegraphics[scale=0.435]{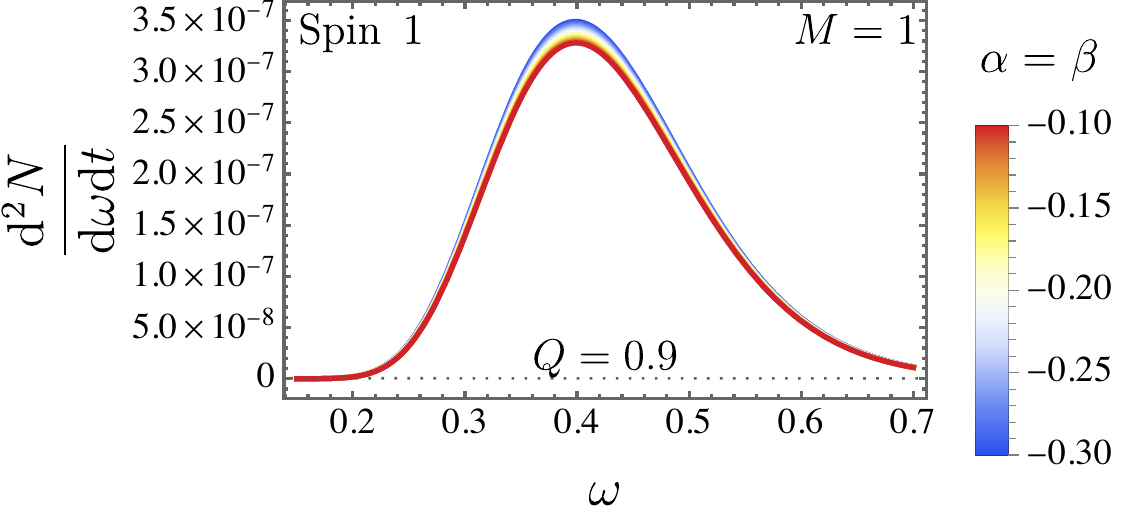}
    \caption{The particle emission rate for spin--$1$ modes is shown. The left panel illustrates how the spectrum varies with $Q$ when $M = 1$ and $\alpha = \beta = -0.01$, whereas the right panel depicts the effect of changing $\alpha = \beta$ while keeping $M = 1$ and $Q = 0.9$ constant. }
    \label{spin1emissionparticle}
\end{figure}

For completeness with the previous sections, we now analyze the particle emission rate for spin--$1$ modes using Eq. (\ref{particlesseee}), replacing $\sigma^{\text{Spin}\,0}_{\ell \omega}$ with $\sigma^{\text{Spin}\,1}_{\ell \omega}$.
Figure \ref{spin1emissionparticle} displays the resulting particle emission rate. As in the scalar case, increasing $Q$ leads to a reduction in the emission rate, whereas decreasing $\alpha = \beta$ produces a slight enhancement.

\begin{figure}
    \centering
      \includegraphics[scale=0.55]{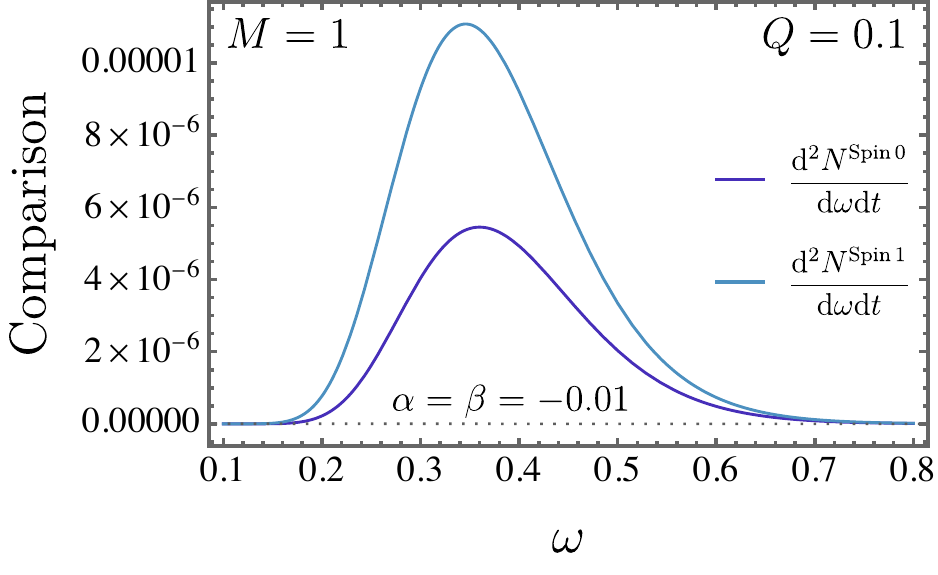}
    \caption{The comparison of the particle emission rates is exhibited for spin--0 and spin--1 cases with $\alpha = \beta = -0.01$, $M = 1$, $Q = 0.1$, and $\ell = 2$.}
    \label{emissioncompparticlesv}
\end{figure}

Moreover, following the same procedure adopted for the energy emission, we now compare the results for the particle emission rate between the spin--0 and spin--1 cases. This comparison is presented in Fig.~\ref{emissioncompparticlesv}. Consistent with the behavior observed for the energy emission rate, the particle emission rate is higher for spin--1 modes than for spin--0 modes. In other words, spin--1 particles are emitted more abundantly. In this manner, we conclude that 
\[\frac{{{\mathrm{d}^2}N^{\text{Spin} \,1}}}{{\mathrm{d}\omega \mathrm{d}t}} > \frac{{{\mathrm{d}^2}N^{\text{Spin} \,0}}}{{\mathrm{d}\omega \mathrm{d}t}}.\]

\begin{table}[!h]
\begin{center}
\begin{tabular}{c c c c || c c c c || c c c c } 
 \hline\hline \hline
 $Q$ & $\alpha=\beta$ & $\omega$ & $t^{\text{V}}_{\text{evap-final}}$ & $Q$ & $\alpha=\beta$ & $\omega$ &  $t^{\text{V}}_{\text{evap-final}}$ & $Q$ & $\alpha=\beta$ & $\omega$ &  $t^{\text{V}}_{\text{evap-final}}$  \\ [0.2ex] 
 \hline
  0.10  & -0.01 & 0.10 & $2.09\times 10^{578}$ & 0.5  & -0.01 & 0.10 & $2.72\times10^{105}$ & 0.5  & -0.01 & 0.10 & $2.72\times 10^{105}$  \\
  
  0.20  & -0.01 & 0.10 & $3.75\times 10^{266}$ & 0.5  & -0.02 & 0.10 & $7.63\times10^{105}$ & 0.5  & -0.01 & 0.20 & $2.61\times 10^{53}$  \\
  
  0.30  & -0.01 & 0.10 & $3.93\times10^{175}$ & 0.5  & -0.03 & 0.10 & $2.24\times 10^{106}$ & 0.5  & -0.01 & 0.30 & $1.45\times 10^{36}$  \\
  
  0.40  & -0.01 & 0.10 & $3.25\times 10^{131}$ & 0.5  & -0.04 & 0.10 & $6.92\times 10^{106}$ & 0.5  & -0.01 & 0.40 & $3.77\times 10^{27}$  \\
  
   0.50  & -0.01 & 0.10 & $2.72\times 10^{105}$ & 0.5  & -0.05 & 0.10 & $2.25\times 10^{107}$ & 0.5  & -0.01 & 0.50 & $2.82\times 10^{22}$  \\
   
 0.60  & -0.01 & 0.10 & $1.59\times10^{88}$ & 0.5  & -0.06 & 0.10 & $7.70\times 10^{107}$ & 0.5  & -0.01 & 0.60 & $1.13\times 10^{19}$  \\ 

 0.70  & -0.01 & 0.10 & $9.38\times10^{75}$ & 0.5  & -0.07 & 0.10 & $2.78\times 10^{108}$ & 0.5  & -0.01 & 0.70 & $4.35\times 10^{16}$  \\
 
 0.80  & -0.01 & 0.10 & $7.05\times10^{66}$ & 0.5  & -0.08 & 0.10 & $1.06\times 10^{109}$ & 0.5  & -0.01 & 0.80 & $6.90\times 10^{14}$ \\
 
 0.90  & -0.01 & 0.10 & $6.11\times10^{59}$  & 0.5  & -0.09 & 0.10 & $4.28\times 10^{109}$ & 0.5  & -0.01 & 0.90 & $2.80\times 10^{13}$ \\
 
 0.99  & -0.01 & 0.10 & $4.69\times10^{54}$ & 0.5  & -0.10 & 0.10 & $1.83\times 10^{110}$ & 0.5  & -0.01 & 0.99 & $2.76\times 10^{12}$  \\ 
 [0.2ex] 
 \hline \hline \hline
\end{tabular}
\caption{\label{evapV} Quantitative values of the evaporation lifetime driven by the emission of spin--1 particle modes, $t^{\text{V}}_{\text{evap-final}}$, for different configurations of the parameters $Q$, $\alpha = \beta$, and $\omega$.
}
\end{center}
\end{table}


\subsection{Spin--$2$ particle modes}

We shall devote our attention to investigate how the spin--$2$ particle are emitted from our black hole under consideration. Initially, as we have accopmlished in our previous sections, we shall begin by studying the evaporation lifetime associated to such a spin. After that, we shall compare it with the previous cases studyed so far. Finally, we investigate the emission process by considering the energy and particle emission; and also again, we compare the outcomes with the previous spin cases.

Then, we consider $\Bar{\Gamma}^{\text{T}}_{\ell \omega}$ as the corresponding version of spin--$2$ modes in Eq. (\ref{slawbotz}). Due to the computational power limitations, we have expanded such an expression in order to obtain feasable outcomes $\mathrm{d}M/\mathrm{d}t$ is too much lengthy for the scalar case, it will not be explicitly presented here.

In Tab.~\ref{evapT}, similarly to the previously studied cases, the final evaporation lifetime $t^{\text{T}}_{\text{evap-final}}$ decreases as $Q$ and $\omega$ increase; the same trend is also observed when $\alpha = \beta$ decreases. A natural question then arises: compared with scalar and vector particle modes, does the tensor mode lead to a faster or slower evaporation? To address this, let us directly compare Tabs.~\ref{evapS} (spin--0), \ref{evapV} (spin--1), and \ref{evapT} (spin--2). The comparison reveals that
\ie
t^{\text{S}}_{\text{evap-final}} > t^{\text{V}}_{\text{evap-final}}>t^{\text{T}}_{\text{evap-final}},
\label{comparisontimesss}
\fe
indicating that a black hole emitting only spin--2 particles evaporates more rapidly than one emitting only spin--0 or spin--1 particles. At this point, one might question the consistency of our approach, since in deriving the greybody bounds for the tensor perturbations in Eq.~(\ref{tensorbgreybody}), we expanded $\mathcal{V}_{\text{T}}(r,\alpha,\beta,Q)/f(r)$ only up to first order in $\alpha$ and up to fourth order in $Q$. Therefore, a natural question arises: is it reasonable to compare the black hole lifetime for the tensor perturbations using this approximation, while no such expansion was applied to the scalar and vector cases? The answer is yes. For the scalar and vector perturbations, the corresponding “filter’’ is introduced through the expansion of $1/(a\, \bar{\Gamma}_{\ell \omega}\, \sigma_{\ell \omega}\, T^{4})$ up to fourth order in $Q$ and first order in $\alpha$. In this manner, all perturbations are treated consistently and can be compared on equal footing, as done in this work.

\begin{figure}
    \centering
      \includegraphics[scale=0.51]{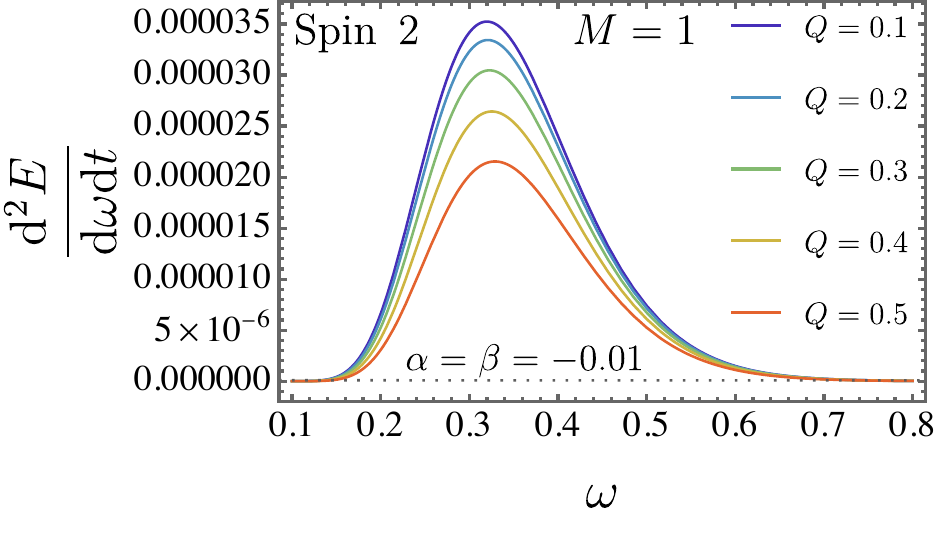}
       \includegraphics[scale=0.51]{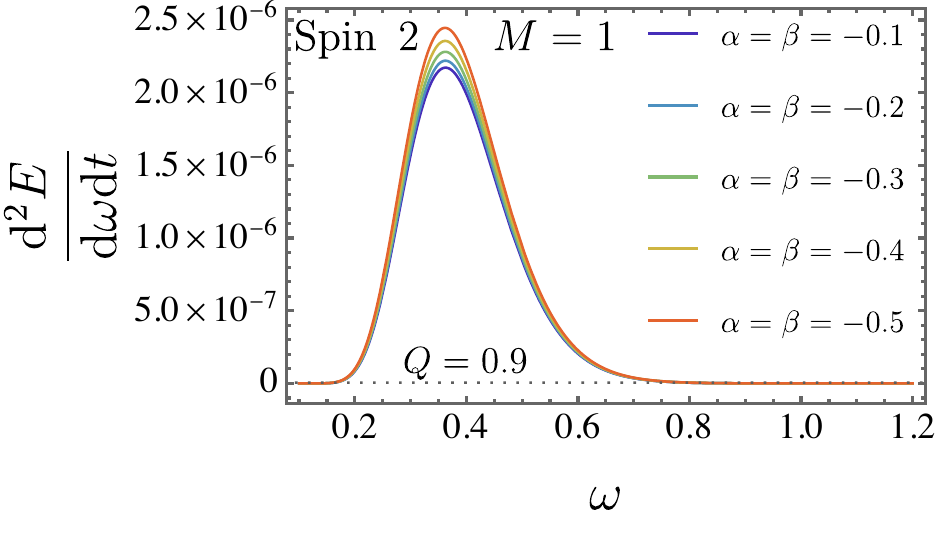}
    \caption{The energy emission rate for spin-$2$ particle modes is displayed. The left panel shows the dependence on the charge $Q$ with $M = 1$ and $\alpha = \beta = -0.01$, while the right panel displays the variation with $\alpha = \beta$ for fixed $M = 1$ and $Q = 0.9$.  }
    \label{spin2emission}
\end{figure}

\begin{figure}
    \centering
      \includegraphics[scale=0.51]{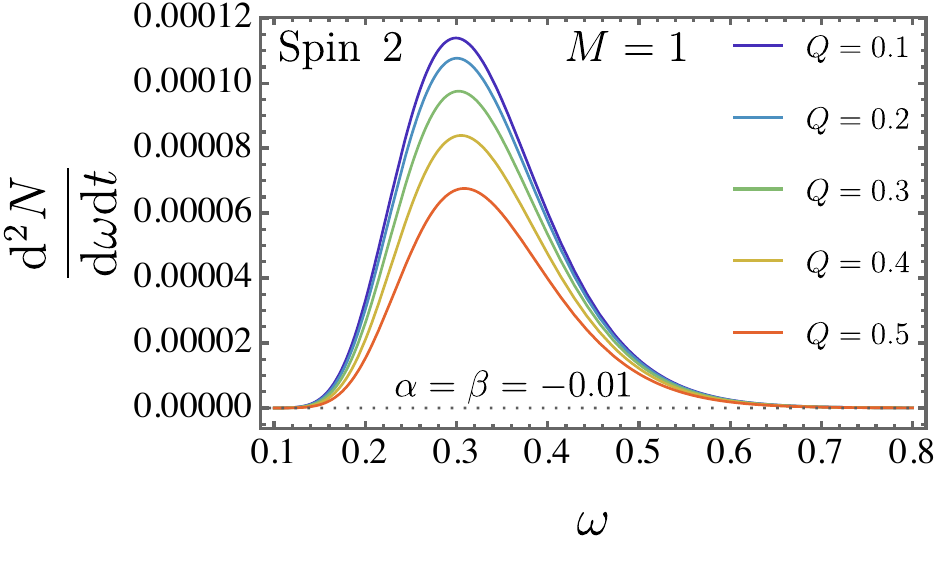}
       \includegraphics[scale=0.51]{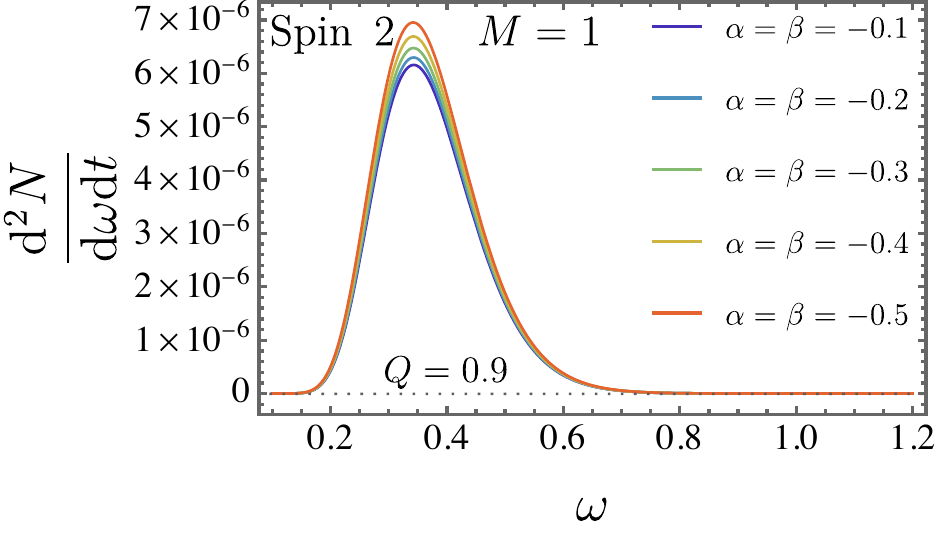}
    \caption{The particle emission rate for spin--$2$ modes is shown. The left panel illustrates how the spectrum varies with $Q$ when $M = 1$ and $\alpha = \beta = -0.01$, whereas the right panel depicts the effect of changing $\alpha = \beta$ while keeping $M = 1$ and $Q = 0.9$ constant. }
    \label{spin2emissionparticle}
\end{figure}

In Fig.~\ref{spin2emission}, we present the energy emission rate for spin--2 particle modes, considering variations of $Q$ (upper panel) and of $\alpha = \beta$ (lower panel). In line with the spin--0 and spin--1 cases, increasing $Q$ (with $\alpha = \beta = -0.01$ fixed) leads to a decrease in the corresponding emission intensity. Conversely, reducing the values of $\alpha = \beta$ (with $Q = 0.9$) results in an enhancement of the emission. Moreover, Fig.~\ref{spin2emissionparticle} displays the particle emission rate. Overall, the same pattern observed in the previous cases is preserved. It is worth mentioning that, in the computation of the emission rate for tensor perturbations, we adopted the complete expression of the corresponding greybody bound—without expanding $\mathcal{V}_{\text{T}}(r,\alpha,\beta,Q)/f(r)$. This approach was chosen because no additional “filter’’ was applied in the analysis of emission rates, unlike in the estimation of the evaporation lifetime. Accordingly, since the greybody factors for the scalar and vector perturbations were also evaluated in their full form, the tensor case was treated in the same manner to guarantee a consistent and fair comparison among all perturbative sectors.

\begin{figure}
    \centering
      \includegraphics[scale=0.55]{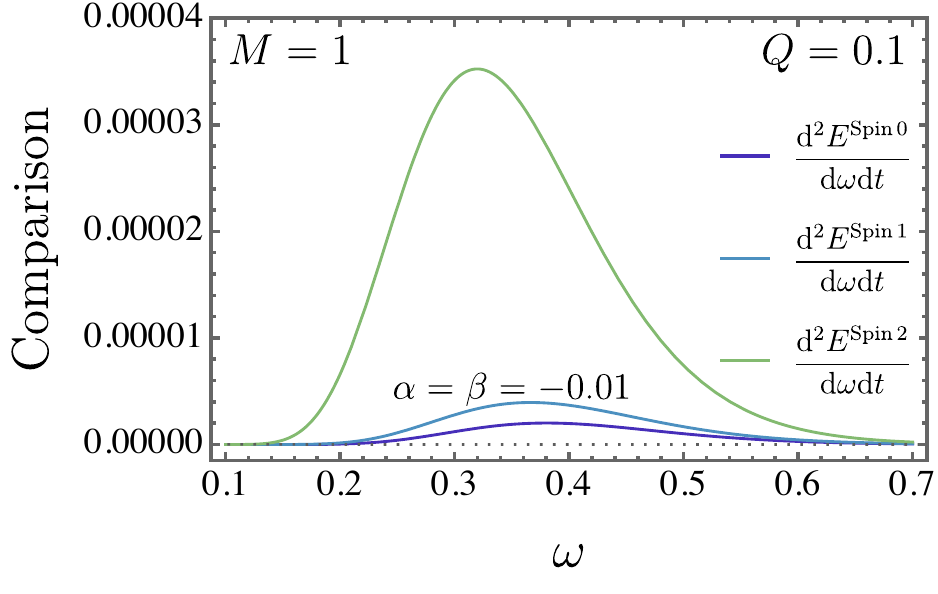}
    \caption{The comparison of the particle emission rates is exhibited for spin--0, spin--1 and spin--2 cases with $\alpha = \beta = -0.01$, $M = 1$, $Q = 0.1$, and $\ell = 2$.}
    \label{emissioncompparticlest}
\end{figure}

Now, let us compare all the emission rates obtained so far, as illustrated in Fig.~\ref{emissioncompparticlest}. Therefore, we can conclude that
\ie
\label{compemissionenergy}
\frac{{{\mathrm{d}^2}E^{\text{Spin} \,2}}}{{\mathrm{d}\omega \mathrm{d}t}} > \frac{{{\mathrm{d}^2}E^{\text{Spin} \,1}}}{{\mathrm{d}\omega \mathrm{d}t}} > \frac{{{\mathrm{d}^2}E^{\text{Spin} \,0}}}{{\mathrm{d}\omega \mathrm{d}t}}.
\fe
In other words, this result shows that the black hole releases more energy through spin--2 modes, followed by spin--1, while spin--0 contributes the least. In this manner, the particle emission can also be compared via Figs. \ref{spin0emissionparticle}, \ref{spin1emissionparticle}, \ref{spin2emissionparticle}. Therefore, we have 
\ie
\label{compparticleemission}
\frac{{{\mathrm{d}^2}N^{\text{Spin} \,2}}}{{\mathrm{d}\omega \mathrm{d}t}} > \frac{{{\mathrm{d}^2}N^{\text{Spin} \,1}}}{{\mathrm{d}\omega \mathrm{d}t}} > \frac{{{\mathrm{d}^2}N^{\text{Spin} \,0}}}{{\mathrm{d}\omega \mathrm{d}t}}.
\fe
Notice that all the results for the emission rate are fully consistent with both the greybody factor comparison in Eq. (\ref{comparisongreybodyall}) and the black hole lifetimes in Eq. (\ref{comparisontimesss}). As expected, higher emission rates—whether expressed in energy (Eq. (\ref{compemissionenergy})) or particle number (Eq. (\ref{compparticleemission}))—are associated with stronger greybody factors and a faster evaporation process of the black hole.

\begin{table}[!h]
\begin{center}
\begin{tabular}{c c c c || c c c c || c c c c } 
 \hline\hline \hline
 $Q$ & $\alpha=\beta$ & $\omega$ & $t^{\text{T}}_{\text{evap-final}}$ & $Q$ & $\alpha=\beta$ & $\omega$ &  $t^{\text{T}}_{\text{evap-final}}$ & $Q$ & $\alpha=\beta$ & $\omega$ &  $t^{\text{T}}_{\text{evap-final}}$  \\ [0.2ex] 
 \hline
  0.10  & -0.01 & 0.10 & $3.45\times 10^{530}$ & 0.5  & -0.01 & 0.10 & $1.96\times 10^{97}$ & 0.5  & -0.01 & 0.10 & $1.96\times 10^{97}$  \\
  
  0.20  & -0.01 & 0.10 & $7.42\times 10^{244}$ & 0.5  & -0.02 & 0.10 & $5.05\times 10^{97}$ & 0.5  & -0.01 & 0.20 & $4.35\times 10^{49}$  \\
  
  0.30  & -0.01 & 0.10 & $3.40\times 10^{161}$ & 0.5  & -0.03 & 0.10 & $1.35\times 10^{98}$ & 0.5  & -0.01 & 0.30 & $6.89\times 10^{33}$  \\
  
  0.40  & -0.01 & 0.10 & $1.46\times 10^{121}$ & 0.5  & -0.04 & 0.10 & $3.82\times 10^{98}$ & 0.5  & -0.01 & 0.40 & $9.58\times 10^{25}$  \\
  
   0.50  & -0.01 & 0.10 & $1.96\times 10^{97}$ & 0.5  & -0.05 & 0.10 & $1.12\times 10^{99}$ & 0.5  & -0.01 & 0.50 & $1.96\times 10^{21}$  \\
   
 0.60  & -0.01 & 0.10 & $3.34\times 10^{81}$ & 0.5  & -0.06 & 0.10 & $3.47\times 10^{99}$ & 0.5  & -0.01 & 0.60 & $1.54\times 10^{18}$  \\ 

 0.70  & -0.01 & 0.10 & $2.16\times 10^{70}$ & 0.5  & -0.07 & 0.10 & $1.12 \times 10^{100}$ & 0.5  & -0.01 & 0.70 & $9.64\times 10^{15}$  \\
 
 0.80  & -0.01 & 0.10 & $9.83\times 10^{61}$ & 0.5  & -0.08 & 0.10 & $3.84\times 10^{100}$ & 0.5  & -0.01 & 0.80 & $2.19\times 10^{14}$ \\
 
 0.90  & -0.01 & 0.10 & $3.43\times 10^{55}$ & 0.5  & -0.09 & 0.10 & $1.38\times 10^{101}$ & 0.5  & -0.01 & 0.90 & $1.18 \times 10^{13}$ \\
 
 0.99  & -0.01 & 0.10 & $7.27\times 10^{50}$ & 0.5  & -0.10 & 0.10 & $5.23\times 10^{101}$ & 0.5  & -0.01 & 0.99 & $1.43\times 10^{12}$  \\ 
 [0.2ex] 
 \hline \hline \hline
\end{tabular}
\caption{\label{evapT} Quantitative values of the evaporation lifetime driven by the emission of spin--2 particle modes, $t^{\text{T}}_{\text{evap-final}}$, for different configurations of the parameters $Q$, $\alpha = \beta$, and $\omega$.
}
\end{center}
\end{table}


\subsection{Spin--$1/2$ particle modes}

Using the same procedure described in the previous section, Eq. (\ref{slawbotz}) is now employed to estimate the evaporation lifetime in the case of spinor perturbations. In this setting, we adopt the greybody factors $\bar{\Gamma}^{\psi}_{\ell \omega}$ and the corresponding cross sections $\sigma^{\psi}_{\ell \omega}$, which account exclusively for spin--$1/2$ modes.

Differently to the other perturbations examined here, and taking into account the assumptions adopted in deriving the greybody factor for spin--$1/2$ particle modes, we are able to obtain an analytical expression for $t^{\psi}_{\text{evap-final}}$. This expression is omitted because, despite relying on the previous approximations, it has 894 terms.

In general lines, three similar features emerge, differing from those observed in the previously analyzed cases. First, increasing the electric charge $Q$ leads to a shorter evaporation lifetime for spin--$1/2$ modes, $t^{\psi}_{\text{evap-final}}$, when $\alpha=\beta$ and $\omega$ are fixed. Second, decreasing $\alpha=\beta$ while keeping $Q$ and $\omega$ constant also reduces the lifetime. Third, higher frequencies $\omega$ cause the black hole to evaporate more rapidly for fixed $Q$ and $\alpha = \beta$. Notably, this behavior closely resembles that observed for the spin--0, spin--1, and spin--2 particle modes. It should be emphasized, however, that a direct comparison with the other perturbations is not fully consistent, since the corresponding greybody factors were also expanded with respect to the mass $M$. For this reason, we have not presented a detailed table listing all numerical values nor compared the resulting evaporation lifetimes among the different perturbative sectors.

\begin{figure}
    \centering
      \includegraphics[scale=0.51]{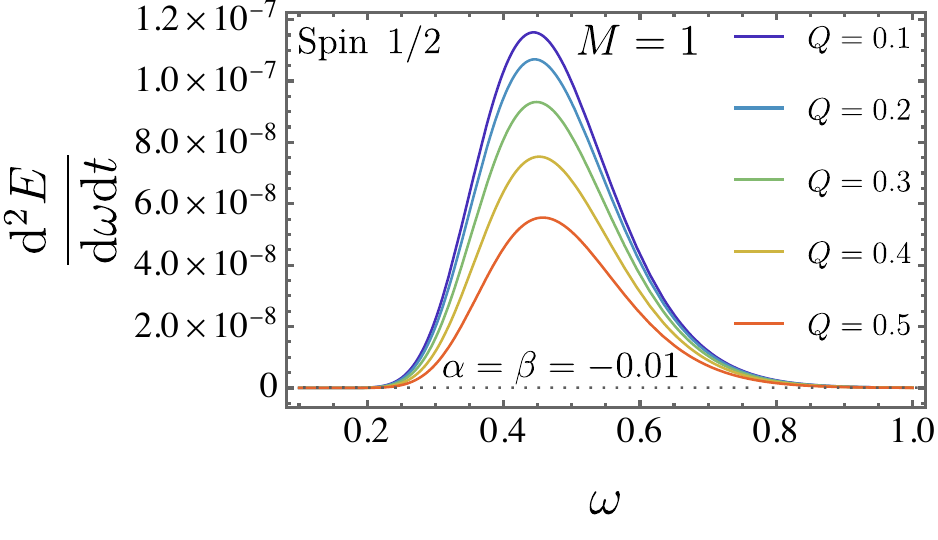}
       \includegraphics[scale=0.51]{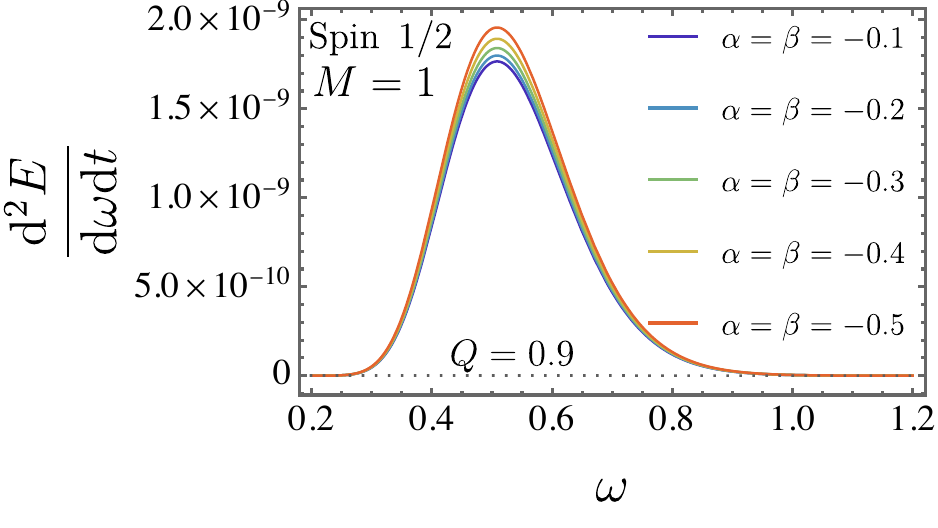}
    \caption{The energy emission rate for spin--$1/2$ particle modes is displayed. The left panel shows the dependence on the charge $Q$ with $M = 1$ and $\alpha = \beta = -0.01$, while the right panel displays the variation with $\alpha = \beta$ for fixed $M = 1$ and $Q = 0.9$. Here, we consider $\ell=5/2$.  }
    \label{spin1dividido2emission}
\end{figure}

\begin{figure}
    \centering
      \includegraphics[scale=0.51]{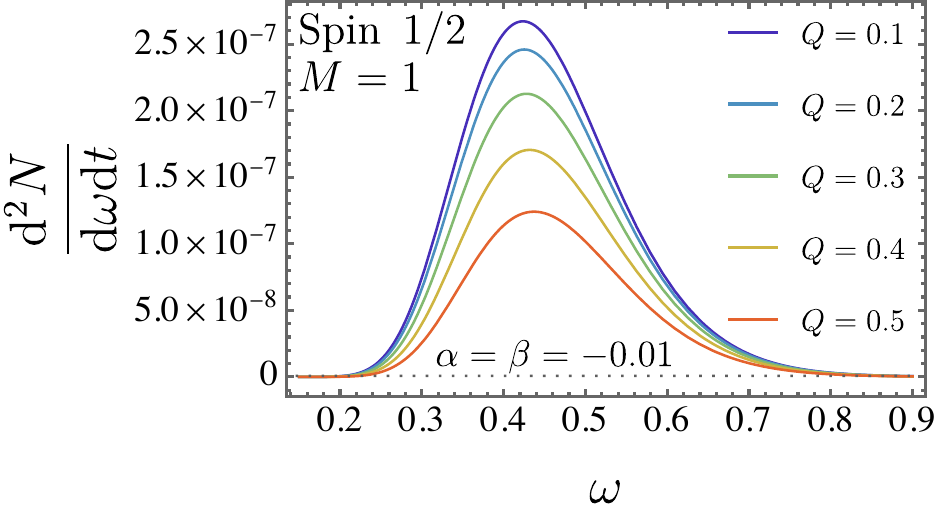}
       \includegraphics[scale=0.51]{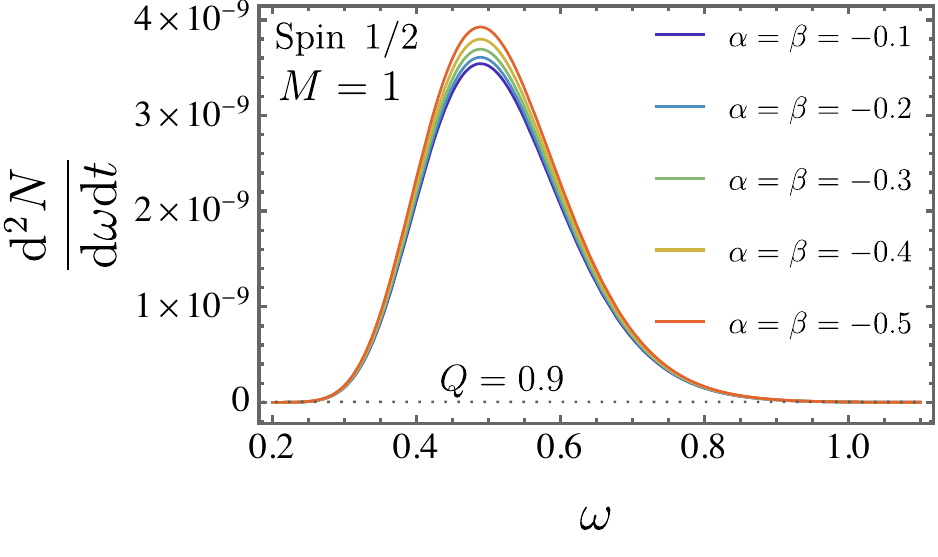}
    \caption{ The particle emission rate for spin--$1/2$ modes is shown. The left panel illustrates how the spectrum varies with $Q$ when $M = 1$ and $\alpha = \beta = -0.01$, whereas the right one depicts the effect of changing $\alpha = \beta$ while keeping $M = 1$ and $Q = 0.9$ constant. Here, we regard $\ell = 5/2$.}
    \label{spin1divide2emissionparticle}
\end{figure}

Apart from this remark, let us now examine the emission rate using the same procedure adopted for the previous spin cases—evaluating its full numerical form without applying any approximations. In Fig.~\ref{spin1dividido2emission}, the left panel shows the energy emission rate, $\mathrm{d}^{2}E/(\mathrm{d}\omega\,\mathrm{d}t)$, for spin–1/2 particle modes as the electric charge $Q$ varies, while the right panel illustrates the corresponding behavior as $\alpha=\beta$ is varied. Figure~\ref{spin1divide2emissionparticle} presents the particle emission rate, following the same patterns observed for the energy emission rate. In general, as $Q$ increases, the emission rate decreases in both the energy and particle channels. We have that a decrease in $\alpha=\beta$ leads to an enhancement of the emission rate, following the same overall trend observed for the other perturbative modes discussed earlier. Here, we considered $\ell =5/2$.

\begin{figure}
    \centering
      \includegraphics[scale=0.51]{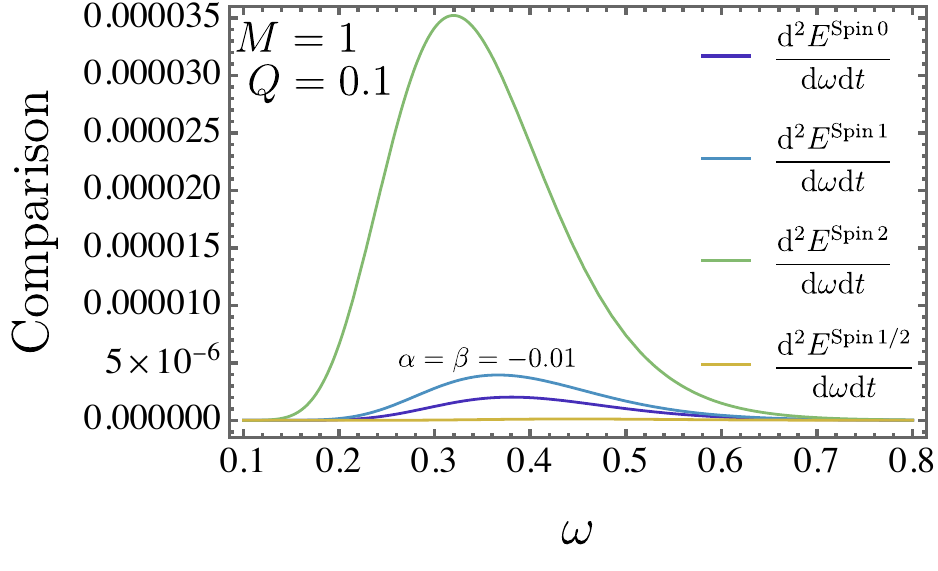}
       \includegraphics[scale=0.51]{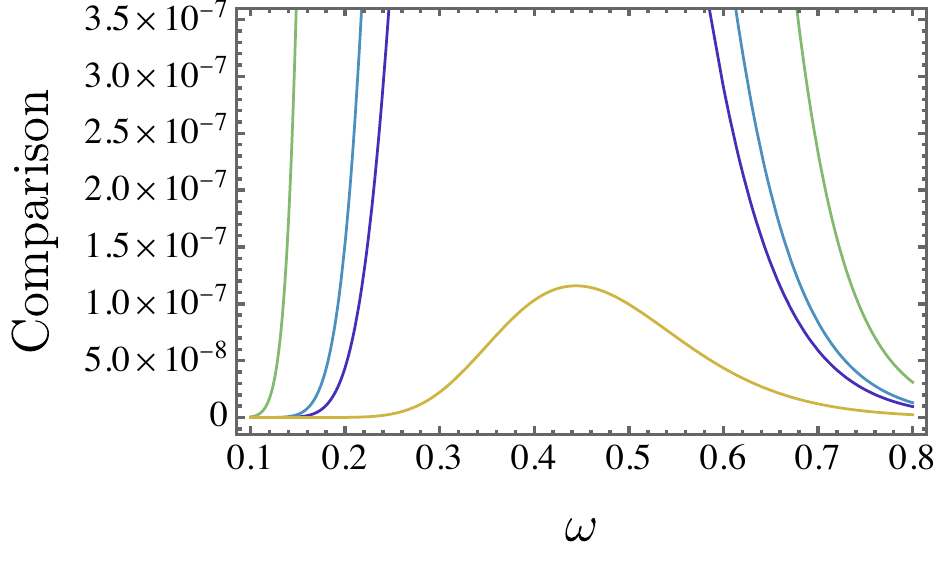}
    \caption{ The comparison of all the energy emission rates regarding spin--0, spin--1, spin--2, and spin--$1/2$ cases with $\alpha = \beta = -0.01$, $M = 1$, $Q = 0.1$, and $\ell = 2$ (for bosons) and $\ell = 5/2$ (for fermions).}
    \label{emissioncompparticlesspinor}
\end{figure}

Moreover, the comparison of energy emission rate for all spins regarded in this paper is shown in Fig. \ref{emissioncompparticlesspinor}. In other words, what we see is the following pattern for the energy emission rate
\ie
 \frac{{{\mathrm{d}^2}E^{\text{Spin} \,2}}}{{\mathrm{d}\omega \mathrm{d}t}} > \frac{{{\mathrm{d}^2}E^{\text{Spin} \,1}}}{{\mathrm{d}\omega \mathrm{d}t}} >
 \frac{{{\mathrm{d}^2}E^{\text{Spin} \,0}}}{{\mathrm{d}\omega \mathrm{d}t}} >  \frac{{{\mathrm{d}^2}E^{\text{Spin} \,1/2}}}{{\mathrm{d}\omega \mathrm{d}t}},
\label{eall}
\fe
and, for the particle emission rate, likewise
\ie
 \frac{{{\mathrm{d}^2}N^{\text{Spin} \,2}}}{{\mathrm{d}\omega \mathrm{d}t}} > \frac{{{\mathrm{d}^2}N^{\text{Spin} \,1}}}{{\mathrm{d}\omega \mathrm{d}t}} >\frac{{{\mathrm{d}^2}N^{\text{Spin} \,0}}}{{\mathrm{d}\omega \mathrm{d}t}} > 
 \frac{{{\mathrm{d}^2}N^{\text{Spin} \,1/2}}}{{\mathrm{d}\omega \mathrm{d}t}} .
\label{nall}
\fe
Analogously to the tensor perturbation case, all results for the emission rate remain consistent with the greybody factor comparison in Eq. (\ref{comparisongreybodyall}). As anticipated, higher emission rates—either in terms of energy (Eq. (\ref{eall})) or particle number (Eq. (\ref{nall}))—correspond to more pronounced greybody factors. In other words, although the evaporation lifetime was not explicitly compared for the spinor perturbations due to reasons presented in the begging of this subsection, the behavior inferred from the emission rate, absorption cross section, and greybody factors allows us to extrapolate that
\ie
t^{\psi}_{\text{evap-final}}> t^{\text{S}}_{\text{evap-final}} > t^{\text{V}}_{\text{evap-final}}>t^{\text{T}}_{\text{evap-final}} ,
\label{comparisontimesss2}
\fe
demonstrating that, compared to all other particle modes analyzed in this work, the tensor mode exhibits the highest emission from the black hole described by Eq. (\ref{balckkk}), thereby resulting in a shorter evaporation lifetime.

A natural question arises from Eqs.~(\ref{eall}) and~(\ref{nall}): why is the emission of radiation more intense for bosonic particles than for fermionic ones? One possible explanation lies in the distinct quantum statistics that govern their occupation probabilities. Bosons, described by Bose–Einstein statistics, can share the same quantum state, which amplifies their collective emission rate. Fermions, on the other hand, obey the Pauli exclusion principle, which forbids multiple particles from occupying an identical state, thereby suppressing their emission. Moreover, the spin–statistics connection affects how each field couples to the background geometry, altering the transmission probability across the potential barrier.

\subsection{High frequency regime}

It should be emphasized that the emitted flux is predominantly composed of neutrinos and photons, both of which are effectively massless particles \cite{hiscock1990evolution,page1976particle}. In this context, the limiting value of the absorption cross section, $\sigma_{\ell \omega} \to \sigma_{\mathrm{lim}}$, approaches $\pi \mathcal{R}^2$, where $\mathcal{R}$ denotes the shadow radius associated with the black hole geometry \cite{heidari2025gravitational}
\ie
\begin{split}
& \mathcal{R} = \sqrt{\frac{\left(\gamma +\frac{1}{2} \left(\sqrt{9 M^2-8 Q^2}+3 M\right)\right)^2}{\frac{Q^2}{\left(\gamma +\frac{1}{2} \left(\sqrt{9 M^2-8 Q^2}+3 M\right)\right)^2}-\frac{2 M}{\gamma +\frac{1}{2} \left(\sqrt{9 M^2-8 Q^2}+3 M\right)}-\frac{\alpha  (2 \beta -1) Q^4}{10 \left(\gamma +\frac{1}{2} \left(\sqrt{9 M^2-8 Q^2}+3 M\right)\right)^6}+1}},
\end{split}
\fe
and, retaining terms only up to first order in $\alpha$, the expressions for $\sigma_{\ell \omega}$ and $T$ can be written as
\ie
\sigma_{lim} \approx \, 27 \pi  M^2 -9 \pi  Q^2 -\frac{\pi  Q^4}{M^2} + \frac{\alpha  \left(\pi  (2 \beta -1) Q^4\right)}{90 M^4},
\fe
and
\ie
T \approx \, \, \frac{1}{8 \pi  M} -\frac{Q^4}{128 \pi  M^5} +\frac{\alpha  \beta  Q^4}{640 \pi  M^7}-\frac{\alpha  Q^4}{1280 \pi  M^7},
\fe
respectively. Moreover, in the high--frequency limit considered here, the greybody factors approach unity, $\Bar{\Gamma}_{\ell \omega} \simeq 1$ \cite{liang2025einstein}. Under these circumstances, Eq.~(\ref{slawbotz}) can be rewritten as
\begin{eqnarray}
	&
\frac{\mathrm{d}M}{\mathrm{d}t} = -\frac{\left(2430 M^6-810 M^4 Q^2-90 M^2 Q^4+\alpha  (2 \beta -1) Q^4\right) \left[160 M^6-10 M^2 Q^4+\alpha  (2 \beta -1) Q^4\right]^4}{241591910400000 \pi ^3 M^{32}} \,.
\end{eqnarray}

The next step consists in evaluating the following integral
\begin{eqnarray}
& \int_{0}^{t_{\text{evap}}} \xi \mathrm{d}\tau = 
	- \int_{M_{i}}^{M_{f}} \mathrm{d}M
\left[ \frac{\left(2430 M^6-810 M^4 Q^2-90 M^2 Q^4+\alpha  (2 \beta -1) Q^4\right) \left[160 M^6-10 M^2 Q^4+\alpha  (2 \beta -1) Q^4\right]^4}{241591910400000 \pi ^3 M^{32}}   \right]^{-1}
	\nonumber \\
& \approx \, \,    \int_{M_{i}}^{M_{f}} \mathrm{d}M
\left[   -\frac{4096}{27} \pi ^3 M^2 -\frac{44032 \pi ^3 Q^4}{729 M^2}-\frac{4096 \pi ^3 Q^2}{81} + \frac{\alpha  \left(126464 \pi ^3 (2 \beta -1) Q^4\right)}{32805 M^4}  \right]    ,
\end{eqnarray}
with $t_{\text{evap}}$ denoting the total duration of the black hole evaporation process and can thus be written as
\ie
\begin{split}
t_{\text{evap}}  = & \, \frac{4096 \pi ^3 M_{i}^3}{81}-\frac{4096 \pi ^3 M_{f}^3}{81} +\frac{44032 \pi ^3 Q^4}{729 M_{f}}-\frac{4096}{81} \pi ^3 M_{f} Q^2 \\
& +\frac{4096}{81} \pi ^3 M_{i} Q^2 -\frac{252928 \pi ^3 \alpha  \beta  Q^4}{98415 M_{f}^3}+\frac{126464 \pi ^3 \alpha  Q^4}{98415 M_{f}^3} \\ & +\frac{252928 \pi ^3 \alpha  \beta  Q^4}{98415 M_{i}^3} 
 -\frac{126464 \pi ^3 \alpha  Q^4}{98415 M_{i}^3}-\frac{44032 \pi ^3 Q^4}{729 M_{i}}.
\end{split}
\fe

Additionally, by requiring the temperature to vanish, $T \to 0$, one obtains the corresponding expression for the remnant mass \cite{heidari2025gravitational}:
\ie
\begin{split}
M_{rem} & =  \frac{\sqrt{\frac{5 \sqrt[3]{3} Q^4+\sqrt[3]{5} \left(9 \alpha  (1-2 \beta ) Q^4+\sqrt{3} \sqrt{Q^8 \left(27 \alpha ^2 (1-2 \beta )^2-25 Q^4\right)}\right)^{2/3}}{\sqrt[3]{9 \alpha  (1-2 \beta ) Q^4+\sqrt{3} \sqrt{Q^8 \left(27 \alpha ^2 (1-2 \beta )^2-25 Q^4\right)}}}}}{2 \sqrt[3]{15}} \\
& \approx \, \frac{Q}{2} + \frac{\alpha  (1-2 \beta )}{20 Q}. 
\end{split}
\fe
From this analysis, it follows that the evaporation process halts before the black hole fully disappears. Consequently, the final mass approaches the remnant value, $M_{f} \to M_{\text{rem}}$, yielding the expression for the total evaporation time
\begin{eqnarray}
	t_{\text{evap-final}} & = & \, \frac{4096 \pi ^3 M_{i}^3}{81}-\frac{44032 \pi ^3 Q^4}{729 M_{i}}+\frac{4096}{81} \pi ^3 M_{i} Q^2-\frac{2560 \pi ^3 Q^3}{81}-\frac{1792}{405} \pi ^3 \alpha  Q 	\nonumber \\
	&& -\frac{126464 \pi ^3 \alpha  Q^4}{98415 M_{i}^3}-\frac{64 \pi ^3 \alpha ^3}{10125 Q^3}-\frac{128 \pi ^3 \alpha ^2}{675 Q}+\frac{3584}{405} \pi ^3 \alpha  \beta  Q 
	\nonumber \\
	&& +\frac{252928 \pi ^3 \alpha  \beta  Q^4}{98415 M_{i}^3}-\frac{256 \pi ^3 \alpha ^3 \beta ^2}{3375 Q^3}+\frac{128 \pi ^3 \alpha ^3 \beta }{3375 Q^3}-\frac{512 \pi ^3 \alpha ^2 \beta ^2}{675 Q}+\frac{512 \pi ^3 \alpha ^2 \beta }{675 Q} 
	\nonumber \\
	&& +\frac{512 \pi ^3 \alpha ^3 \beta ^3}{10125 Q^3}+\frac{202342400 \pi ^3 \alpha  Q^7}{19683 \left(-2 \alpha  \beta +\alpha +10 Q^2\right)^3}-\frac{404684800 \pi ^3 \alpha  \beta  Q^7}{19683 \left(-2 \alpha  \beta +\alpha +10 Q^2\right)^3} 
	\nonumber \\
	&& +\frac{880640 \pi ^3 Q^5}{729 \left(-2 \alpha  \beta +\alpha +10 Q^2\right)}.
\end{eqnarray}

\begin{figure}
    \centering
      \includegraphics[scale=0.43]{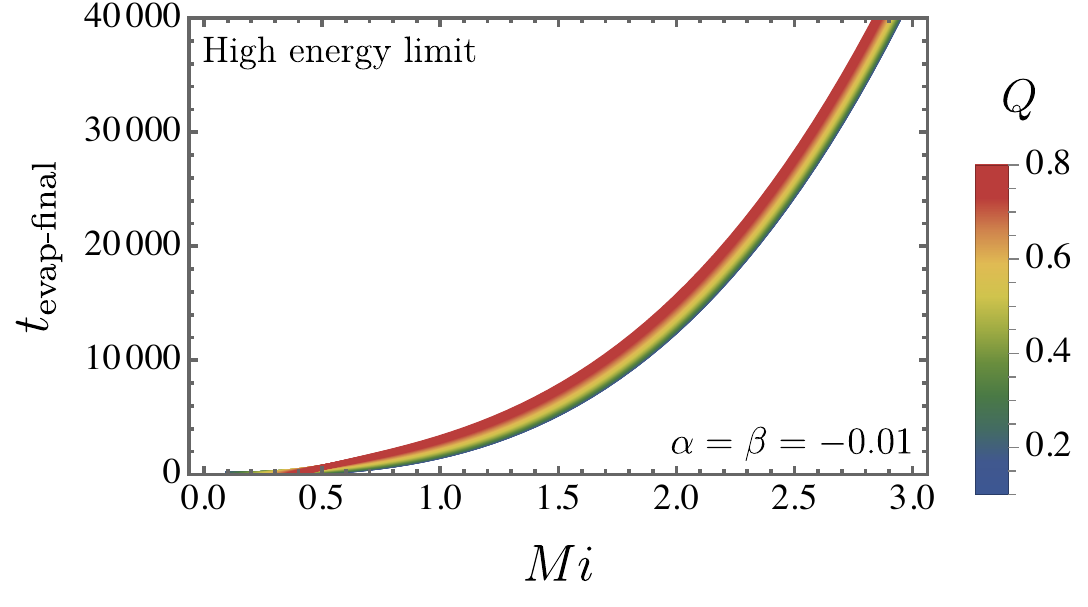}
       \includegraphics[scale=0.43]{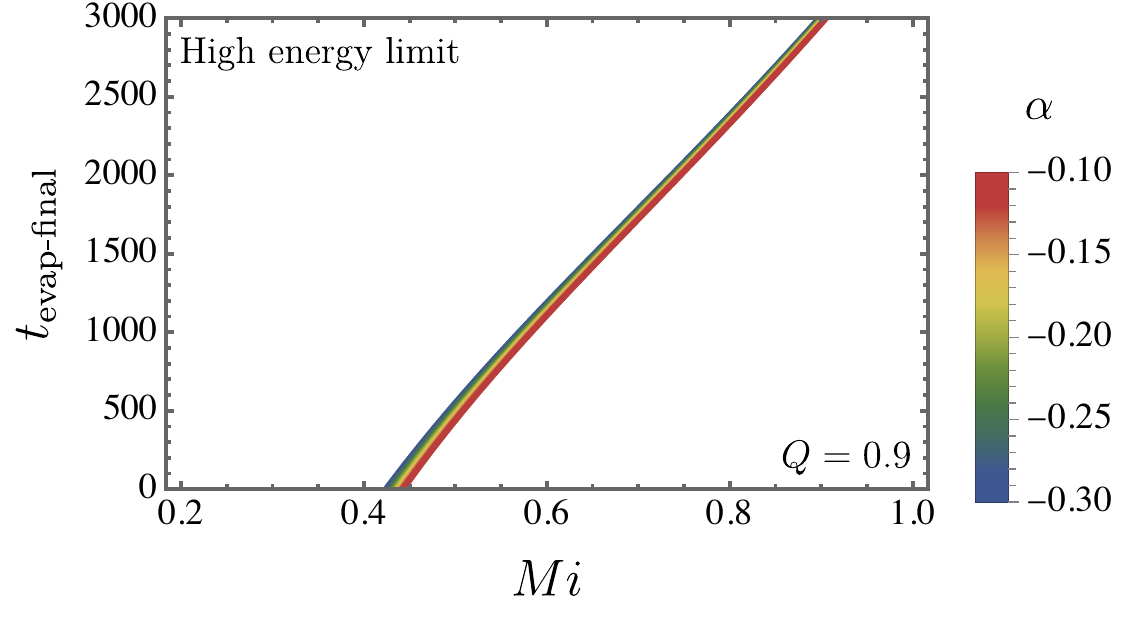}
    \caption{In the high--energy regime, the final evaporation time $t_{\text{evap-final}}$, obtained in the limit $M_f \to M_{\text{rem}}$, is depicted as a function of the initial mass $M_i$. The left panel illustrates the dependence on different charge values $Q$ for fixed $\alpha = \beta = -0.01$, whereas the right panel presents the variation with $\alpha = \beta$ for a constant charge $Q = 0.9$. }
    \label{evaporationhighlimit}
\end{figure}

To highlight the physical interpretation of the results, Fig.~\ref{evaporationhighlimit} is presented. Unlike the trend identified for vector perturbations, larger values of $Q$ result in a notable extension of the final evaporation time, $t_{\text{evap-final}}$.

Next, we turn our attention to the analysis of the emission rate. Quantum fluctuations occurring in the vicinity of the event horizon continuously produce and annihilate particle–antiparticle pairs. In this scenario, one of the particles may tunnel through the potential barrier, escaping to infinity while its partner falls into the black hole, thereby reducing its mass over time—a process known as Hawking radiation, as previously outlined. For an observer located far from the source, the shadow region of the black hole is associated with the high–energy absorption cross section, which approaches a constant asymptotic value, $\sigma_{\text{lim}}$. Following Refs.~\cite{decanini2011universality,araujo2024asdasdeffects,papnoi2022rotating}, the corresponding energy emission rate takes the form
\ie
\label{emission}
	\frac{{{\mathrm{d}^2}E}}{{\mathrm{d}\omega \mathrm{d}t}} = \frac{{2{\pi ^2}\sigma_{lim}}}{{{e^{\frac{\omega }{T}}} - 1}} {\omega ^3},
\fe
where $\omega$ denotes the frequency associated with the emitted photons. After inserting the explicit forms of the shadow radius and the Hawking temperature into the expression, the resulting formula for the energy emission rate is obtained as
\ie
\frac{\mathrm{d}^{2}E}{\mathrm{d}\omega \mathrm{d} t} =  \frac{\pi ^3 \omega ^3 \left(2430 M^6-810 M^4 Q^2-90 M^2 Q^4+\alpha  (2 \beta -1) Q^4\right)}{45 M^4 \left(e^{\frac{1280 \pi  M^7 \omega }{160 M^6-10 M^2 Q^4+\alpha  (2 \beta -1) Q^4}}-1\right)} .
\fe

\begin{figure}
    \centering
      \includegraphics[scale=0.435]{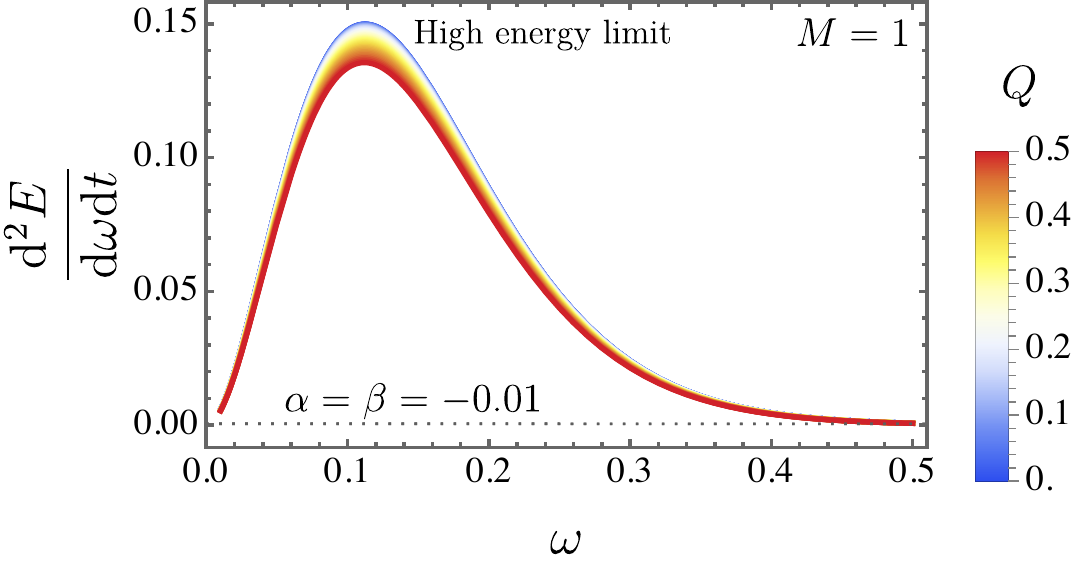}
       \includegraphics[scale=0.435]{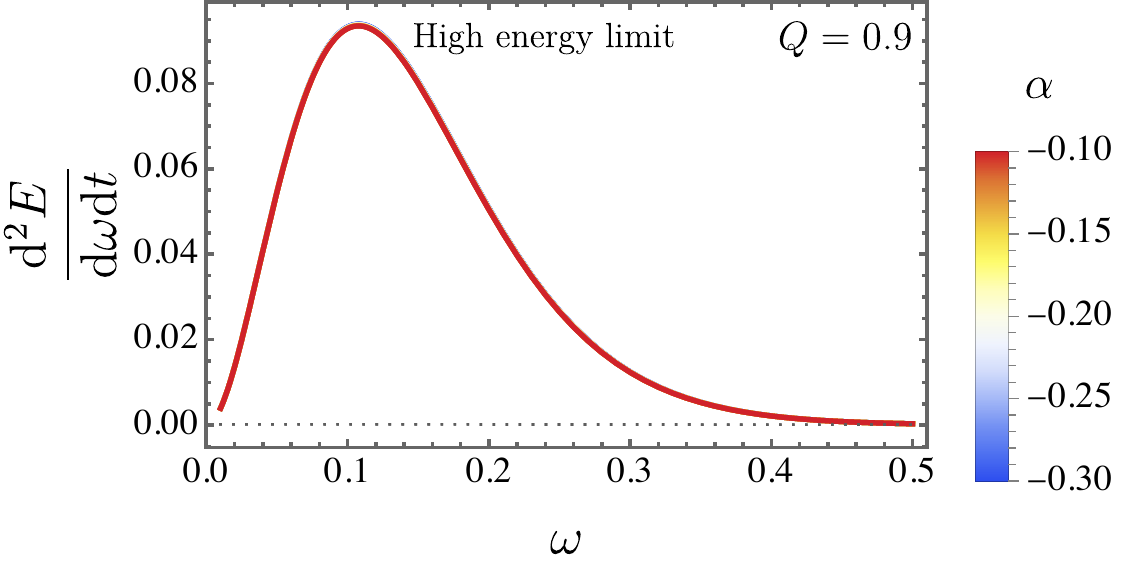}
    \caption{Energy emission rate in the high–energy regime. The left panel shows the dependence of the spectrum on the charge $Q$ for fixed parameters $M = 1$ and $\alpha = \beta = -0.01$, while the right one displays the variation with $\alpha = \beta$ at fixed $M = 1$ and $Q = 0.9$.}
    \label{energyemsissionratehigh}
\end{figure}

Figure~\ref{energyemsissionratehigh} illustrates the behavior of the energy emission rate. The results show that, even in the high–energy regime, the overall trend remains consistent with the patterns identified in the preceding sections.

Furthermore, in a similar manner, the corresponding particle emission rate can also be determined
\begin{equation}
\frac{\mathrm{d}^{2}N}{\mathrm{d}\omega \mathrm{d}t}
= \frac{2\pi^{2}\,\sigma_{lim}\,\omega^{2}}
       {{{e^{\frac{\omega }{T}}} - 1}} = \frac{\pi ^3 \omega^2 \left(2430 M^6-810 M^4 Q^2-90 M^2 Q^4+\alpha  (2 \beta -1) Q^4\right)}{45 M^4 \left(e^{\frac{1280 \pi  M^7 \omega }{160 M^6-10 M^2 Q^4+\alpha  (2 \beta -1) Q^4}}-1\right)}.
\end{equation}

\begin{figure}
    \centering
      \includegraphics[scale=0.435]{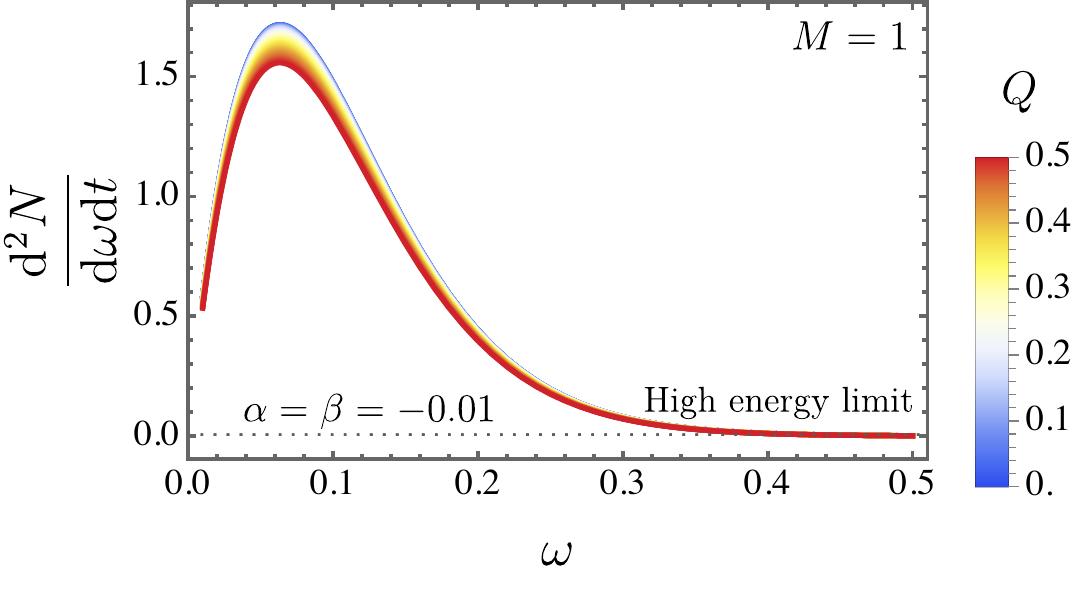}
       \includegraphics[scale=0.435]{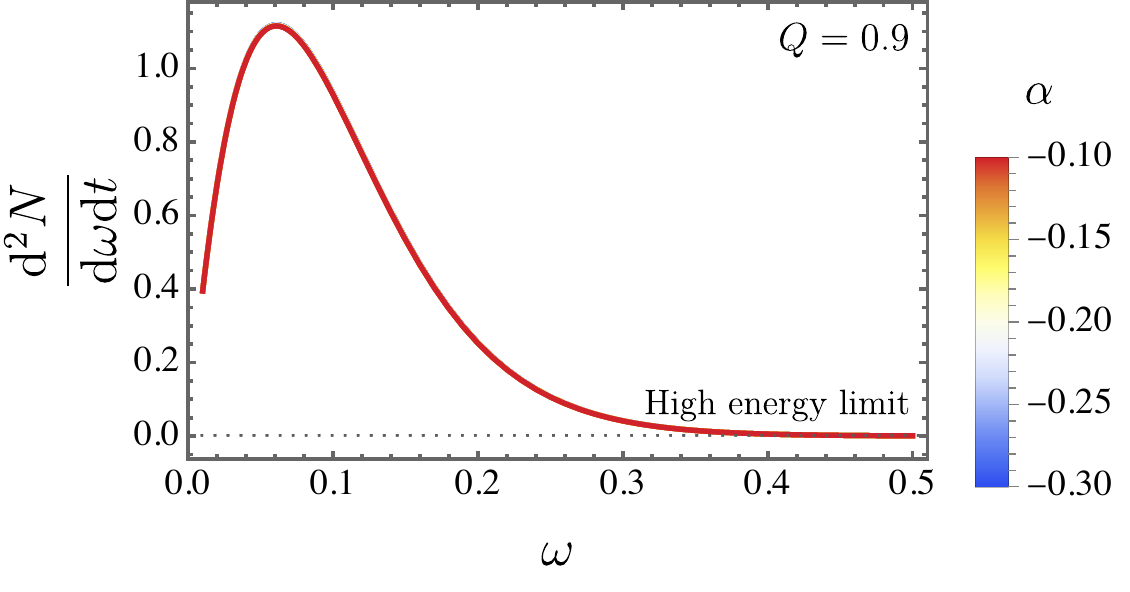}
    \caption{The particle emission rate for high energy limit case is shown. The left panel illustrates how the spectrum varies with $Q$ when $M = 1$ and $\alpha = \beta = -0.01$, whereas the right one depicts the effect of changing $\alpha = \beta$ while keeping $M = 1$ and $Q = 0.9$ constant. }
    \label{emissionparticlehigh}
\end{figure}

Conversely, Fig.~\ref{emissionparticlehigh} displays the particle emission rate. The results confirm that the overall behavior remains consistent with the trends identified in the previous analyses.


\section{The correspondence of QNMs and greybody factors}\label{SectionVIII}

The quasinormal mode frequencies can be efficiently estimated using the semi–analytical WKB method. Owing to the complexity of the lapse function in the numerical evaluation, the third–order formulation of the WKB approach is adopted, providing the spectrum of quasinormal modes according to the relation \cite{iyer1987black,iyer1987black2,konoplya2019higher,konoplya2011quasinormal,cardoso2001quasinormal}
\begin{equation}\label{3wkb}
\begin{split}
    {\omega ^2}=&  \,\,{V_0} + \sqrt { - 2{V_0}^{\prime \prime }} \Lambda (n) - i\left(n + \frac{1}{2}\right)\sqrt { - 2{V_0}^{\prime \prime }} (1 + \Omega (n)),
\end{split}
 \end{equation}
where 
\begin{equation}
\begin{split}
    \Lambda (n) =   \frac{1}{{\sqrt { - 2{V_0}^{\prime \prime }} }}\left[\frac{1}{8}\left(\frac{{V_0^{\left(4\right)}}}{{{V_0}^{\prime \prime }}}\right)\left(\frac{1}{4} + {\alpha ^2}\right) - \frac{1}{{288}}{\left(\frac{{{V_0}^{\prime \prime \prime }}}{{{V_0}^{\prime \prime }}}\right)^2}\left(7 + 60{\alpha ^2}\right)\right],
\end{split}
\end{equation}
and 
\begin{eqnarray}
    \Omega (n) & = & \left( {\frac{1}{{ - 2{V_0}^{\prime \prime }}}} \right)
  \frac{5}{{6912}}{\left(\frac{{{V_0}^{\prime \prime \prime }}}{{{V_0}^{\prime \prime }}}\right)^4}\left( {77 + 188 \times {\alpha ^2}} \right)  - \frac{1}{{384}}  \left( {\frac{{{V_0}{{^{\prime \prime \prime }}^2}V_0^{(4)}}}{{{V_0}{{^{\prime \prime }}^3}}}} \right)\left( {51 + 100{\alpha ^2}} \right)  
  	\nonumber  \\ 
   && + \frac{1}{{2304}}{\left(\frac{{V_0^{(4)}}}{{{V_0}^{\prime \prime }}}\right)^2}\left( {67 + 68{\alpha ^2}} \right)  + \frac{1}{{288}}\left( {\frac{{{V_0}^{\prime \prime \prime }V_0^{(5)}}}{{{V_0}{{^{\prime \prime }}^2}}}} \right)\left( {19 + 28{\alpha ^2}} \right) 
   		\nonumber  \\ 
    && - \frac{1}{{288}}\left(\frac{{V_0^{(6)}}}{{{V_0}^{\prime \prime }}}\right)\left( {5 + 4{\alpha ^2}} \right). 
\end{eqnarray}
In this relation, $\alpha = n + \tfrac{1}{2}$, with $n$ representing the overtone index, which satisfies the constraint $n \leq \ell$.

A recent study \cite{konoplya2024correspondence} established an approximate link between quasinormal modes and greybody factors, which becomes exact in the eikonal or high–frequency regime. In this limit, the greybody factors of spherically symmetric black holes take on a simpler form, being mainly governed by the fundamental quasinormal mode. For smaller values of the angular momentum number $\ell$, however, additional terms associated with higher overtones start to play a role. Within the WKB framework, the corresponding transmission and reflection coefficients are obtained from the following relation \cite{iyer1987black}
\begin{equation}
{\left| R \right|^2} = \frac{1}{{1 + {e^{ - 2\pi i{\mathcal{K}}}}}},
\end{equation}
\begin{equation}\label{Trans}
{\left| T \right|^2} = \frac{1}{{1 + {e^{  2\pi i{\mathcal{K}}}}}}.
\end{equation}

Following the formulation in Ref.~\cite{konoplya2024correspondence}, the parameter $\mathcal{K}$ is constructed from the two dominant quasinormal frequencies, $\omega_0$ and $\omega_1$, which correspond to the fundamental mode ($n = 0$) and the first overtone ($n = 1$). Each frequency $\omega$ can be decomposed into its real and imaginary components, $\omega_R$ and $\omega_I$, describing the oscillatory behavior and the damping rate, respectively
\begin{equation}\label{Tfactor}
     - i{\mathcal{K}} =  - \frac{{{\omega ^2} - {\omega _{0R}}^2}}{{4{\omega _{0R}}{\omega _{0I}}}} + {\Delta _1} + {\Delta _2} + {\Delta _f},
\end{equation}
where
\ie
{\Delta _1}  = \frac{{{\omega _{0R}} - {\omega _{1R}}}}{{16{\omega _{0I}}}},
\fe
\ie
\begin{split}
    \Delta_2 & =  - \frac{{{\omega ^2} - \omega_{0R}^2}}{{32{\omega _{0R}}{\omega _{0I}}}}\left[\frac{{{{({\omega _{0R}} - {\omega _{R1}})}^2}}}{{4{\omega _{0I}}^2}} - \frac{{3{\omega _{0I}} - {\omega _{1I}}}}{{3{\omega _{0I}}}}\right] + \frac{{{{({\omega ^2} - \omega _{0R}^2)}^2}}}{{16\omega _{0R}^3{\omega _{0I}}}}\left[1 + \frac{{{\omega _{0R}}({\omega _{0R}} - {\omega _{1R}})}}{{4\omega _{0I}^2}}\right], 
\end{split}
\fe
and
\ie
\begin{split}
\Delta_f &=  - \frac{{{{({\omega ^2} - \omega _{0R}^2)}^3}}}{{32\omega _{0R}^5{\omega _{0I}}}}\left\{1 + \frac{{{\omega _{0R}}({\omega _{0R}} - {\omega _{1R}})}}{{4{\omega _{0I}}^2}}  + \omega _{0R}^2\left[\frac{{{{({\omega _{0R}} - {\omega _{1R}})}^2}}}{{16\omega _{0I}^4}} - \frac{{3{\omega _{0I}} - {\omega _{1I}}}}{{12{\omega_{0I}}}}\right]\right\}.
\end{split}
\fe

In what follows, the developed formalism is employed to evaluate the greybody factors corresponding to scalar, vector, tensor, and spinor perturbations. The parameter $\mathcal{K}$ is computed using Eq.~(\ref{Trans}), while the dominant quasinormal frequencies involved in this calculation are obtained through the third--order WKB approach defined in Eq.~(\ref{3wkb}). For clarity in the forthcoming figures, the greybody factors are denoted as $\Bar{\Gamma}(\omega) = |T_{b}|$.


\subsection{Spin--$0$ particle modes}

\begin{figure}
    \centering
      \includegraphics[scale=0.6]{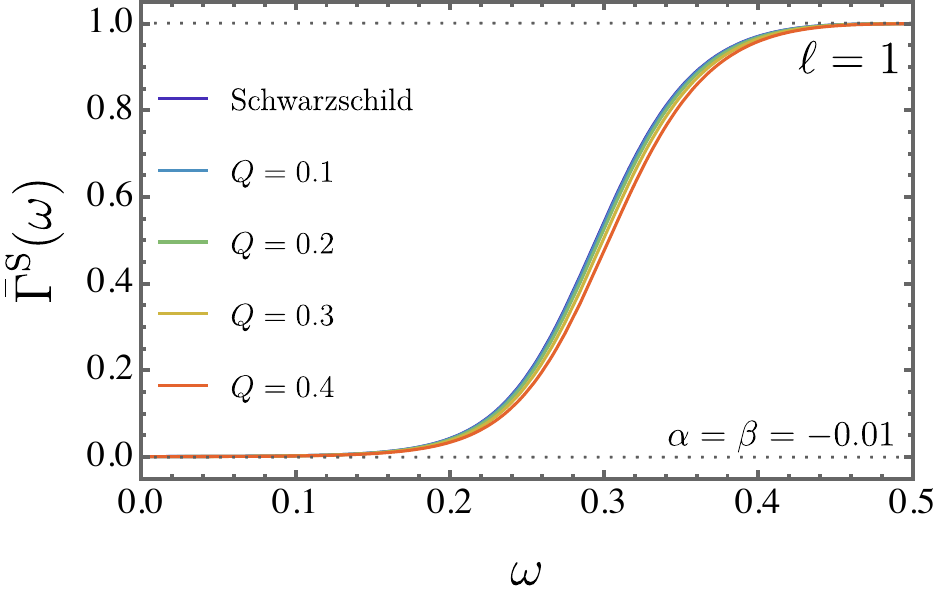}
    \caption{Correlation between quasinormal modes and greybody factors for scalar perturbations, evaluated with $\alpha = \beta = -0.01$ and angular momentum number $\ell = 1$.}
    \label{scalarcorrespondence}
\end{figure}

\begin{figure}
    \centering
      \includegraphics[scale=0.7]{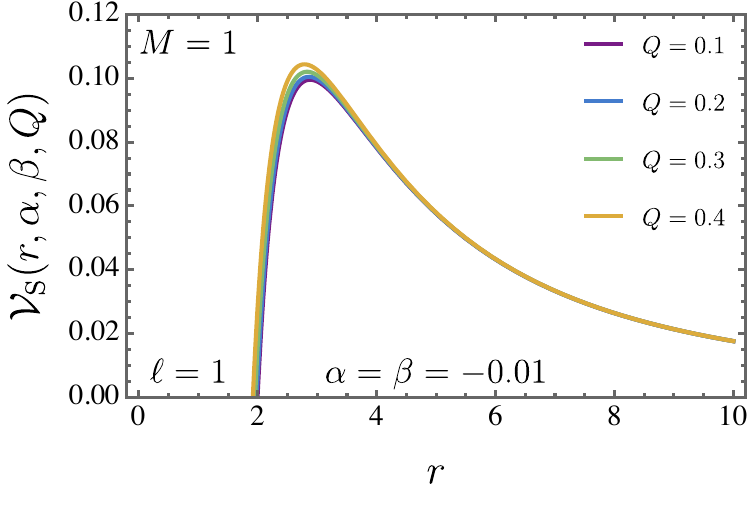}
    \caption{The profile of the scalar effective potential $\mathcal{V}_{\text{S}}(r,\alpha,\beta,Q)$ is plotted against the radial coordinate $r$, considering $M=1$, $\ell=1$, and $\alpha=\beta=-0.01$, for various choices of the charge parameter $Q$. }
    \label{scalarpotential}
\end{figure}

Figure~\ref{scalarcorrespondence} illustrates the relationship between the quasinormal modes and the greybody factors for scalar perturbations, considering $\alpha = \beta = -0.01$ and $\ell = 1$. As the electric charge $Q$ grows, the amplitude of $\bar{\Gamma}^{\text{S}}(\omega)$ diminishes relative to the Schwarzschild configuration. This trend is consistent with the quasinormal mode spectrum reported in Ref.~\cite{heidari2025gravitational}, where increasing $Q$ results in enhanced damping and slightly elevated oscillation frequencies, implying that the corresponding effective potential barrier becomes steeper and less transparent, as illustrated in Fig.~\ref{scalarpotential}.

\subsection{Spin--$1$ particle modes}

\begin{figure}
    \centering
      \includegraphics[scale=0.6]{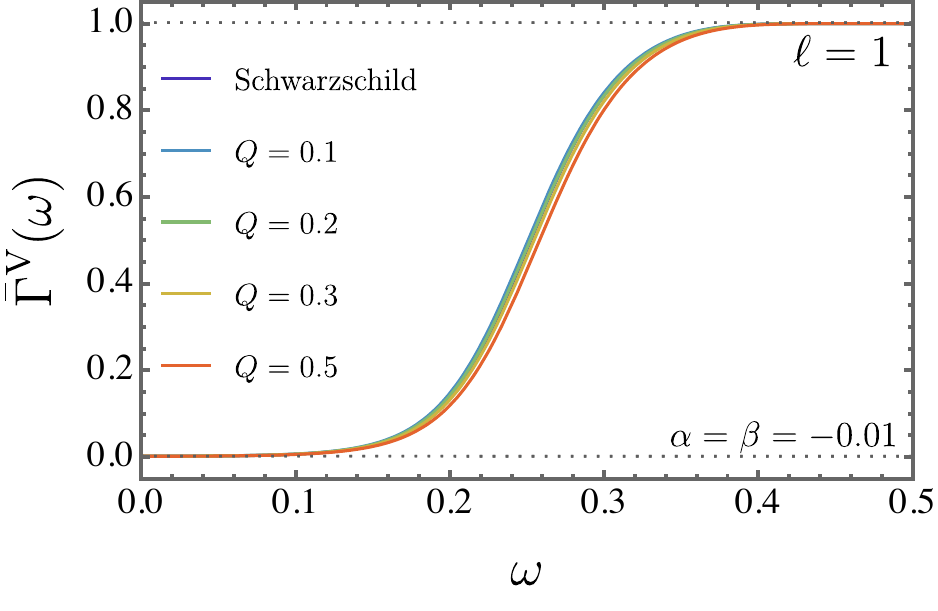}
    \caption{Correlation between quasinormal modes and greybody factors for vector perturbations, evaluated with parameters $\alpha = \beta = -0.01$ and angular momentum number $\ell = 1$.}
    \label{vectorrespondence}
\end{figure}

\begin{figure}
    \centering
      \includegraphics[scale=0.7]{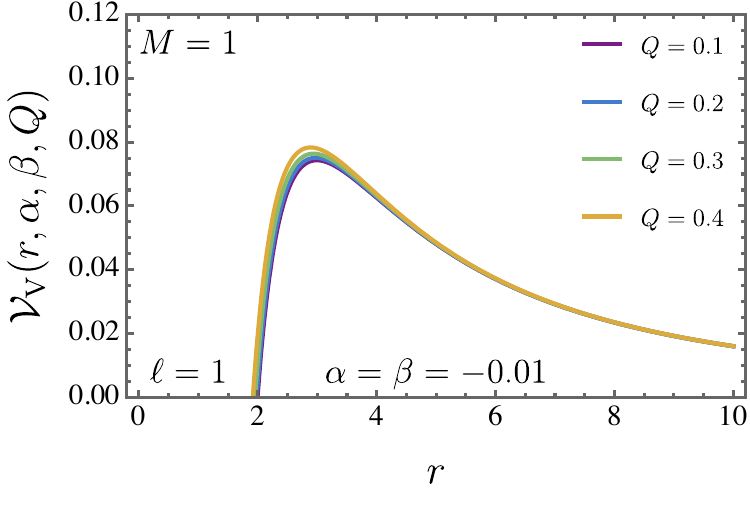}
    \caption{The variation of the vector effective potential $\mathcal{V}_{\text{V}}(r,\alpha,\beta,Q)$ with respect to the radial coordinate $r$ is presented for several charge values $Q$, considering the fixed parameters $M = 1$, $\ell = 1$, and $\alpha = \beta = -0.01$.}
    \label{vectorpotential}
\end{figure}

Figure~\ref{vectorrespondence} depicts the relationship between the quasinormal modes and greybody factors for vector perturbations, evaluated for $\alpha = \beta = -0.01$ and $\ell = 1$. As the charge $Q$ increases, the amplitude of $\bar{\Gamma}^{\text{V}}(\omega)$ declines relative to the Schwarzschild background, following a trend analogous to that of the scalar sector. This behavior agrees with the quasinormal mode spectra reported in Ref.~\cite{heidari2025gravitational}, where higher $Q$ values produce modes with greater damping. The outcome indicates that, for spin--1 perturbations, the effective potential barrier heightens with increasing $Q$, thereby enhancing wave reflection and suppressing oscillations, as illustrated in Fig.~\ref{vectorpotential}.

\subsection{Spin--$2$ particle modes}

\begin{figure}
    \centering
      \includegraphics[scale=0.6]{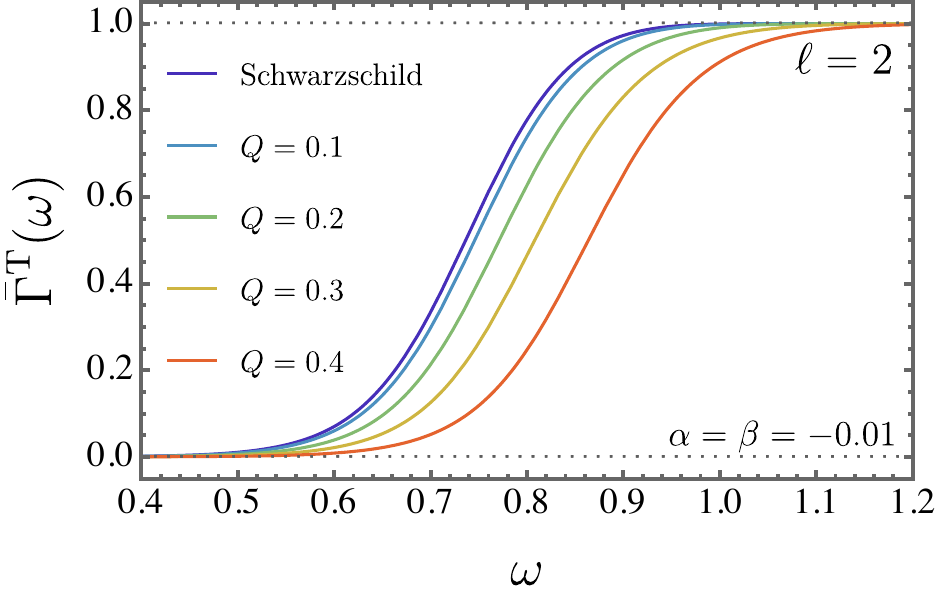}
    \caption{Relation between quasinormal modes and greybody factors for tensor perturbations, evaluated with parameters $\alpha = \beta = -0.01$ and angular momentum number $\ell = 2$.}
    \label{tensorrespondence}
\end{figure}

\begin{figure}
    \centering
      \includegraphics[scale=0.7]{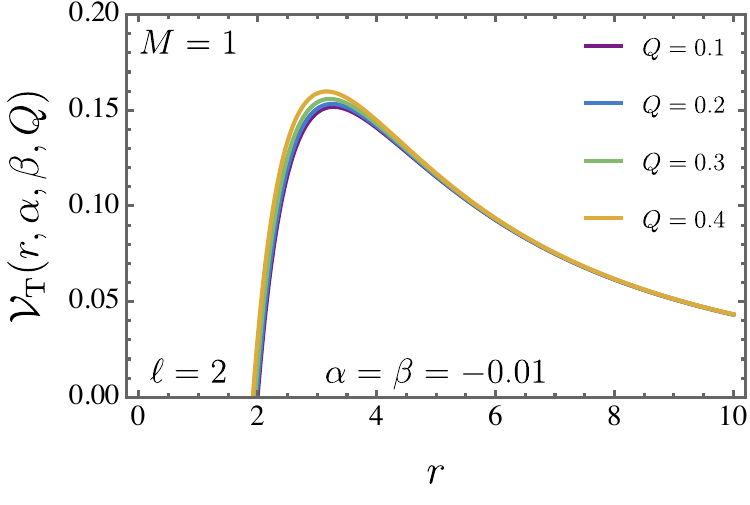}
    \caption{The dependence of the tensor effective potential $\mathcal{V}_{\text{T}}(r,\alpha,\beta,Q)$ on the radial coordinate $r$ is illustrated for several charge values $Q$, with the fixed parameters $M = 1$, $\ell = 2$, and $\alpha = \beta = -0.01$. }
    \label{tensorpotential}
\end{figure}

Figure~\ref{tensorrespondence} displays the relationship between quasinormal modes and greybody factors for tensor perturbations, taking $\alpha = \beta = -0.01$ and $\ell = 2$. As the charge $Q$ grows, the amplitude of $\bar{\Gamma}^{\text{T}}(\omega)$ diminishes relative to the Schwarzschild scenario—a tendency also observed for scalar and vector modes. This result agrees with the quasinormal mode spectrum discussed in Ref.~\cite{heidari2025gravitational}, where larger $Q$ values correspond to more heavily damped oscillations. In both analyses, the increase in $Q$ heightens the effective potential barrier (see Fig.~\ref{tensorpotential}), resulting in stronger wave reflection and greater suppression of oscillatory behavior.

\subsection{Spin--$1/2$ particle modes}

\begin{figure}
    \centering
      \includegraphics[scale=0.6]{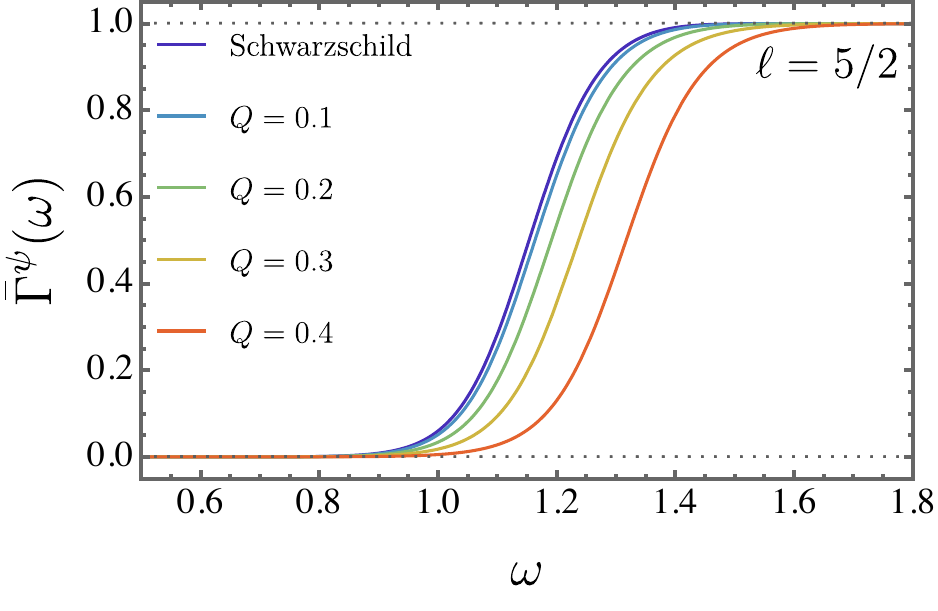}
    \caption{Connection between quasinormal modes and greybody factors for spinor perturbations, obtained for $\alpha = \beta = -0.01$ and angular momentum number $\ell = 5/2$.}
    \label{spinorialcorrespondence}
\end{figure}

\begin{figure}
    \centering
      \includegraphics[scale=0.7]{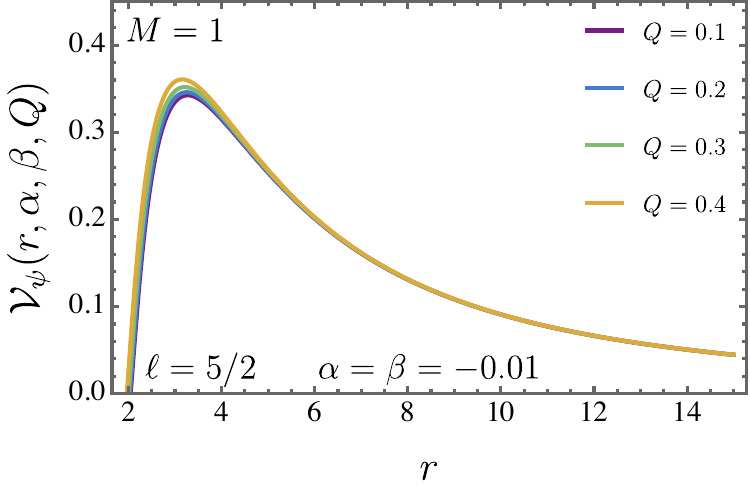}
    \caption{ The radial profile of the spinorial effective potential $\mathcal{V}_{\psi}(r,\alpha,\beta,Q)$ is plotted for several charge values $Q$, taking fixed parameters $M = 1$, $\ell = 5/2$, and $\alpha = \beta = -0.01$.}
    \label{spinorpotential}
\end{figure}

Figure~\ref{spinorialcorrespondence} illustrates the relationship between quasinormal modes and greybody factors for spinor perturbations, evaluated for $\alpha = \beta = -0.01$ and $\ell = 2$. As the charge $Q$ grows, the amplitude of $\bar{\Gamma}^{\psi}(\omega)$ diminishes in comparison with the Schwarzschild case. This trend mirrors the behavior observed for scalar, vector, and tensor modes and agrees with the quasinormal mode spectrum reported in Ref.~\cite{heidari2025gravitational}, where increasing $Q$ leads to stronger damping. In this situation, the rise in $Q$ enhances the effective potential barrier (see Fig.~\ref{spinorpotential}), thereby amplifying wave reflection and suppressing the oscillatory response.

\section{Conclusion}\label{Sec:Conclusion}

In this work, we investigated the fundamental role of particle spin in several fundamental processes related to black hole thermodynamics—namely, particle creation, greybody factors, absorption cross sections, and evaporation—within the framework of $f(R,T)$ gravity coupled to a modified electrodynamics. The analysis was motivated by the quest to understand how modifications of the underlying gravitational and electromagnetic interactions influence quantum emission phenomena near the event horizon. The study began, therefore, with a characterization of the black hole geometry derived from modified electrodynamics. Particular attention was given to the structure of the event horizon, $r_{h}$, and to the conditions ensuring its positivity, which depended explicitly on the new coupling parameters $\alpha$ and $\beta$ as well as on the electric charge $Q$. In this manner, confirming the physical consistency required by the solution with respect to these parameters was fundamental for obtaining a coherent description of the related quantum effects, since they depend on the event horizon conditions.

The particle creation mechanism was analyzed for both bosonic and fermionic fields within the modified black hole background. In the bosonic sector, massless scalar perturbations were employed to probe the spacetime geometry; by decomposing the field modes and evaluating the corresponding Bogoliubov coefficients, the Hawking temperature and the spectrum of emitted particles were obtained. Interpreting Hawking radiation as a tunneling process through Painlevé--Gullstrand coordinates enabled the calculation of the imaginary part of the classical action, $\text{Im}\,\mathcal{S}_{(\alpha,\beta)}$, via the residue method, leading to the particle creation density $n(\omega,\alpha,\beta,Q)$. A similar procedure applied to fermionic perturbations, based on the Dirac equation, yielded the fermionic creation density $n_{f}(\omega,\alpha,\beta,Q)$. In both cases, the densities were found to decrease with increasing electric charge $Q$ and coupling parameters $\alpha$ and $\beta$, indicating that stronger electromagnetic or gravitational coupling suppressed particle emission and confirming the universal pattern of this effect across different spin sectors.

Subsequently, we examined the behavior of the greybody factors $|T_b|$ associated with the various perturbative modes. For scalar and vector perturbations, exact analytical expressions for the transmission coefficients $|T_b^{\text{S}}|$ and $|T_b^{\text{V}}|$ were successfully derived. In contrast, tensor and spinorial perturbations required approximate methods due to the increasing complexity of the effective potential. In the tensor case, the potential $\mathcal{V}_{\text{T}}(r,\alpha,\beta,Q)/f(r)$ was expanded up to fourth order in the electric charge $Q$ and first order in the coupling parameter $\alpha$. For the spinorial modes, an additional expansion in the mass parameter $M$ up to first order was required to obtain tractable results. In all cases, it was observed that increasing $Q$ led to a reduction in the magnitude of the corresponding greybody factors. A comparative numerical study performed without any approximations, for fixed parameters $\alpha=\beta=-0.01$, $M=1$, $Q=0.1$, and $\ell=2$ (for bosons) and $\ell =5/2$ (for fermions), revealed the hierarchy $|T_b^{\text{T}}| > |T_b^{\text{V}}| > |T_b^{\text{S}}| > |T_b^{\psi}|$, indicating that tensor modes are transmitted more efficiently through the potential barrier than fields of lower spin.

The absorption cross sections were then analyzed for all spin sectors using the WKB approximation, and the results were found to be fully consistent with those obtained for the greybody factors. Specifically, the absorption cross sections $\sigma^{\text{S}}_{\text{abs}}$, $\sigma^{\text{V}}_{\text{abs}}$, $\sigma^{\text{T}}_{\text{abs}}$, and $\sigma^{\psi}_{\text{abs}}$ all decreased with increasing $Q$ and $\alpha=\beta$, following the hierarchy $\sigma^{\text{T}}_{\text{abs}} > \sigma^{\text{V}}_{\text{abs}}  > \sigma^{\text{S}}_{\text{abs}} > \sigma^{\psi}_{\text{abs}}$ .

Furthermore, we examined the black hole evaporation process by computing the corresponding emission rates of energy and particles for each spin sector. The rate of mass loss was estimated through the Stefan--Boltzmann relation $\mathrm{d}M/\mathrm{d}t = -a\,\bar{\Gamma}_{\ell\omega}\,\sigma_{\ell\omega}\,T^4$, where $\bar{\Gamma}_{\ell\omega}$ denotes the averaged transmission probability and $\sigma_{\ell\omega}$ the effective absorption area. The evolution of the black hole was found to culminate in a finite remnant characterized by the mass $M_{f} \approx Q/2 + \alpha(1 - 2\beta)/(20Q)$. In general, an increase in the electric charge $Q$ reduced the evaporation lifetime, although this tendency was inverted in the high--frequency regime, where larger $Q$ values led to longer--lived configurations. The comparative analysis among spin sectors showed the relation $t^{\psi}_{\text{evap-final}} > t^{\text{S}}_{\text{evap-final}}  > t^{\text{V}}_{\text{evap-final}} > t^{\text{T}}_{\text{evap-final}}$, revealing that scalar fields contribute to a slower energy loss, while tensor perturbations dominate the emission at late stages.

The emission rates of energy and particles were computed from the standard expressions $\mathrm{d}^2E/(\mathrm{d}\omega\,\mathrm{d}t) = 2\pi^2\sigma_{\ell\omega}\omega^3/[e^{\omega/T} - 1]$ and $\mathrm{d}^2N/(\mathrm{d}\omega\,\mathrm{d}t) = 2\pi^2\sigma_{\ell\omega}\omega^2/[e^{\omega/T} - 1]$. As the electric charge $Q$ increased, both the energy and particle emission rates decreased for all types of perturbations, including in the high--frequency regime. In other words, the relative intensities among spins followed the relations given by Eqs. (\ref{eall}) and (\ref{nall}), indicated that higher--spin fields radiated more efficiently than those with lower spin. One possible explanation for that laid in the distinct quantum statistics that governed their occupation probabilities. Bosons, described by Bose–Einstein statistics, could share the same quantum state, which amplified their collective emission rate. Fermions, on the other hand, obeyed the Pauli exclusion principle, which forbade multiple particles from occupying an identical state, thereby suppressing their emission. Moreover, the spin–statistics connection affected how each field coupled to the background geometry, altering the transmission probability across the potential barrier.

Finally, we explored the correlation between quasinormal modes and greybody factors across all spin sectors. The same qualitative tendency was observed: as the electric charge $Q$ increased, the amplitude and intensity of the correlated greybody factors diminished, reinforcing the suppressive influence of charge on quantum emission processes.

As a natural continuation of this work, future investigations may extend the present analysis to newly derived black hole solutions in higher--order curvature--scalar gravity models~\cite{AraujoFilho:2025vgb,Nashed:2025ebr}. These theories, which generalize the $f(R,T)$ framework by incorporating nonlinear curvature invariants and couplings between scalar and geometric fields, provide a fertile ground for exploring the quantum and thermodynamic properties of gravitating systems beyond the standard semiclassical regime. In such contexts, the inclusion of additional curvature terms or scalar degrees of freedom may significantly alter the structure of the effective potential governing perturbations, thereby modifying the associated greybody factors, absorption cross sections, and emission spectra. These and other ideas are now under development.


\section*{Acknowledgments}
\hspace{0.5cm} A.A.A.F. is supported by Conselho Nacional de Desenvolvimento Cient\'{\i}fico e Tecnol\'{o}gico (CNPq) and Fundação de Apoio à Pesquisa do Estado da Paraíba (FAPESQ), project numbers 150223/2025-0 and 1951/2025. N.H. is supported by Conselho
Nacional de Desenvolvimento Científico e Tecnológico (CNPq) with grant number 152891/2025-0. Furthermore, she would like to acknowledge the contribution of the COST Action CA21106 - COSMIC WISPers in the Dark Universe: Theory, astrophysics and experiments (CosmicWISPers), the COST Action CA21136 - Addressing observational tensions in cosmology with systematics and fundamental physics (CosmoVerse), the COST Action CA23130 - Bridging high and low energies in search of quantum gravity (BridgeQG). F.S.N.L. acknowledges support from the Fundação para a Ciência e a Tecnologia (FCT) Scientific Employment Stimulus contract with reference CEECINST/00032/2018, and funding through the research grants UIDB/04434/2020, UIDP/04434/2020 and PTDC/FIS-AST/0054/2021.

\section*{Data Availability Statement}

Data Availability Statement: No Data associated with the manuscript

\bibliographystyle{ieeetr}
\bibliography{main}

\end{document}